\documentclass[11pt,a4j]{article}
\usepackage{amsmath,amssymb}
\usepackage{bm}
\usepackage{graphicx}
\usepackage{verbatim}
\usepackage{wrapfig}
\usepackage{ascmac}
\usepackage{relsize,amsmath,amssymb}
\usepackage{subcaption}
\usepackage{float}
\usepackage{mathrsfs}
\usepackage{fancybox}
\usepackage{here}
\usepackage{cite}
\newcommand{\eps}{\epsilon}

\newcommand\la{\langle}
\newcommand\ra{\rangle}

\newcommand\beq{\begin{eqnarray}}
\newcommand\eeq{\end{eqnarray}}
\newcommand\beqs{\begin{eqnarray*}}
\newcommand\eeqs{\end{eqnarray*}}

\def\Pslash{\rlap/{\mkern-1mu P}}

\def\pslash{\rlap/{\mkern-1mu p}}

\def\kslash{\rlap/{\mkern-1mu k}}

\def\Pxslash{\rlap/{\mkern-1mu P_{\it X}}}
\def\Phslash{\rlap/{\mkern-1mu P_{h}}}

\def\Nhat{\widehat{N}}

\def\Hhat{\widehat{H}}

\def\Ghat{\widehat{G}}

\def\nn{\nonumber}
\def\del{\partial}

\def\oz{\frac{1}{z}}
\def\ozd{\frac{1}{z'}}
\def\ozo{\frac{1}{z_1}}
\def\ozt{\frac{1}{z_2}}

\def\Nhat{\widehat{N}}
\def\Ohat{\widehat{O}}

\def\Hhat{\widehat{H}}

\def\Ghat{\widehat{G}}
\def\DGhat{\Delta\widehat{G}}
\def\Hhat{\widehat{H}}

\def\3Tb{3\bar{T}}

\def\Dtilde{\widetilde{D}}

\allowdisplaybreaks[1]

\begin{document}
\begin{flushright}
\today
\end{flushright}
\vspace*{5mm}
\begin{center}
{\Large \bf Twist-3 Gluon Fragmentation Contribution to
Polarized \\[5pt]
Hyperon Production in Unpolarized Proton-Proton Collision}
\vspace{1.5cm}\\
 {\sc Yuji Koike$^1$, Kenta Yabe$^2$ and Shinsuke Yoshida$^{3,4}$}
\\[0.7cm]
\vspace*{0.1cm}
{\it $^1$ Department of Physics, Niigata University, Ikarashi, Niigata 950-2181, Japan}

\vspace{0.2cm}

{\it $^2$ Graduate School of Science and Technology, Niigata University,
Ikarashi, Niigata 950-2181, Japan}

\vspace{0.2cm}

{\it $^3$ Guangdong Provincial Key Laboratory of Nuclear Science, Institute of Quantum Matter, South
China Normal University, Guangzhou 510006, China}

\vspace{0.2cm}

{\it $^4$ Guangdong-Hong Kong Joint Laboratory of Quantum Matter, Southern Nuclear Science Computing Center, South China Normal University, Guangzhou 510006, China}
\\[3cm]

{\large \bf Abstract} \end{center}
\vspace{0.2cm}

Understanding the origin and mechanism
of the transverse polarization of hyperons produced in unpolarized proton-proton
collision, $pp\to \Lambda^\uparrow X$, 
has been one of the long-standing issues in high-energy spin physics.  
In the framework of the collinear factorization applicable to large-$p_T$ hadron
productions, this phenomenon is a twist-3 observable which is caused by multi-parton correlations
either in the initial protons or in the process of fragmentation into the hyperon.  
We derive the twist-3 gluon fragmentation function (FF) contribution to this process in the leading order
(LO) with respect to the QCD coupling constant.  
Combined with the known results for the contribution from the twist-3 distribution function 
and the twist-3 quark FF, this completes the LO twist-3 cross section.
We also found that the model independent relations among the twist-3 gluon FFs 
based on the QCD equation of motion and the Lorentz invariance property of the correlation functions
guarantee the color gauge invariance and the frame-independence of the cross section.

%
%
\newpage
\section{Introduction}
It has been known that the hyperons produced in unpolarized proton-proton
collisions are polarized perpendicularly to the scattering plane, $pp\to \Lambda^\uparrow X$
\footnote{Here $\Lambda$ collectively denotes spin-1/2 hyperons such as $\Lambda$ $\Sigma$, $\Xi$, etc.}
\cite{Bunce:1976yb,Heller:1978ty,Erhan:1979xm,Heller:1983ia,Lundberg:1989hw,
Yuldashev:1990az,Ramberg:1994tk,Fanti:1998px,Abt:2006da,Aaij:2013oxa,ATLAS:2014ona}.
The observed polarizations show a tendency that they become larger in the forward rapidity region, where
the asymmetry is as large as 30\%. 
Hyperon polarization was also observed in other reactions such as 
$\gamma p\to \Lambda^\uparrow X$\,\cite{Aston:1981em,Abe:1983jy},
quasi-real photo-production of $\Lambda$'s in lepton scattering\,\cite{Airapetian:2007mx,Airapetian:2014tyc}
and in electron-positron collisions, 
$e^+e^-\to\Lambda^\uparrow X$\,\cite{Ackerstaff:1997nh,Abdesselam:2016nym}.  
These transverse polarizations in unpolarized collisions are the 
examples of the transverse single spin asymmetries (SSAs), where only one particle 
participating in the scattering process is polarized.  
Another well known SSA is the asymmetry with regard to the
initial (transverse) spin such as $p^\uparrow p\to h X$ ($h=\pi,\,K, \eta,\,{\rm jet}$, etc.)
\cite{Adams:1991rw,Adams:1991cs,Adams:2003fx,Adamczyk:2012xd,:2008mi,Adare:2013ekj} and 
$ep^\uparrow \to ehX$\,\cite{Airapetian:2013bim,Allada:2013nsw}. 
For the last several decades, 
lots of efforts have been made to 
understand the origin and mechanism for these large SSAs,   
since perturbative QCD at twist-2 level gives almost zero SSAs\,\cite{Kane:1978nd}.  

For a high-energy collision in which particles 
with large transverse momentum are produced, the cross section 
can be computed in the framework of the collinear factorization of
perturbative QCD.  
In this framework, SSAs appear as twist-3 observables to which nonperturbative multi-parton correlation functions
contribute instead of collinear twist-2 parton distribution functions (PDFs) and/or 
fragmentation functions (FFs).  
Through the studies of SSAs, the technique of calculating the twist-3 cross section has made much progress
and has been applied to many relevant processes in the leading order (LO) with respect to the QCD coupling.  
For example, the complete LO cross section for
$p^\uparrow p\to h X$  ($h=\pi,\,D,\,\gamma,\,\gamma^*$)
has been derived
\cite{Qiu:1991wg,Qiu:1998ia,Kanazawa:2000hz,Ji:2006vf,Kouvaris:2006zy,Koike:2007rq,Koike:2009ge,
Koike:2011b,Koike:2011nx,Kang:2010zzb,Kanazawa:2011er,Metz:2012ct,Beppu:2013uda,Kanazawa:2014nea}, 
and the RHIC data has been analyzed and interpreted, 
which suggests the main source of the asymmetry is the twist-3
fragmentation contribution\,\cite{Kanazawa:2014dca,Gamberg:2017gle}.

In this paper we study
$pp\to\Lambda^\uparrow X$ in the collinear twist-3 factorization.  
Two kinds of twist-3 cross section contribute to this process:  
(i) Twist-3 unpolarized PDF in one of the initial proton convoluted with
the twist-2 ``transversity'' FF for the final hyperon and
the twist-2 unpolarized PDF in another proton, and
(ii) Twist-3 FFs for the polarized hyperon convoluted with the twist-2 unpolarized PDFs
in the initial protons.
The complete LO cross section for (i) was derived in 
\cite{Kanazawa:2001a,Zhou:2008,Koike:2015zya}.
The second one (ii) can be further classified into two, depending on whether 
the twist-3 FF is of (ii-a) quark-gluon correlation type or of (ii-b) gluon correlation type.
The complete LO cross section for (ii-a) was derived in \cite{Koike:2017fxr}, 
while (ii-b) has not been studied so far. 
In this paper we focus on the (ii-b) contribution, and derive the
corresponding cross section, which completes the LO twist-3 cross section for
this process.  (A short version of the present work was
presented in \cite{Yabe:2019awq,Kenta:2019bxd}.)
We develop a formalism for deriving the gauge and frame independent
contribution to this twist-3 cross section
from the purely gluonic FFs, and present the result for $pp\to \Lambda^\uparrow X$.

The remainder of this paper is organized as follows.
In section 2, we introduce the complete set of the gluonic 
FFs for spin-1/2 hadron up to twist-3
defined from correlators of two- and three- gluon field strengths, which are
necessary to derive the twist-3 cross section.  
We also recall from \cite{Koike:2019zxc} 
the exact relations among those FFs based on the QCD equation of motion 
and the
Lorentz invariance, 
which play a crucial
role to guarantee the gauge and Lorentz invariance of the cross 
section\footnote{Importance of these relations for the frame independence of
the twist-3 cross sections have been realized for the twist-3 quark distribution 
functions and FFs\cite{Koike:2017fxr,Kanazawa:2014tda,
Kanazawa:2015jxa,Kanazawa:2015ajw}.}.
In section 3, we develop a formalism to derive the twist-3
gluon FF contribution 
to the twist-3 cross section, and present the corresponding LO cross section 
for $pp\to \Lambda^\uparrow X$.  We will discuss how the Lorentz invariance of
the twist-3 cross section is realized, using the relations introduced in section 2.  
Gauge invariance of the cross section is discussed in Appendix B.  
Section 4 is devoted to a brief summary.  
In other Appendices, we discuss some technical aspects of the actual calculations.


\section{Gluon Fragmentation Functions}

\subsection{Three types of twist-3 gluon fragmentation functions}
In this section we introduce twist-3
gluon FFs for a spin-1/2 baryon
relevant to 
$pp\to\Lambda^\uparrow X$\,\cite{Mulders:2000sh,Gamberg:2018fwy,Koike:2019zxc} 
and summarize their basic properties
derived in \cite{Koike:2019zxc}.  
They are classified into three types; {\it intrinsic}, {\it kinematical} and {\it dynamical} FFs.  
The
{\it intrinsic} twist-3 gluon FFs 
are defined as the Fourier transform of the lightcone correlator of the gluon's 
field strength
$F^{\mu\nu}_a$:
\begin{eqnarray}
&&
\hat{\Gamma}^{\alpha\beta}(z)\nonumber\\
&&={1 \over {N^2-1}}\sum_X\!\int \!\frac{d\lambda}{2\pi} 
e^{-i{\lambda \over z}}\la 0|\left([\infty w,0]F^{w\beta}(0)\right)_a| h(P_h,S_h)X\ra
\la h(P_h,S_h)X|\left(F^{w\alpha}(\lambda w)[\lambda w,\infty w]\right)_a| 0\ra \nn\\
&&
=-g_{\perp}^{\alpha\beta}\widehat{G}(z)-iM_h\eps^{P_h w\alpha\beta}(S_h\cdot w)\Delta\widehat{G}(z)
-iM_h\eps^{P_h w S_{\perp} [\alpha}w^{\beta]}\Delta\widehat{G}_{3T}(z)
+M_h\eps^{P_h w S_{\perp}\{\alpha}w^{\beta\}}\Delta\widehat{G}_{3\bar{T}}(z),\nn\\
\label{gFraI}
\end{eqnarray}
where $N=3$ is the number of colors, 
$|h(P_h,S_h)\ra$ is the baryon state with the four momentum $P_h$  ($P_h^2=M_h^2$)
and the spin vector $S_h$ ($S_h^2=-M_h^2$), and $[\lambda w, \infty w]$ is the gauge link
in the adjoint representation connecting $\lambda w$ and $\infty w$. 
For the transversely polarized baryon,
we use the spin vector $S_\perp$ normalized as $S_\perp^2=-1$.  
In the twist-3 accuracy
$P_h$ can be regarded as lightlike.  For a baryon with large momentum, 
$P_h\simeq (|\vec{P}_h|, \vec{P}_h)$,
another lightlike vector $w$ is defined as $w=1/(2|\vec{P}_h|^2)(|\vec{P}_h|, -\vec{P}_h)$ 
which satisfies $P_h\cdot w=1$.  
In (\ref{gFraI}), we use the notation $F^{w \beta}\equiv F^{\mu\beta}w_\mu$,
and $\{\ \ \}$ ($[\ ]$) implies the symmetrization (anti-symmetrization) of Lorentz indices, i.e. for
arbitrary four vectors $a^\alpha$ and
$b^\beta$,  
$a^{\{\alpha} b^{\beta\}}\equiv a^\alpha b^\beta + a^\beta b^\alpha$ and
$a^{[\alpha} b^{\beta]}\equiv a^\alpha b^\beta - a^\beta b^\alpha$.
$\widehat{G}(z)$ and  $\Delta \widehat{G}(x)$ are twist-2, and 
$\Delta \widehat{G}_{3T}(z)$ and $\Delta \widehat{G}_{3\bar{T}}(z)$ are twist-3.  
We also note 
$\Delta \widehat{G}_{3\bar{T}}(z)$ is naively $T$-odd, contributing to SSAs.  
Each function in (\ref{gFraI}) has a support on $0<z<1$.

The {\it kinematical} FFs contain the transverse derivative of the
correlation functions of the field strengths: 
\begin{eqnarray}
&&\hat{\Gamma}_{\del}^{\alpha\beta\gamma}(z)
=\frac{1}{N^2-1}\sum_X\!\int \!\!\frac{d\lambda}{2\pi}\!e^{-i{\lambda\over z}}\la 0|
\left([\infty w,0]
F^{w\beta}(0)\right)_a| h(P_h,S_\perp)X\ra\nn\\
&&\qquad\qquad\qquad\qquad\qquad
\times\la h(P_h,S_\perp)X|\left(F^{w\alpha}(\lambda w)
[\lambda w,\infty w]\right)_a| 0\ra\overleftarrow{\del}^\gamma\nn\\
&&\qquad\qquad= -i\frac{M_h}{2}g_{\perp}^{\alpha\beta}\eps^{P_h w S_\perp \gamma}\hat{G}_T^{(1)}(z)
+\frac{M_h}{2}\eps^{P_h w\alpha\beta}S_\perp^\gamma\Delta\hat{G}_T^{(1)}(z)\nn\\
&&\qquad\qquad\qquad\qquad
-i\frac{M_h}{8}\left(\eps^{P_h w S_\perp\{\alpha}g_{\perp}^{\beta\}\gamma}
+\eps^{P_h w\gamma\{\alpha}S_\perp^{\beta\}}\right)\Delta\hat{H}_T^{(1)}(z),
\label{gFraK}
\end{eqnarray}
where each function is defined to be real.   The kinematical FFs are related to the $k_T^2/M_h^2$-moment of the
transverse-momentum-dependent (TMD) FFs~\cite{Mulders:2000sh}.
Each function in (\ref{gFraK}) has a support on $0<z<1$.   


To define the {\it dynamical} FFs, we introduce
the lightcone correlation functions of three field strengths:
\begin{eqnarray}
&&\hat{\Gamma}_{F,abc}^{\alpha\beta\gamma}(\ozo,\ozt)\nn\\ 
&&=\frac{1}{N^2-1}\sum_X\!\int \!\!\frac{d\lambda}{2\pi}\! \int \!\!\frac{d\mu}{2\pi}
 e^{-i{\lambda\over z_1}}e^{-i\mu({1\over z_2}-{1\over z_1})}\la 0|F^{w\beta}_b(0)| h(P_h,S_\perp)X
\ra\la h(P_h,S_\perp)X|F^{w\alpha}_a(\lambda w)
 gF^{w\gamma}_c(\mu w)| 0\ra,
\label{gFraD}\nn\\
\end{eqnarray}
where the gauge link operators are suppressed for simplicity. 
The color indices of this correlator can be expanded in terms of the anti-symmetric and symmetric 
structure constants of color SU(N), $-if^{abc}$ and $d^{abc}$, as
\begin{eqnarray}
\hat{\Gamma}_{F,abc}^{\alpha\beta\gamma}(\ozo,\ozt)=-\frac{if^{abc}}{N}
\hat{\Gamma}_{FA}^{\alpha\beta\gamma}(\ozo,\ozt)
+d^{abc}\frac{N}{N^2-4}
\hat{\Gamma}_{FS}^{\alpha\beta\gamma}(\ozo,\ozt).
\label{gFraDexpand}
\end{eqnarray}
The dynamical FFs can be defined from 
$\hat{\Gamma}_{FA}^{\alpha\beta\gamma}\left(\ozd,\oz\right)$
and $\hat{\Gamma}_{FS}^{\alpha\beta\gamma}\left(\ozd,\oz\right)$:
\begin{eqnarray}
&&\hat{\Gamma}_{FA}^{\alpha\beta\gamma}(\ozo,\ozt)\nn\\ 
&&=\frac{-if_{abc}}{N^2-1}\sum_X\!\int \!\!\frac{d\lambda}{2\pi}\! \int \!\!\frac{d\mu}{2\pi}
 e^{-i{\lambda\over z_1}}e^{-i\mu({1\over z_2}-{1\over z_1})}\la 0|F^{w\beta}_b(0)| h(P_h,S_\perp)X
\ra\la h(P_h,S_\perp)X|F^{w\alpha}_a(\lambda w)
 gF^{w\gamma}_c(\mu w)| 0\ra\nn\\ 
&&=-{M_h}\biggl(\Nhat_{1}\left(\ozo,\ozt\right)g^{\alpha\gamma}_{\perp}
\eps^{P_h w S_\perp\beta}\hspace{-0.05cm}+\Nhat_{2}\left(\ozo,\ozt\right)g^{\beta\gamma}_{\perp}
\eps^{P_h w S_\perp\alpha}\hspace{-0.05cm}-\hspace{-0.05cm}
\Nhat_{2}\left(\ozt-\ozo,\ozt\right)g^{\alpha\beta}_{\perp}\eps^{P_h w S_\perp\gamma} \biggr),
\label{gFraDA}\nn\\
\end{eqnarray}
\begin{eqnarray}
&&\hat{\Gamma}_{FS}^{\alpha\beta\gamma}(\ozo,\ozt)\nn\\ 
&&=\frac{d_{abc}}{N^2-1}\sum_X
\!\int \!\!\frac{d\lambda}{2\pi}\! \int \!\!\frac{d\mu}{2\pi} e^{-i{\lambda\over z_1}}
e^{-i\mu({1\over z_2}-{1\over z_1})}\la 0|F^{w\beta}_b(0)| h(P_h,S_\perp)
X\ra\la h(P_h,S_\perp)X|F^{w\alpha}_a(\lambda w)gF^{w\gamma}_c(\mu w)| 0\ra\nn\\
&&=-{M_h}\biggl(\Ohat_{1}\left(\ozo,\ozt\right)g^{\alpha\gamma}_{\perp}
\eps^{P_h w S_\perp\beta}\hspace{-0.05cm}+\Ohat_{2}\left(\ozo,\ozt\right)g^{\beta\gamma}_{\perp}
\eps^{P_h w S_\perp\alpha}\hspace{-0.05cm}+
\Ohat_{2}\left(\ozt-\ozo,\ozt\right)g^{\alpha\beta}_{\perp}\eps^{P_h w S_\perp\gamma} \biggr).\nn\\
\label{gFraDS}
\end{eqnarray}
Correlation functions (\ref{gFraDA}) and (\ref{gFraDS}), respectively, define
two independent set of the {\it complex} functions 
$\left\{\Nhat_{1},
\Nhat_{2}
\right\}$
and $\left\{\Ohat_{1}
, \Ohat_{2}
\right\}$ due to the exchange symmetry of the field strengths.  
Functions $\Nhat_1$ and $\Ohat_1$
satisfy the relations
\beq
\Nhat_{1}\left(\ozo,\ozt\right)&=&-\Nhat_{1}\left(\ozt-\ozo,\ozt\right),\nn\\
\Ohat_{1}\left(\ozo,\ozt\right)&=&\Ohat_{1}\left(\ozt-\ozo,\ozt\right).  
\label{swi_nat}
\eeq
The real parts of these four FFs are $T$-even and the imaginary parts are $T$-odd, 
the latter being the sources of SSAs.  
$\Nhat_{1,2}\left(\ozo,\ozt\right)$ and $\Ohat_{1,2}\left(\ozo,\ozt\right)$
have a support on $\ozt>1$ and $\ozt>\ozo>0$.  

For the derivation of the twist-3 gluon FF contribution to $pp\to\Lambda^\uparrow  X$,
one also needs another dynamical FF,
\beq
\widetilde{\Delta}_{ij}^\alpha\left({1\over z_1},{1\over z_2}\right)
&=&{1\over N}\sum_X\int{d\lambda\over 2\pi}\int{d\mu\over 2\pi}e^{-i{\lambda\over z_1}}
e^{-i\mu({1\over z_2}-{1\over z_1})}
\la 0| F_a^{w\alpha}(\mu w)|hX\ra\la hX|\bar{\psi}_j(\lambda w)t^a\psi_i(0)| 
0\ra\nonumber\\
&=&M_h\left[ e^{\alpha P_hwS_\perp}(\Pslash_h)_{ij} \widetilde{D}_{FT}
\left({1\over z_1},{1\over z_2}\right)+
iS_\perp^\alpha \left(\gamma_5\Pslash_h\right)_{ij}\widetilde{G}_{FT}
\left({1\over z_1},{1\over z_2}\right)
\right], 
\label{FFtilde}
\eeq
where we explicitly wrote spinor indices $i,\, j$.  
$\widetilde{D}_{FT}$ and $\widetilde{G}_{FT}$ are also complex functions, with the $T$-even real part and 
the $T$-odd  imaginary part.  They have a support on $\ozo>0$, $\ozt<0$ and $\ozo-\ozt>1$.  
As was shown in 
\cite{Koike:2019zxc}, the constraint relations for the twist-3 gluon FFs involve
these quark-gluon correlation functions.
We collectively call the functions in (\ref{gFraDA}),
(\ref{gFraDS}) and (\ref{FFtilde}) {\it dynamical} twist-3 FFs.

\subsection{Constraint relations among the twist-3 gluon FFs }\label{sec:gRel}
Three types of the twist-3 gluon FFs defined in the previous subsection are not
independent from each other, but are related by the
QCD equation-of-motion (EOM relations) and the Lorentz invariance properties
of the correlation functions (LIRs).  
In \cite{Koike:2019zxc} the complete set of those relations have been derived.   
Here we recall those relations, 
which, as we will see in the next section, play a crucial role to guarantee
the Lorentz- and gauge- invariance of the twist-3 cross sections for $pp\to\Lambda^\uparrow X$.  
Here we quote from \cite{Koike:2019zxc} the relevant relations.  

First the intrinsic FFs $\DGhat_{\3Tb}(z)$ can be written in terms of the kinematical FFs and 
the dynamical ones as (See eq. (50) of \cite{Koike:2019zxc}):
\beq
&&{1\over z}\DGhat_{\3Tb}(z)
= - \Im \Dtilde_{FT}(z)
\nonumber\\
&&\qquad+\int_0^{1/z} d\left({1\over z'}\right)  
\left( {1\over {1\over z}- {1\over z'}}\right)\Im
\left\{ 
2\Nhat_{1} \left( {1\over z'},{1\over z}\right)+ \Nhat_{2}\left({1\over z'},{1\over z}\right) 
-\Nhat_{2}\left( {1\over z}-{1\over z'},{1\over z}\right)
\right\}\nonumber\\
&&\qquad+{1\over 2}\left( \Ghat_T^{(1)}(z) + \Delta\Hhat_T^{(1)}(z)\right), 
\label{FFDFodd1}
\eeq
where $\Dtilde_{FT}(z)$ is defined as
\beq
\Dtilde_{FT}(z)\equiv {2\over C_F}\int_0^{1/z}d\left({1\over z_1}\right)
\Dtilde_{FT}\left({1\over z_1},{1\over z_1}-{1\over z}\right), 
\label{DtildeFT}
\eeq
with
$C_F={N^2-1\over 2N}$ and $\Im\Dtilde_{FT}$ indicates the imaginary part of
$\Dtilde_{FT}$.  
The kinematical FFs can be expressed in terms of the dynamical ones as (see eqs. (74) and (75)
of \cite{Koike:2019zxc}): 
\beq
&&\widehat{G}_T^{(1)}(z)=
-{2\over z^2}\int_1^{1/z}d\left({1\over z_2}\right)z_2^3\,\Im \Dtilde_{FT}(z_2)\nonumber\\
&&\quad+{4\over z^2}\int_1^{1/z}d\left({1\over z_2}\right)z_2^3
\int_0^{1/z_2}d\left({1\over z_1}\right){1\over 1/z_2-1/z_1}\Im\left[ 
\Nhat_1\left({1\over z_1},{1\over z_2}\right) - \Nhat_2\left( {1\over z_2}-{1\over z_1} ,{1\over z_2}\right) 
\right]\nonumber\\
&&\quad+{2\over z^2}\int_1^{1/z}d\left({1\over z_2}\right)z_2^2 
\int_0^{1/z_2}d\left({1\over z_1}\right){1\over \left(1/z_2-1/z_1\right)^2}\Im\left[ 
\Nhat_1\left({1\over z_1},{1\over z_2}\right) + \Nhat_2\left({1\over z_1},{1\over z_2}\right) \right.\nonumber\\
&&\left.\hspace{9cm}-2  \Nhat_2\left( {1\over z_2}-{1\over z_1} ,{1\over z_2}\right) 
\right],
\label{kinG(1)}
\eeq
and
\beq
&&\Delta\widehat{H}_T^{(1)}(z)=
-{4\over z^4}\int_1^{1/z}d\left({1\over z_2}\right)z_2^5\,\Im \Dtilde_{FT}(z_2)\nonumber\\
&&\quad+{8\over z^4}\int_1^{1/z}d\left({1\over z_2}\right)z_2^5\int_0^{1/z_2}
d\left({1\over z_1}\right){1\over 1/z_2-1/z_1}\Im\left[ 
\Nhat_1\left({1\over z_1},{1\over z_2}\right) + \Nhat_2\left({1\over z_1},{1\over z_2}\right) 
\right]\nonumber\\
&&\quad+{4\over z^4}\int_1^{1/z}d\left({1\over z_2}\right)z_2^4 \int_0^{1/z_2}
d\left({1\over z_1}\right){1\over \left(1/z_2-1/z_1\right)^2}\Im\left[ 
\Nhat_1\left({1\over z_1},{1\over z_2}\right) + \Nhat_2\left({1\over z_1},{1\over z_2}\right) 
\right].  
\label{kinH(1)}
\eeq
From (\ref{FFDFodd1}), (\ref{kinG(1)}) and (\ref{kinH(1)}), the intrinsic
FF $\DGhat_{\3Tb}(z)$ is also written by
the dynamical ones.  
For the derivation of the twist-3 cross section for $pp\to \Lambda^\uparrow X$,
one needs derivatives of the kinematic FFs.  From (\ref{kinG(1)}) and (\ref{kinH(1)}), we can
obtain those derivatives in terms of the kinematical FFs themselves and the dynamical ones as
\beq
\frac{1}{z}\frac{\del}{\del(1/z)}[ \Ghat_T^{(1)}(z)]&=&
-2\,\Im \Dtilde_{FT}(z)
+2\Ghat_T^{(1)}(z)\nn\\
&+&
4\int_0^{1/z} d\left({1\over z'}\right) {1\over {{1\over z}-{1\over z'}}}\Im
\left\{ 
\Nhat_{1} \left( {1\over z'},{1\over z}\right)
-\Nhat_{2}\left( {1\over z}-{1\over z'},{1\over z}\right)
\right\}\nn\\
&+&
{2\over z}\int_0^{1/z} d\left({1\over z'}\right) {1\over \left({1\over z}-{1\over z'}\right)^2}\Im
\left\{ 
\Nhat_{1} \left( {1\over z'},{1\over z}\right)
+\Nhat_{2}\left({1\over z'},{1\over z}\right) 
-2\Nhat_{2}\left( {1\over z}-{1\over z'},{1\over z}\right)
\right\},\nn\\
\label{FFrel_G}
\eeq
and
\beq
\frac{1}{z}\frac{\del}{\del(1/z)}[\Delta\Hhat_T^{(1)}(z)]&=&
-4\,\Im \Dtilde_{FT}(z)
+4\Delta\Hhat_T^{(1)}(z)\nn\\
&+&
8\int_0^{1/z} d\left({1\over z'}\right) {1\over {{1\over z}-{1\over z'}}}\Im
\left\{ 
\Nhat_{1} \left( {1\over z'},{1\over z}\right)
+\Nhat_{2}\left({1\over z'},{1\over z}\right) 
\right\}\nn\\
&+&
{4\over z}\int_0^{1/z} d\left({1\over z'}\right) {1\over \left({1\over z}-{1\over z'}\right)^2}\Im
\left\{ 
\Nhat_{1} \left( {1\over z'},{1\over z}\right)
+\Nhat_{2}\left({1\over z'},{1\over z}\right) 
\right\}.
\label{FFrel_H}
\eeq
In the next section, we will use (\ref{FFDFodd1}), (\ref{FFrel_G}) and (\ref{FFrel_H})
to write the twist-3 gluon FF contribution to $pp\to\Lambda^\uparrow X$
in a frame-independent form.

\section{Twist-3 gluon fragmentation contribution to $pp\to\Lambda^\uparrow X$}

\begin{figure}[b]
\begin{center}
  \includegraphics[width=14cm]{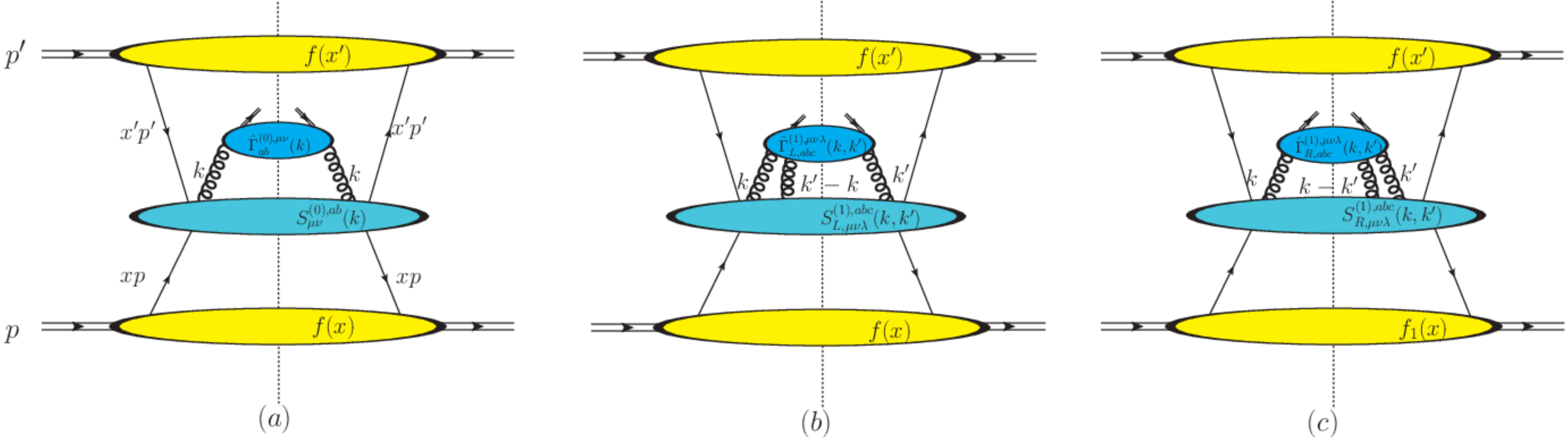}\hspace{1cm}
\end{center}
 \caption{Diagrams representing the twist-3 gluon fragmentation contribution to 
 $pp\to\Lambda^\uparrow X$.}
\label{fig3}
\end{figure}

In this section, we develop a formalism for calculating the twist-3 gluon fragmentation contribution to
\begin{eqnarray}
p(p)+p(p')\to\Lambda^\uparrow(P_h,S_\perp)+X,
\label{pplambdax}
\end{eqnarray}
where $p$, $p'$ and $P_h$ are the momenta of the particles and $S_\perp$ is the transverse 
spin vector of $\Lambda^\uparrow$.  
We work in Feynman gauge so that one can check the appearance of the 
gauge invariant FFs explicitly.  
As in the case of the twist-3 quark FF\,\cite{Koike:2017fxr,Kanazawa:2013uia}, 
naively $T$-odd FFs give rise to
the cross section as a nonpole contribution.  
The twist-3 FF contribution to (\ref{pplambdax}) can be written as
\begin{eqnarray}
E_{h}\frac{d\sigma(p,p',P_h;S_\perp)}{d^3P_h}=\frac{1}{16\pi^2S_{E}}\!\int\!\!
\frac{dx}{x}f_1(x)\!\int\frac{dx'}{x'}f_1(x')
W_{q,g}(xp,x'p',P_h/z,S_\perp),
\label{s21}
\end{eqnarray}
where $S_{E}=(p+p')^2$ is the center-of-mass energy squared, 
$E_h=\sqrt{M_h^2+\vec{P}_h^2}$ is the energy of the hyperon, 
$x,\, x'$ are the momentum fractions of the partons coming out of
the initial nucleons and $f_1(x)$ is the unpolarized quark or gluon distributions 
in the nucleon.  
$W_{q}$ ($W_g$) is the hadronic tensor representing the partonic hard scattering
followed by the fragmentation of a quark (gluon) into the final $\Lambda^\uparrow$. 
In (\ref{s21}), summation over all possible channels is implied.   
The LO cross section for $W_q$ 
was derived in \cite{Koike:2017fxr}.  
Here we focus on 
the twist-3  gluon fragmentation contribution $W_g$ in (\ref{s21}) 
which is diagrammatically shown in Fig. \ref{fig3}.  
$W_g$ consists of three terms corresponding to Figs. \ref{fig3} (a), (b), (c): 
\begin{eqnarray}
\hspace{-1cm}&&W_{g}(xp,x'p',P_h;S_\perp) \equiv  W^{(a)}_g+W^{(b)}_g+W^{(c)}_g\nn\\
&&\qquad=
\int\frac{d^4k}{(2\pi)^4}
\left[
\Gamma^{(0),\mu\nu}_{ab}(k)
S^{(0),ab}_{\mu\nu}(k)
\right]\nn\\
&&\qquad+
{1\over 2} \int\frac{d^4k}{(2\pi)^4}\int\frac{d^4k'}{(2\pi)^4}
\biggl[
\Gamma^{(1),\mu\nu\lambda}_{L,abc}(k,k')
S^{(1),abc}_{L,\mu\nu\lambda}(k,k')+
\Gamma^{(1),\mu\nu\lambda}_{R,abc}(k,k')
S^{(1),abc}_{R,\mu\nu\lambda}(k,k')
\biggr],
\label{s97}
\end{eqnarray}
where the factor $1/2$ in front of
$W_g^{(b)}$ and $W_g^{(c)}$ takes into account
the exchange symmetry of the gluon fields in the fragmentation matrix elements.  
The four momenta
$k$ and  $k'$are those of the partons fragmenting into the final $\Lambda$.   
$S^{(0),ab}_{\mu\nu}(k)$, $S^{(1),abc}_{L,\mu\nu\lambda}(k,k')$ and $S^{(1),abc}_{R,\mu\nu\lambda}(k,k')$
are the partonic hard scattering parts, and 
$\hat{\Gamma}^{(0),\mu\nu}_{ab}(k)$, 
$\hat{\Gamma}^{(1),\mu\nu\lambda}_{L,abc}(k,k')$ and $\hat{\Gamma}^{(1),\mu\nu\lambda}_{R,abc}(k,k')$
are the corresponding hadronic matrix elements representing fragmentation 
of partons into $h$ ($h=\Lambda^\uparrow$).  
The upper indices $(0)$ and $(1)$ represent the number of extra gluon lines
compared with the lowest order
gluon fragmentation contribution to the cross section.   
Hadronic matrix elements are defined as 
\begin{eqnarray}
&&\Gamma^{(0),\mu\nu}_{ab}(k)=\sum_X\int{d^4\xi}\,{e^{-ik\xi}}
{\la0|}A^\nu_b(0){|hX 
\ra}{\la hX|}A^\mu_a(\xi){|0\ra},
\\[0.4cm]
&&\Gamma^{(1),\mu\nu\lambda}_{L,abc}(k,k')=\sum_X\int{d^4\xi}\int{d^4\eta}\,{e^{-ik\xi}}{e^{-i(k'-k)\eta}}
{\la0|}A^\nu_b(0){|hX 
\ra}{\la hX|}A^\mu_a(\xi)gA^\lambda_c(\eta){|0\ra},
\\[0.4cm]
&&\Gamma^{(1),\mu\nu\lambda}_{R,abc}(k,k')=\sum_X\int{d^4\xi}\int{d^4\eta}\,{e^{-ik\xi}}{e^{-i(k'-k)\eta}}
{\la0|}A^\nu_b(0)gA^\lambda_c(\eta){|hX 
\ra}{\la hX|}A^\mu_a(\xi){|0\ra},
\end{eqnarray}
where the gauge coupling $g$ associated with the attachment of 
the extra gluon line to the hard part
is included in 
$\hat{\Gamma}^{(1)}_L(k,k')$ and $\hat{\Gamma}^{(1)}_{R}(k,k')$.  Therefore
the hard parts $S^{(0)}$, $S_L^{(1)}$ and $S_R^{(1)}$ are of $O(g^4)$ in the LO
calculation.    
From hermiticity, one has $\hat{\Gamma}^{(1),\mu\nu\lambda}_{L,abc}(k,k') ^\star
=\hat{\Gamma}^{(1)\nu\mu\lambda}_{R,bac}(k',k)$ and
$S^{(1),abc}_{L,\mu\nu\lambda}(k,k')^\star = S^{(1),bac}_{R,\nu\mu\lambda}(k',k)$,
which guarantees $W_g$ in (\ref{s97}) is real.  
One can extract the twist-3 effect from (\ref{s97}) by applying the 
collinear expansion.  
The collinear expansion of 
$S^{(0)}$, $S_L^{(1)}$ and $S_R^{(1)}$ 
with respect to the parton momenta $k^\mu$ and $k'^\mu$ around the parent hadron's 
momentum $P_h^\mu$
reads
\begin{eqnarray}
S^{(0),ab}_{\mu\nu}(k)
&=&S^{(0),ab}_{\mu\nu}(z)
+\left.\frac{\del S^{(0),ab}_{\mu\nu}(k)}{\del k^\alpha}\right|_{c.l.}\Omega^\alpha_{\ \sigma} k^\sigma
\nn\\
&+&\frac{1}{2}\left.\frac{\del^2 S^{(0),ab}_{\mu\nu}(k)}{\del k^\alpha\del k^\beta}\right|_{c.l.}
\Omega^\alpha_{\ \sigma} k^\sigma\Omega^\beta_{\ \rho} k^\rho
+\frac{1}{6}\left.\frac{\del^3 S^{(0),ab}_{\mu\nu}(k)}{\del k^\alpha\del k^\beta\del k^\gamma}
\right|_{c.l.}\Omega^\alpha_{\ \sigma} k^\sigma\Omega^\beta_{\ \rho} 
k^\rho\Omega^\gamma_{\ \tau} k^\tau+\cdots, 
\end{eqnarray}
and
\begin{eqnarray}
&&
S^{(1),abc}_{L,\mu\nu\lambda}(k,k')
=S^{(1),abc}_{L,\mu\nu\lambda}(z,z')
+\left.\frac{\del S^{(1),abc}_{L,\mu\nu\lambda}(k,k')}{\del k^\alpha}\right|_{c.l.}
\Omega^\alpha_{\ \sigma} k^\sigma
+\left.\frac{\del S^{(1),abc}_{L,\mu\nu\lambda}(k,k')}{\del k'^\alpha}\right|_{c.l.}
\Omega^\alpha_{\ \sigma} k'^\sigma
\nn\\
&&+\frac{1}{2}\left.\frac{\del^2 S^{(1),abc}_{L,\mu\nu\lambda}(k,k')}{\del k^\alpha\del k^\beta}\right|_{c.l.}\Omega^\alpha_{\ \sigma} k^\sigma\Omega^\beta_{\ \rho} k^\rho
+\frac{1}{2}\left.\frac{\del^2 S^{(1),abc}_{L,\mu\nu\lambda}(k,k')}{\del k'^\alpha\del k'^\beta}\right|_{c.l.}
\Omega^\alpha_{\ \sigma} k'^\sigma\Omega^\beta_{\ \rho} k'^\rho
\nn\\
&&+\left.\frac{\del^2 S^{(1),abc}_{L,\mu\nu\lambda}(k,k')}{\del k^\alpha\del k'^\beta}\right|_{c.l.}
\Omega^\alpha_{\ \sigma} k^\sigma\Omega^\beta_{\ \rho} k'^\rho
\nn\\
&&+\frac{1}{6}\left.\frac{\del^3 S^{(1),abc}_{L,\mu\nu\lambda}(k,k')}{\del k^\alpha\del k^\beta\del k^\gamma}\right|_{c.l.}
\Omega^\alpha_{\ \sigma} k^\sigma\Omega^\beta_{\ \rho} k^\rho\Omega^\gamma_{\ \tau} k^\tau
+\frac{1}{6}\left.\frac{\del^3 S^{(1),abc}_{L,\mu\nu\lambda}(k,k')}{\del k'^\alpha\del k'^\beta\del k'^\gamma}\right|_{c.l.}
\Omega^\alpha_{\ \sigma} k'^\sigma\Omega^\beta_{\ \rho} k'^\rho\Omega^\gamma_{\ \tau} k'^\tau
\nn\\
&&+\frac{1}{2}\left.\frac{\del^3 S^{(1),abc}_{L,\mu\nu\lambda}(k,k')}{\del k^\alpha\del k^\beta\del k'^\gamma}\right|_{c.l.}
\Omega^\alpha_{\ \sigma} k^\sigma\Omega^\beta_{\ \rho} k^\rho\Omega^\gamma_{\ \tau} k'^\tau
+\frac{1}{2}\left.\frac{\del^3 S^{(1),abc}_{L,\mu\nu\lambda}(k,k')}{\del k^\alpha\del k'^\beta\del k'^\gamma}\right|_{c.l.}
\Omega^\alpha_{\ \sigma} k^\sigma\Omega^\beta_{\ \rho} k'^\rho\Omega^\gamma_{\ \tau} k'^\tau\
+\cdots,\nn\\
\end{eqnarray}
where $\Omega^\alpha_{\ \mu}\equiv g^\alpha_{\ \mu}-P_h^\alpha w_{\mu}$, 
$S^{(0),ab}_{\mu\nu}(z)\equiv S^{(0),ab}_{\mu\nu}(P_h/z)$, 
$S^{(1),abc}_{L,\mu\nu\lambda}(z,z')\equiv S^{(1),abc}_{L,\mu\nu\lambda}(P_h/z,P_h/z')$,  
``$|_{c.l.}$'' indicates taking the collinear limit, i.e., $k\to P_h/z$ and $k'\to P_h/z'$ and
$\cdots$ denotes the contribution of
twist-4 or higher.   The collinear expansion for $S_R^{(1)}$ can be performed similarly.  
We also decompose the gauge field $A^\mu$ as
\begin{eqnarray}
A^\mu=(A\cdot w)P_h^\mu+\Omega^\mu_{\ \nu} A^\nu.  
\end{eqnarray}
Inserting this decomposition into the expression for 
$\hat{\Gamma}^{(0)}$, $\hat{\Gamma}^{(1),}_{L}$ and $\hat{\Gamma}^{(1)}_{R}$, one obtains
\begin{eqnarray}
\Gamma^{(0),\mu\nu}_{ab}&=&{P_h^\mu}{P_h^\nu}\Gamma^{(0),ww}_{ab}+
{P_h^\mu}{\Omega^\nu_{\ \rho}}\Gamma^{(0),w\rho}_{ab}
+{P_h^\nu}{\Omega^\mu_{\ \sigma}}\Gamma^{(0),\sigma w}_{ab}+
{\Omega^\mu_{\ \sigma}}{\Omega^\nu_{\ \rho}}\Gamma^{(0),\sigma \rho}_{ab},\\
\Gamma^{(1),\mu\nu\lambda}_{L,abc}&=&{P_h^\mu}{P_h^\nu}{P_h^\lambda}\Gamma^{(1),www}_{L,abc}
+{P_h^\mu}{P_h^\nu}{\Omega^\lambda_{\ \tau}}\Gamma^{(1),ww\tau}_{L,abc}+
{P_h^\mu}{P_h^\lambda}{\Omega^\nu_{\ \rho}}\Gamma^{(1),w\rho w}_{L,abc}
+{P_h^\nu}{P_h^\lambda}{\Omega^\mu_{\ \sigma}}\Gamma^{(1),\sigma ww}_{L,abc}
\nn\\
&+&{P_h^\mu}{\Omega^\nu_{\ \rho}}{\Omega^\lambda_{\ \tau}}\Gamma^{(1),w\rho\tau}_{L,abc}
+{P_h^\nu}{\Omega^\mu_{\ \sigma}}{\Omega^\lambda_{\ \tau}}\Gamma^{(1),\sigma w\tau}_{L,abc}
+{P_h^\lambda}{\Omega^\mu_{\ \sigma}}{\Omega^\nu_{\ \rho}}\Gamma^{(1),\sigma\rho w}_{L,abc}
\nn\\
&+&{\Omega^\mu_{\ \sigma}}{\Omega^\nu_{\ \rho}}{\Omega^\lambda_{\ \tau}}
\Gamma^{(1),\sigma\rho\tau}_{L,abc},
\end{eqnarray}
and likewise for $\Gamma^{(1),\mu\nu\lambda}_{R,abc}$.  
Inserting the above expansion into (\ref{s97}) and keeping the terms with two or three
$\Omega^\mu_{\ \nu}$\,s in the product of the hard parts 
$S^{(0)}$, $S_L^{(1)}$, $S_R^{(1)}$ and the fragmentation matrix elements
$\hat{\Gamma}^{(0)}$, $\hat{\Gamma}_L^{(1)}$, $\hat{\Gamma}_R^{(1)}$,  
one can obtain the twist-2 and 3 contributions 
to the cross section.  
To get a gauge invariant cross section, one needs to fully utilize the following
Ward identities for the hard parts (see Appendix A):
\begin{eqnarray}
{k}^{\mu}S^{ab}_{\mu\nu}(k)={k}^{\nu}S^{ab}_{\mu\nu}(k)=0, 
\label{ward1}
\end{eqnarray}
\begin{eqnarray}
{(k'-k)}^{\lambda}S^{abc}_{L\,\mu\nu\lambda}(k,k')&=&\frac{-if^{abc}}{N^2-1}S_{\mu\nu}(k')
+G^{abc}_{\mu\nu}(k,k'),
\label{ward2}\\
{k}^{\mu}S^{abc}_{L\,\mu\nu\lambda}(k,k')&=&\frac{if^{abc}}{N^2-1}S_{\lambda\nu}(k')
+G^{cab}_{\lambda\nu}(k'-k,k'), 
\label{ward3}\\
{k'}^{\nu}S^{abc}_{L\,\mu\nu\lambda}(k,k')&=&0, 
\label{ward4}
\end{eqnarray}
\begin{eqnarray}
{(k'-k)}^{\lambda}S^{abc}_{R\,\mu\nu\lambda}(k,k')&=&\frac{if^{abc}}{N^2-1}
S_{\mu\nu}(k)+\left(G^{bac}_{\nu\mu}(k',k)\right)^\star,
\label{ward5}\\
{k'}^{\nu}S^{abc}_{R\,\mu\nu\lambda}(k,k')&=&\frac{if^{abc}}{N^2-1}S_{\mu\lambda}(k)+
\left(G^{cab}_{\lambda\mu}(k-k',k)\right)^\star,
\label{ward6}\\
{k}^{\mu}S^{abc}_{R\,\mu\nu\lambda}(k,k')&=&0.  
\label{ward7}
\end{eqnarray}
Here and below we suppress the upper indices 
$(0)$ and $(1)$ from the hard parts $S^{(0)}$, 
$S^{(1)}_L$ and $S^{(1)}_R$
for simplicity and 
$S^{\mu\nu}(k)\equiv S^{\mu\nu,ab}(k)\delta_{ab}$.   The $G$-terms appear 
due to the off-shellness of the parton momenta entering the fragmentation matrix elements.
We present the actual forms of those ``ghost terms'' in Appendix A.
Here we only mention that they are proportional to $f^{abc}$ and satisfy the relation
\beq
k^\mu G^{abc}_{\mu\nu}(k,k') = k'^\nu G^{abc}_{\mu\nu}(k,k') =0.  
\eeq
We have found that the ghost terms do not contribute to the twist-3 cross 
section\footnote{This was also the case for the twist-3 quark FF contribution to
$ep^\uparrow\to ehX$\,\cite{Kanazawa:2013uia}
and the three-gluon distribution function contribution to
$\vec{p}p^\uparrow \to DX$\,\cite{Hatta:2013wsa}.}.
We thus discard them in the following.  
To get the twist-3 cross section in the gauge invariant form, we use
the collinear limit of these identities 
as well as the first, second and third derivatives with respect to $k$ and  $k'$ of these Ward identities
in the collinear limit.


After very lengthy calculation 
one eventually obtains the twist-2 and -3 contributions from  Fig. \ref{fig3} (a), (b) and (c) to $W_g$
in the following form:  
\begin{eqnarray}
W^{(a)}_g
&=&
{\Omega_{\ \alpha}^\mu}{\Omega_{\ \beta}^\nu}\sum_X\int d\left(\frac{1}{z}\right)z^2\Biggl(\Gamma^{(0)\alpha\beta}(z)-\Gamma^{(1)\alpha\beta}_{AA}(z)-\Gamma^{(1)\alpha\beta}_{AAR}(z)\Biggr)S_{\mu\nu}(z)
\nn\\
&-&i
{\Omega_{\ \alpha}^\mu}{\Omega_{\ \beta}^\nu}{\Omega_{\ \gamma}^\lambda}\sum_X
\int d\left(\frac{1}{z}\right)z^2\Biggl(\Gamma^{(0)\alpha\beta\gamma}_\del(z)-
\Gamma^{(1)\alpha\beta\gamma}_{AA\del}(z)-
\Gamma^{(1)\alpha\beta\gamma}_{AAR\del}(z)\Biggr) 
\left.\frac{\del S_{\mu\nu}(k)}{\del k^\lambda}\right|_{c.l.},
\label{wga}
\end{eqnarray}
\begin{eqnarray}
W^{(b)}_g
&=&
{\Omega_{\ \alpha}^\mu}{\Omega_{\ \beta}^\nu}\sum_X
\int d\left(\frac{1}{z}\right)z^2\Biggl(\Gamma^{(1)\alpha\beta}_{[A]}(z)
+\Gamma^{(1)\alpha\beta}_{AA}(z)\Biggr)S_{\mu\nu}(z)
\nn\\
&-&
i{\Omega_{\ \alpha}^\mu}{\Omega_{\ \beta}^\nu}{\Omega_{\ \gamma}^\lambda}\sum_X
\int d\left(
\frac{1}{z}\right)z^2
\Biggl(\Gamma^{(1)\alpha\beta\gamma}_{\del[F]}(z)+\Gamma^{(1)\alpha\beta\gamma}_{\del[\del A]}(z)
+
\Gamma^{(1)\alpha\beta\gamma}_{\del[A]}(z)
+\Gamma^{(1)\alpha\beta\gamma}_{AA\del}(z)
\Biggr)\left.\frac{\del S_{\mu\nu}(k)}{\del k^\lambda}\right|_{c.l.}
\nn\\
&+&
{i \over 2}{\Omega_{\ \alpha}^\mu}{\Omega_{\ \beta}^\nu}{\Omega_{\ \gamma}^\lambda}
\sum_X\int d\left(\frac{1}{z}\right)\int d\left(\frac{1}{z'}\right)\frac{zz'}{1/z'-1/z}
\Gamma^{(1)\alpha\beta\gamma}_{F,abc}(\oz,\ozd)S^{L,abc}_{\mu\nu\lambda}(z,z'),
\label{wgb}
\end{eqnarray}
\begin{eqnarray}
W^{(c)}_g
&=&
{\Omega_{\ \alpha}^\mu}{\Omega_{\ \beta}^\nu}\int d\left(\frac{1}{z}\right)z^2
\Biggl(\Gamma^{(1)\alpha\beta}_{[AR]}(z)+\Gamma^{(1)\alpha\beta}_{AAR}(z)\Biggr) S_{\mu\nu}(z)
\nn\\
&-&
i{\Omega_{\ \alpha}^\mu}{\Omega_{\ \beta}^\nu}{\Omega_{\ \gamma}^\lambda}
\sum_X\int d\left(\frac{1}{z}\right)z^2
\Biggl(
\,\Gamma^{(1)\alpha\beta\gamma}_{\del[A\del]}(z)
+\Gamma^{(1)\alpha\beta\gamma}_{AAR\del}(z)
\Biggr)
\left.\frac{\del S_{\mu\nu}(k)}{\del k^\lambda}\right|_{c.l.}
\nn\\
&+&
{i \over 2}{\Omega_{\ \alpha}^\mu}{\Omega_{\ \beta}^\nu}{\Omega_{\ \gamma}^\lambda}
\sum_X\int d\left(\frac{1}{z}\right)\int d\left(\frac{1}{z'}\right)\frac{zz'}{1/z'-1/z}
\Gamma^{(1)\alpha\beta\gamma}_{FR,abc}(\oz,\ozd) S^{R,abc}_{\mu\nu\lambda}(z,z').  
\label{wgc}
\end{eqnarray}
Each fragmentation matrix elements $\Gamma$'s appearing in the above expression 
are defined as follows:  
\begin{eqnarray}
&&\Gamma^{(0)\alpha\beta}(z)=
{\delta^{ab} \over {N^2-1}}\sum_X\int\frac{d\lambda}{2\pi} e^{-i{\lambda \over z}}\la 0|F^{\beta w}_b(0)| hX\ra
\la hX|F^{\alpha w}_a(\lambda w)| 0\ra,\\
&&\Gamma^{(0)\alpha\beta\gamma}_\del(z)=
{\delta^{ab} \over {N^2-1}}\sum_X\int\frac{d\lambda}{2\pi} e^{-i{\lambda \over z}}\la 0|F^{\beta w}_b(0)| hX\ra
\la hX|F^{\alpha w}_a(\lambda w)\overleftarrow{\del}^\gamma| 0\ra,
\end{eqnarray}
\begin{eqnarray}
&&\Gamma^{(1)\alpha\beta}_{AA}(z)=
{1 \over {N^2-1}}\sum_X\int\frac{d\lambda}{2\pi} e^{-i{\lambda \over z}}\la 0|F^{\beta w}_a(0)| hX\ra\la hX|gf_{abc}A^{\alpha}_b(\lambda w)A^{w}_c(\lambda w)| 0\ra,\\
&&\Gamma^{(1)\alpha\beta}_{AAR}(z)=
{1 \over {N^2-1}}\sum_X\int\frac{d\lambda}{2\pi} e^{-i{\lambda \over z}}\la 0|gf_{abc}A^{\beta}_b(0)A^{w}_c(0)| hX\ra
\la hX|F^{\alpha w}_a(\lambda w)| 0\ra,\\
&&\Gamma^{(1)\alpha\beta\gamma}_{AA\del}(z)=
{1 \over {N^2-1}}\sum_X\int\frac{d\lambda}{2\pi} e^{-i{\lambda \over z}}\la 0|F^{\beta w}_a(0)| hX\ra\la hX|gf_{abc}A^{\alpha}_b(\lambda w)A^{w}_c(\lambda w)\overleftarrow{\del}^\gamma| 0\ra,\\
&&\Gamma^{(1)\alpha\beta\gamma}_{AAR\del}(z)=
{1 \over {N^2-1}}\sum_X\int\frac{d\lambda}{2\pi} e^{-i{\lambda \over z}}\la 0|gf_{abc}A^{\beta}_b(0)A^{w}_c(0)| hX\ra
\la hX|F^{\alpha w}_a(\lambda w)\overleftarrow{\del}^\gamma| 0\ra,
\end{eqnarray}
\begin{eqnarray}
&&\Gamma^{(1)\alpha\beta}_{[A]}(z)=
{(-if^{abc}) \over {N^2-1}}\sum_X\int\frac{d\lambda}{2\pi} e^{-i{\lambda \over z}}
\la 0|F^{\beta w}_b(0)| hX\ra\la hX|F^{\alpha w}_a(\lambda w)ig{\int^\lambda_\infty}d\mu A^w_c(\mu w)| 0\ra, \\
&&\Gamma^{(1)\alpha\beta\gamma}_{\del[F]}(z)=
{(-if^{abc}) \over {N^2-1}}\sum_X\int\frac{d\lambda}{2\pi} e^{-i{\lambda \over z}}
\la 0|F^{\beta w}_b(0)| hX\ra\la hX|F^{\alpha w}_a(\lambda w)ig{\int^\lambda_\infty}d\mu F^{\gamma w}_c(\mu w)| 0
\ra, \\
&&\Gamma^{(1)\alpha\beta\gamma}_{\del[\del A]}(z)=
{(-if^{abc}) \over {N^2-1}}\sum_X\int\frac{d\lambda}{2\pi} e^{-i{\lambda \over z}}
\la 0|F^{\beta w}_b(0)| hX\ra\la hX|F^{\alpha w}_a(\lambda w)\overleftarrow{\del}^\gamma ig{\int^\lambda_\infty}
d\mu A^w_c(\mu w)| 0\ra,\nn\\
\\
&&\Gamma^{(1)\alpha\beta\gamma}_{\del[A]}(z)=
{(-if^{abc}) \over {N^2-1}}\sum_X\int\frac{d\lambda}{2\pi} e^{-i{\lambda \over z}}
\la 0|F^{\beta w}_b(0)| hX\ra\la hX|F^{\alpha w}_a(\lambda w)igA^\gamma_c(\lambda w)| 0\ra,
\\
&&\Gamma^{(1)\alpha\beta\gamma}_{F,abc}(\oz,\ozd)=
\int\frac{d\lambda}{2\pi}\int\frac{d\mu}{2\pi} e^{-i{\lambda \over z}}e^{-i\mu({1\over z'}-{1\over z})}
\la 0|F^{\beta w}_b(0)| hX\ra
\la hX|F^{\alpha w}_a(\lambda w)gF^{\gamma w}_c(\mu w)| 0\ra,
\label{dyFL}
\end{eqnarray}
\begin{eqnarray}
&&\Gamma^{(1)\alpha\beta}_{[AR]}(z)=
{(-if^{abc}) \over {N^2-1}}\sum_X\int\frac{d\lambda}{2\pi} e^{-i{\lambda \over z}}
\la 0| ig{\int^\infty_0}d\mu A^w_c(\mu w) F^{\beta w}_b(0)
| hX\ra\la hX|F^{\alpha w}_a(\lambda w)| 0\ra,\\
&&\Gamma^{(1)\alpha\beta\gamma}_{\del[A\del]}(z)=
{(-if^{abc}) \over {N^2-1}}\sum_X\int\frac{d\lambda}{2\pi} e^{-i{\lambda \over z}}
\la 0|  ig{\int^\infty_0}d\mu A^w_c(\mu w)  F^{\beta w}_b(0) | hX\ra
\la hX|F^{\alpha w}_a(\lambda w)\overleftarrow{\del}^\gamma| 0\ra,\nn\\
\\
&&\Gamma^{(1)\alpha\beta\gamma}_{FR,abc}(\oz,\ozd)=
\sum_X
\int\frac{d\lambda}{2\pi}\int\frac{d\mu}{2\pi} e^{-i{\lambda \over z}}e^{-i\mu({1\over z'}-{1\over z})}
\la 0| gF^{\gamma w}_c(\mu w)  F^{\beta w}_b(0)
| hX\ra\la hX|F^{\alpha w}_a(\lambda w)| 0\ra.\nn\\
\label{dyFR}
\end{eqnarray}
In the LO calculation of Figs. 1 (a), (b) and (c), 
only $O(1)$ and $O(g)$ contributions from the hadronic matrix elements are produced.  
We thus note that, in (\ref{wga})-(\ref{wgc}),  
one can identify the correlation functions of the field strength 
$F_{\mu\nu}^a=\partial_\mu A_\nu^a -\partial_\nu A_\mu^a +g f^{abc}A_\mu^b A_\nu^c$
to this accuracy.  
The first term in 
$W_g^{(a)}$ is the $O(1)$ contribution from $\hat{\Gamma}(z)$ in (\ref{gFraI}), 
ignoring $O(g^2)$ terms $\sim \la 0| gf^{abc}A^b A^c |hX\ra 
\la hX| gf^{ab'c'} A^{b'} A^{c'}|0\ra$.  
The first terms in
$W_g^{(b)}$ and $W_g^{(c)}$ 
are the $O(g)$ terms arising from the expansion of the gauge link 
and the $O(g)$ part of the field strength in
$\hat{\Gamma}(z)$.
These terms contain both twist-2 and intrinsic twist-3 FFs.  
Likewise the second term in $W_g^{(a)}$ is the $O(1)$ contribution from 
the kinematical twist-3 FFs $\hat{\Gamma}_\del(z)$ in (\ref{gFraK}).  
The second terms in $W_g^{(b)}$ and $W_g^{(c)}$ are the $O(g)$ terms 
arising from the expansion of the gauge link 
and the $O(g)$ part of the field strength
in $\hat{\Gamma}_\del(z)$.  
The third term in $W_g^{(b)}$ is the $O(g)$ contribution from
$\hat{\Gamma}_F$ defined in (\ref{gFraDA}) and (\ref{gFraDS}).
Likewise the third term in $W_g^{(c)}$ is associated with 
$\hat{\Gamma}_{FR}\sim 
(\hat{\Gamma}_F)^\star$.  
This way we have obtained the sum of
$W_g^{(a)}$, $W_g^{(b)}$ and $W_g^{(c)}$ in the color gauge invariant form 
in terms of the $intrinsic$, $kinematical$ and $dynamical$ FFs.  
Inserting these expressions into (\ref{s21}), one can eventually
express the twist-3 gluon FF contribution to the cross section as
\begin{eqnarray}
&&\hspace{-0.9cm} E_{h}\frac{d\sigma(p,p',P_h;S_\perp)}{d^3P_h}=
\frac{1}{16\pi^2S_E}\int^1_0\frac{dx}{x}f_1(x)
\int^1_0 \frac{dx'}{x'}f_1(x')\biggl[
{\Omega^\mu_{\ \alpha}}{\Omega^\nu_{\ \beta}}
\int^1_0 dz\, {\rm Tr}\left[\hat{\Gamma}^{\alpha\beta}(z)S_{\mu\nu}(P_h/z)\right]
\nn\\
&&\hspace{-0.8cm}-i\,
{\Omega^\mu_{\ \alpha}}{\Omega^\nu_{\ \beta}}{\Omega^\lambda_{\ \gamma}}
\int^1_0 dz\, {\rm Tr}\left[\hat{\Gamma}_{\del}^{\alpha\beta\gamma}(z)
\left.\frac{\del S_{\mu\nu}(k)}{\del k^\lambda}\right|_{c.l.}\right]
+
{\Re}\Bigl\{i\,
{\Omega^\mu_{\ \alpha}}{\Omega^\nu_{\ \beta}}{\Omega^\lambda_{\ \gamma}}
\int^1_0 \frac {dz}{z}\int^\infty_z \frac{dz'}{z'}\,
\left(\frac{1}{1/z-1/z'}\right)\nn\\
&&\qquad
\times
{\rm Tr}\left[
\left(
-\frac{if^{abc}}{N}
\hat{\Gamma}_{FA}^{\alpha\beta\gamma}\left(\ozd,\oz\right)
+d^{abc}\frac{N}{N^2-4}
\hat{\Gamma}_{FS}^{\alpha\beta\gamma}\left(\ozd,\oz\right)
\right)
S^L_{\mu\nu\lambda,abc}(z',z)\right]
\Bigr\}\biggr],
\label{PPgfragForma}
\end{eqnarray}
where we have used the expression (\ref{gFraDexpand}), and 
$\left. \right|_{c.l.}$ implies the collinear limit, $k\to P_h/z$.  
In writing down the contribution from the dynamical FFs in
(\ref{PPgfragForma}), we have interchanged the role of the variables, 
$z$ and $z'$, from (\ref{wgb}) and (\ref{wgc}) for later convenience.

\begin{figure}[h]
\begin{center}
  \includegraphics[width=5cm]{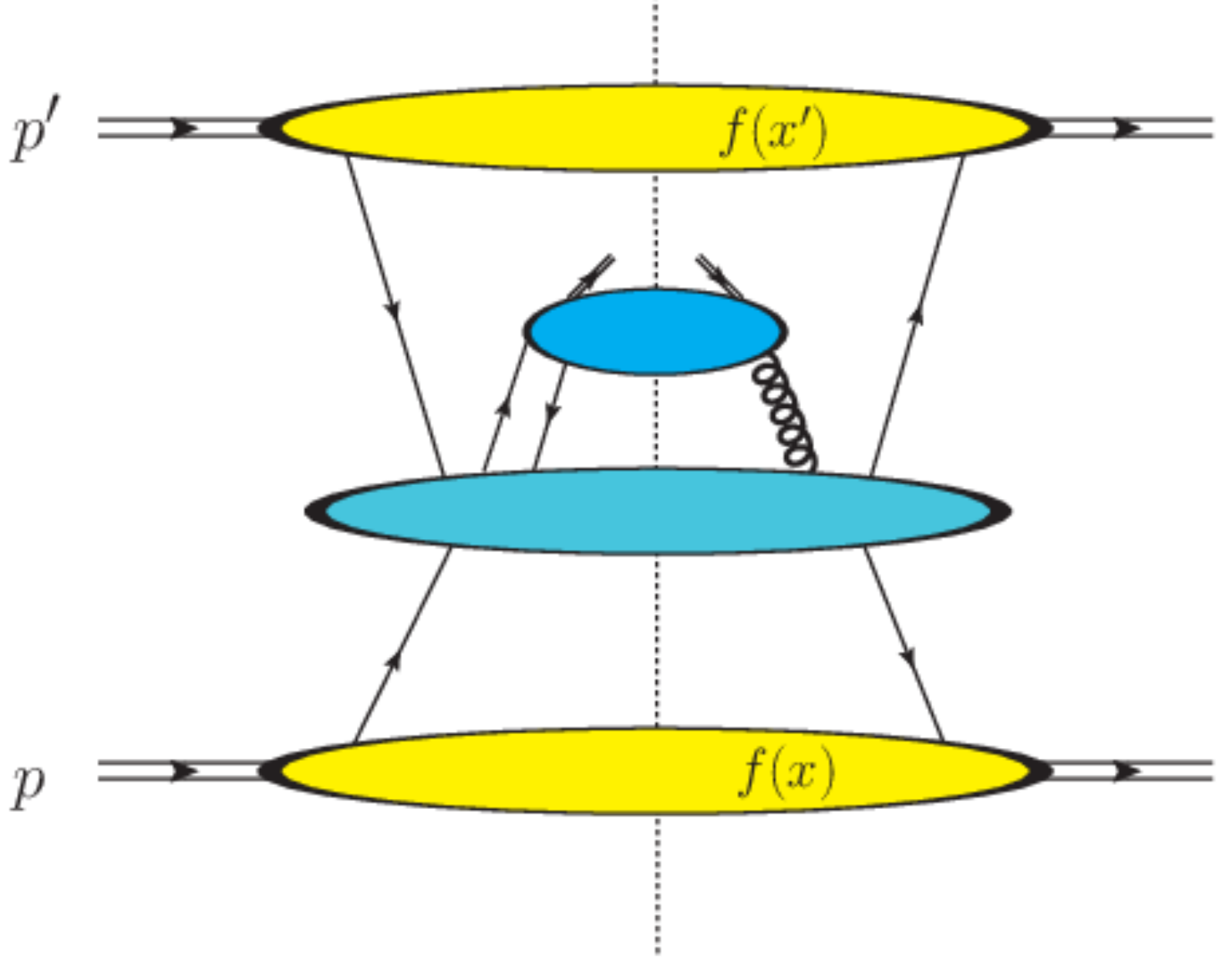}
\end{center}
 \caption{Diagram for the $q\bar{q}g$-type correlation function in $pp\to \Lambda^\uparrow X$.  
 Mirror diagram also contributes. }
\label{PP_qqg}
\end{figure}

Next we substitute (\ref{gFraI}), (\ref{gFraK}), (\ref{gFraDA}), (\ref{gFraDS}) into 
(\ref{PPgfragForma}).   
We recall that the $q\bar{q}g$-type FFs (\ref{FFtilde}) are related to 
the purely gluonic FFs, as was shown in Sec. \ref{sec:gRel}.  Therefore we
consider the contribution shown in Fig. \ref{PP_qqg} together.  
Note also that each hard part contains the factor 
$\delta\left( (xp+x'p'-k)^2\right)\sim \delta\left( (xp+x'p'-P_h/z)^2\right)$
corresponding to the on-shell condition for the final unobserved parton,
and its derivative with respect to $k$ causes the derivative of the kinematical FFs
by partial integration with respect to $1/z$.
Separating various contributions 
based on the $z'$-dependence of the hard cross sections for the dynamical FFs (See Appendix C), 
we can write the cross section as
\begin{eqnarray}
&&\hspace{-0.9cm}  E_{h}\frac{d\sigma(p,p',P_h;S_\perp)}{d^3P_h} \nonumber\\
&=&
\frac{M_h\alpha_s^2}{S_E}\int^1_0\frac{dx}{x}f_1(x)
\int^1_0\frac{dx'}{x'}f_1(x')\int^1_0dz\,
\delta((xp+x'p'-P_h/z)^2)\nn\\
&\times&
\Biggl\{
\frac{\Delta\hat{G}_{3\bar{T}}(z)}{z}\hat{H}_{int}
+
\left(\hat{G}_T^{(1)}(z)\hat{H}_{NDG}
+\frac{1}{z}\frac{\del\,\hat{G}_T^{(1)}(z)}{\del (1/z)}\hat{H}_{DG}\right)\nn\\
&&\qquad\qquad\qquad\qquad+
\left(\Delta\hat{H}_T^{(1)}(z)\hat{H}_{NDH}
+\frac{1}{z}\frac{\del\,\Delta\hat{H}_T^{(1)}(z)}{\del (1/z)}\hat{H}_{DH}\right)
\nn\\
&+&
\int^{1/z}_0 d\left(\frac{1}{z'}\right)
\Biggl[
\sum_{i=1}^3 \Im\Nhat_{i}\left(\frac{1}{1/z-1/z'}\hat{H}_1^{N_i}+\frac{1}{z}
\left(\frac{1}{1/z-1/z'}\right)^2\hat{H}_2^{N_i}+z'\hat{H}_3^{N_i}+\frac{z'^2}{z}\hat{H}_4^{N_i}\right)
\nn\\
&+&
\sum_{i=1}^3 \Im\Ohat_{i}\left(\frac{1}{1/z-1/z'}\hat{H}_1^{O_i}+\frac{1}{z}
\left(\frac{1}{1/z-1/z'}\right)^2\hat{H}_2^{O_i}+z'\hat{H}_3^{O_i}+\frac{z'^2}{z}\hat{H}_4^{O_i}\right)
\Biggr]\nn\\
&+&
\int^{1/z}_0 d\left(\frac{1}{z'}\right)\, \frac{2}{C_F}
\Bigl[
\Im\widetilde{D}_F\left(\ozd,\ozd-\oz\right)
(\hat{H}_{DF1}+\frac{1}{z}\frac{1}{1/z-1/z'}\hat{H}_{DF2}+\frac{z'}{z}\hat{H}_{DF3})\nn\\
&+&
\Im\widetilde{G}_F\left(\ozd,\ozd-\oz\right)
(\hat{H}_{GF1}+\frac{1}{z}\frac{1}{1/z-1/z'}\hat{H}_{GF2}+\frac{z'}{z}\hat{H}_{GF3})
\Bigr]
\Biggr\},
\label{PPgfragForma3}
\end{eqnarray}
where
$\hat{H}_{int}$, $\hat{H}_{NDG}$, $\hat{H}_{DG}$, etc. represent the partonic hard cross sections
for each FF (after separating the $z'$-dependence for dynamical FFs) 
and they are the functions of the Mandelstam variables in the parton level, 
$s=(xp+x'p')^2=2xx'p\cdot p'$, $t=(xp-P_h/z)^2=-2(x/z)p\cdot P_h$ 
and $u=(x'p'-P_h/z)^2=-2(x'/z)p'\cdot P_h$,
multiplied by kinematic factors
with $\epsilon^{pP_h w S_\perp}$ and $\epsilon^{p'P_h w S_\perp}$.  
(See eq.(\ref{hard_example}) below as an example.)
In (\ref{PPgfragForma3}), 
we have used the shorthand notation, 
$\Nhat_{3}\equiv-\Nhat_{2}\left(\oz-\ozd,\oz\right)$
and $\Ohat_{3}\equiv\Ohat_{2}\left(\oz-\ozd,\oz\right)$.  
Change of the variable
$\frac{1}{z'}\leftrightarrow \frac{1}{z}-\frac{1}{z'}$ in the 
$z'$ and $z'^2/z$ terms in the contribution from $\Nhat_{i}$ and $\Ohat_{i}$
leads to the following form
owing to the
exchange symmetry (\ref{swi_nat}) of $\Nhat_{1}$ and $\Ohat_{1}$:
\begin{eqnarray}
&&\hspace{-0.5cm} E_{h}\frac{d\sigma(p,p',P_h;S_\perp)}{d^3P_h}=
\frac{M_h\alpha_s^2}{S_E}\int^1_0\frac{dx}{x}f_1(x)
\int^1_0\frac{dx'}{x'}f_1(x')\int^1_0 dz\,
\delta(s+t+u)\nn\\
&&\times
\Biggl\{
\frac{\Delta\hat{G}_{3\bar{T}}(z)}{z}\hat{H}_{int}+
\hat{G}_T^{(1)}(z)\hat{H}_{NDG}
+\frac{1}{z}\frac{\del\,\hat{G}_T^{(1)}(z)}{\del (1/z)}\hat{H}_{DG}
+
\Delta\hat{H}_T^{(1)}(z)\hat{H}_{NDH}
+\frac{1}{z}\frac{\del\,\Delta\hat{H}_T^{(1)}(z)}{\del (1/z)}\hat{H}_{DH}
\nn\\
&&+
\int^{1/z}_0 d\left(\frac{1}{z'}\right)
\Biggl[
\frac{1}{1/z-1/z'}\Im
\left\{\Nhat_{1}\left(\ozd,\oz\right)\left(\hat{H}_1^{N_1}-\hat{H}_3^{N_1}\right)
+\Nhat_{2}\left(\ozd,\oz\right)\left(\hat{H}_1^{N_2}-\hat{H}_3^{N_3}\right)\right.
\nn\\
&&\left.\qquad\qquad\qquad-\Nhat_{2}\left(\oz-\ozd,\oz\right)
\left(\hat{H}_1^{N_3}-\hat{H}_3^{N_2}\right)\right\}\nn\\
&&\qquad+
\frac{1}{z}\left(\frac{1}{1/z-1/z'}\right)^2\Im
\left\{\Nhat_{1}\left(\ozd,\oz\right)\left(\hat{H}_2^{N_1}-\hat{H}_4^{N_1}\right)
+\Nhat_{2}\left(\ozd,\oz\right)\left(\hat{H}_2^{N_2}-\hat{H}_4^{N_3}\right)\right.\nn\\
&&\left.
\qquad\qquad\qquad-\Nhat_{2}\left(\oz-\ozd,\oz\right)\left(\hat{H}_2^{N_3}-\hat{H}_4^{N_2}\right)\right\}
\nn\\
&&\qquad+
\frac{1}{1/z-1/z'}\Im
\left\{\Ohat_{1}\left(\ozd,\oz\right)\left(\hat{H}_1^{O_1}+\hat{H}_3^{O_1}\right)
+\Ohat_{2}\left(\ozd,\oz\right)\left(\hat{H}_1^{O_2}+\hat{H}_3^{O_3}\right)\right.
\nn\\
&&\left.\qquad\qquad\qquad+\Ohat_{2}\left(\oz-\ozd,\oz\right)\left(\hat{H}_1^{O_3}
+\hat{H}_3^{O_2}\right)\right\}\nn\\
&&\qquad+
\frac{1}{z}\left(\frac{1}{1/z-1/z'}\right)^2\Im
\left\{\Ohat_{1}\left(\ozd,\oz\right)\left(\hat{H}_2^{O_1}+\hat{H}_4^{O_1}\right)
+\Ohat_{2}\left(\ozd,\oz\right)\left(\hat{H}_2^{O_2}+\hat{H}_4^{O_3}\right)\right.\nn\\
&&\left.\qquad\qquad\qquad
+\Ohat_{2}\left(\oz-\ozd,\oz\right)\left(\hat{H}_2^{O_3}+\hat{H}_4^{O_2}\right)\right\}
\Biggr]
\nn\\
&&+
\int^{1/z}_0 d\left(\frac{1}{z'}\right)\, \frac{2}{C_F}
\Bigl[
\Im\widetilde{D}_F\left(\ozd,\ozd-\oz\right)
\left(\hat{H}_{DF1}+\frac{1}{z}\frac{1}{1/z-1/z'}\hat{H}_{DF2}+\frac{z'}{z}\hat{H}_{DF3}\right)\nn\\
&&\qquad\qquad+
\Im\widetilde{G}_F\left(\ozd,\ozd-\oz\right)
\left(\hat{H}_{GF1}+\frac{1}{z}\frac{1}{1/z-1/z'}\hat{H}_{GF2}+\frac{z'}{z}\hat{H}_{GF3}\right)
\Bigr]
\Biggr\}.  
\label{PPgfragForma4}
\end{eqnarray}
Here
we remind that the gauge invariance and the frame independence of 
the cross section (\ref{PPgfragForma4}) are realized in a very nontrivial manner.  
We show that those properties are guaranteed by the
EOM relation and the LIRs introduced in the previous
section.  
The gauge invariance is satisfied by the EOM relation (\ref{FFDFodd1}), which is 
discussed  in detail in
Appendix B.  Here we demonstrate how the frame independence
of the cross section is achieved.  

To make clear the issue of frame dependence, we pick up
the hard cross sections 
$\hat{H}_2^{N_1}-\hat{H}_4^{N_1}$, $\hat{H}_{DG}$ and $\hat{H}_{DH}$ in
(\ref{PPgfragForma4}) in the
$qg\to gq$-channel, as an example.  They
can be computed to be
\beq
\hat{H}_2^{N_1}-\hat{H}_4^{N_1}&=&{C_F\over N}\Bigl((2t+u)x\epsilon^{pP_hwS_{\perp}}
+tx'\epsilon^{p'P_hwS_{\perp}}\Bigr)\Bigl({s^2+t^2\over s^2 t^2}\Bigr)
\nonumber\\
&&-{1\over 2}\Bigl(u(2t+u)x\epsilon^{pP_hwS_{\perp}}
+t(2t+3u)x'\epsilon^{p'P_hwS_{\perp}}\Bigr){s^2+t^2\over s t u^3},
\nn\\
\hat{H}_{DG}&=&-{C_F\over N}\Bigl(x\epsilon^{pP_hwS_{\perp}}
+x'\epsilon^{p'P_hwS_{\perp}}\Bigr)\Bigl({s^2+t^2\over s^2 t}\Bigr)
\nonumber\\
&&+\Bigl(x\epsilon^{pP_hwS_{\perp}}
+x'\epsilon^{p'P_hwS_{\perp}}\Bigr)\Bigl({s^2+t^2\over s u^2}\Bigr),
\nonumber\\
\hat{H}_{DH}&=&0.  
\label{hard_example}
\eeq
We note that each cross section contains the lightlike vector $w^\mu$.  
On the other hand, the physical cross section
should be able to be represented in terms of the vectors $p$, $p'$ $P_h$ and $S_\perp$
in a Lorentz invariant form.  
Since $w^\mu$ is defined from $P_h^\mu$, its actual form
depends on the frame.  One can express
the vector $w$ in terms of $p$, $p'$ and $P_h$ as\cite{Kanazawa:2015ajw}
\beq
w^{\mu}=\alpha{p^{\mu}\over p\cdot P_h}
+(1-\alpha){p^{\mu}\over p'\cdot P_h}
+\left\{ -\alpha(1-\alpha){p\cdot p'\over p\cdot P_h p'\cdot P_h}
+\beta^2 p\cdot p' p\cdot P_h p'\cdot P_h\right\} P_h^{\mu}
+\beta \epsilon^{\mu p p' P_h},\nn\\
\eeq 
which satisfies $P_h\cdot w=1$ and $w^2=0$.    
The values of $\alpha$ and $\beta$ specify the frame we choose, and 
the above form of the hard cross sections leads to $\alpha$- and 
$\beta$-dependent cross sections.  
However, use of the EOM relation, (\ref{FFDFodd1}), and the
LIRs, (\ref{FFrel_G}) and (\ref{FFrel_H}), leads to
the cross section independent from $\alpha$ and $\beta$ as will be seen below.   
In the twist-3 cross section (\ref{PPgfragForma4}), 
we eliminate the intrinsic FF $\Delta\widehat{G}_{3\bar{T}}(z)/z$ and the
derivative of the two kinematical FFs by using those relations.  
Then the resulting cross section is written in terms of
the (nonderivative) kinematical FFs and the dynamical FFs.  
If we pick up the hard cross section for
\beq
{1\over z}\int_0^{1/z}\,d\left({1\over z'}\right){1\over (1/z-1/z')^2}\Im \Nhat_1\left(\ozd,\oz\right),  
\eeq
we have the combination
\beq
\hat{H}_2^{N_1}-\hat{H}_4^{N_1}+2\hat{H}_{DG}+4\hat{H}_{DH}
={\cal E}\hat{\sigma}_{DN1},
\label{sigmaN1}
\eeq
where 
\beq
&&\hat{\sigma}_{DN1}\equiv -\frac{C_F}{N}(\frac{1}{s^2}+\frac{1}{t^2})-
\frac{(2 t - u) (s^2 + t^2)}{2 s t u^3},\\
&&{\cal E}\equiv x't{\,}\epsilon^{p'P_hwS_\perp}-xu{\,}\epsilon^{pP_hwS_\perp}=
-2{xx'\over z}\epsilon^{pp'P_hS_\perp}.
\label{Epstensor}
\eeq 
One should note that the kinematical factor ${\cal E}$ appearing in this
combination is free from $\alpha$ and $\beta$, which is written after the last equal sign
in (\ref{Epstensor}). 
We have found that all the coefficient hard cross sections for all FFs with the same 
$z'$-dependence define the frame-independent cross section with the common
kinematic factor ${\cal E}$.  
This shows that the frame dependence has
been removed from the twist-3 
cross section thanks to the EOM relations and the LIRs.
This way we can define the following set of 
frame independent hard cross sections (in addition to (\ref{sigmaN1})):  
\beq
{\cal E}\hat{\sigma}_G&&=\hat{H}_{NDG}+\frac{1}{2}\hat{H}_{int}+2\hat{H}_{DG},\\
{\cal E}\hat{\sigma}_H&&=\hat{H}_{NDH}+\frac{1}{2}\hat{H}_{int}+4\hat{H}_{DH},\\
{\cal E}\hat{\sigma}_{N1}&&=\hat{H}_1^{N_1}-\hat{H}_3^{N_1}+2\hat{H}_{int}+4\hat{H}_{DG,}+8\hat{H}_{DH},\\
{\cal E}\hat{\sigma}_{N2}&&=\hat{H}_1^{N_2}-\hat{H}_3^{N_3}+\hat{H}_{int}+8\hat{H}_{DH},\\
{\cal E}\hat{\sigma}_{N3}&&=-\hat{H}_1^{N_3}+\hat{H}_3^{N_2}-\hat{H}_{int}-4\hat{H}_{DG},\\
{\cal E}\hat{\sigma}_{DN2}&&=\hat{H}_2^{N_2}-\hat{H}_4^{N_3}+2\hat{H}_{DG}+4\hat{H}_{DH},\\
{\cal E}\hat{\sigma}_{DN3}&&=-\hat{H}_2^{N_3}+\hat{H}_4^{N_2}-4\hat{H}_{DG},\\
{\cal E}\hat{\sigma}_{O1}&&=\hat{H}_1^{O_1}+\hat{H}_3^{O_1},\\
{\cal E}\hat{\sigma}_{O2}&&=\hat{H}_1^{O_2}+\hat{H}_3^{O_3},\\
{\cal E}\hat{\sigma}_{O3}&&=\hat{H}_1^{O_3}+\hat{H}_3^{O_2},\\
{\cal E}\hat{\sigma}_{DO1}&&=\hat{H}_2^{O_1}+\hat{H}_4^{O_1},\\
{\cal E}\hat{\sigma}_{DO2}&&=\hat{H}_2^{O_2}+\hat{H}_4^{O_3},\\
{\cal E}\hat{\sigma}_{DO3}&&=\hat{H}_2^{O_3}+\hat{H}_4^{O_2},\\
{\cal E}\hat{\sigma}_{DF1}&&=\hat{H}_{DF1}-\hat{H}_{int}-2\hat{H}_{DG}-4\hat{H}_{DH},\\
{\cal E}\hat{\sigma}_{DF2}&&=\hat{H}_{DF2},\\
{\cal E}\hat{\sigma}_{DF3}&&=\hat{H}_{DF3},\\
{\cal E}\hat{\sigma}_{GF1}&&=\hat{H}_{GF1},\\
{\cal E}\hat{\sigma}_{GF2}&&=\hat{H}_{GF2},\\
{\cal E}\hat{\sigma}_{GF3}&&=\hat{H}_{GF3}.
\eeq
With these hard cross sections,
the manifestly frame-independent twist-3 cross section is given by
\begin{eqnarray}
&&\hspace{-0.9cm} E_{h}\frac{d\sigma(p,p',P_h;S_\perp)}{d^3P_h}\nonumber\\
&=&
\frac{M_h\alpha_s^2}{S_E}\int^1_0\frac{dx}{x}f_1(x)
\int^1_0\frac{dx'}{x'}f_1(x')\int^{1}_0 dz\, 
\delta(s+t+u)\left(-2{xx'\over z}\epsilon^{pp'P_hS_\perp}\right)
\nn\\
&\times&
\Biggl\{
\hat{G}_T^{(1)}(z)\hat{\sigma}_{G}
+
\Delta\hat{H}_T^{(1)}(z)\hat{\sigma}_{H}
\nn\\
&+&
\int^{1/z}_0 d\left(\frac{1}{z'}\right)
\Biggl[
\frac{1}{1/z-1/z'}\Im
\left(\Nhat_{1}\left(\ozd,\oz\right)\hat{\sigma}_{N1}
+\Nhat_{2}\left(\ozd,\oz\right)\hat{\sigma}_{N2}
+\Nhat_{2}\left(\oz-\ozd,\oz\right)\hat{\sigma}_{N3}\right)\nn\\
&+&
\frac{1}{z}\left(\frac{1}{1/z-1/z'}\right)^2\Im
\left(\Nhat_{1}\left(\ozd,\oz\right)\hat{\sigma}_{DN1}
+\Nhat_{2}\left(\ozd,\oz\right)\hat{\sigma}_{DN2}
+\Nhat_{2}\left(\oz-\ozd,\oz\right)\hat{\sigma}_{DN3}\right)
\nn\\
&+&
\frac{1}{1/z-1/z'}\Im
\left(\Ohat_{1}\left(\ozd,\oz\right)\hat{\sigma}_{O1}
+\Ohat_{2}\left(\ozd,\oz\right)\hat{\sigma}_{O2}
+\Ohat_{2}\left(\oz-\ozd,\oz\right)\hat{\sigma}_{O3}\right)\nn\\
&+&
\frac{1}{z}\left(\frac{1}{1/z-1/z'}\right)^2\Im
\left(\Ohat_{1}\left(\ozd,\oz\right)\hat{\sigma}_{DO1}
+\Ohat_{2}\left(\ozd,\oz\right)\hat{\sigma}_{DO2}
+\Ohat_{2}\left(\oz-\ozd,\oz\right)\hat{\sigma}_{DO3}\right)
\Biggr]
\nn\\
&+&
\int^{1/z}_0 d\left(\frac{1}{z'}\right)\, \frac{2}{C_F}
\Bigl[
\Im\widetilde{D}_F\left(\ozd,\ozd-\oz\right)
\left(\hat{\sigma}_{DF1}+\frac{1}{z}\frac{1}{1/z-1/z'}\hat{\sigma}_{DF2}+\frac{z'}{z}\hat{\sigma}_{DF3}\right)\nn\\
&+&
\Im\widetilde{G}_F\left(\ozd,\ozd-\oz\right)
\left(\hat{\sigma}_{GF1}+\frac{1}{z}\frac{1}{1/z-1/z'}\hat{\sigma}_{GF2}+\frac{z'}{z}\hat{\sigma}_{GF3}\right)
\Bigr]
\Biggr\}.
\label{PPgfragForma5}
\end{eqnarray}\\
Equation  (\ref{PPgfragForma5})
is the final result for the twist-3 gluon FF contribution to $pp\to\Lambda^\uparrow X$.

Below we give the LO Feynman diagrams for the hard part in each channel 
and present the results for hard cross sections, using the partonic Mandelstam variables, $s$, 
$t$, and $u$.  

\vspace{0.5cm}

\begin{figure}[H]
    \begin{tabular}{cc}
      \begin{minipage}[t]{0.43\hsize}
        \centering
       \includegraphics[width=7.0cm]{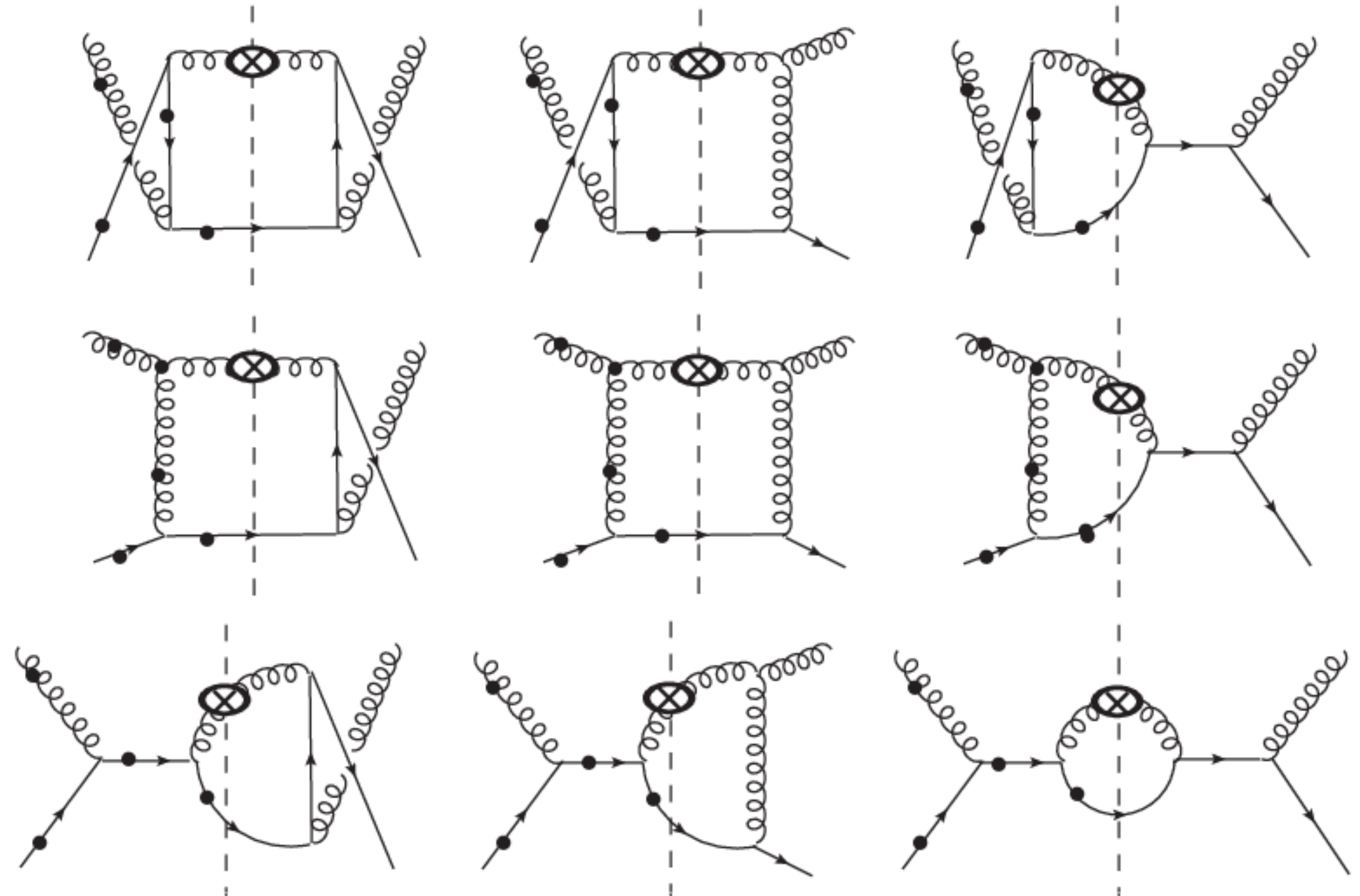}
        \caption{Diagrams
        in the $qg\to gq$ channel for the hard part $S(k)$ and $S_{L}(z',z)$ in eq. (\ref{PPgfragForma}).
          For $S_{L}(z',z)$, an extra gluon line connecting $\otimes$ and
        each of the
        black dots should be added for each diagram.  }
        \label{fig_qggq1}
      \end{minipage} &\hspace{0.3cm}
      \begin{minipage}[t]{0.43\hsize}
        \centering
  \includegraphics[width=8.0cm]{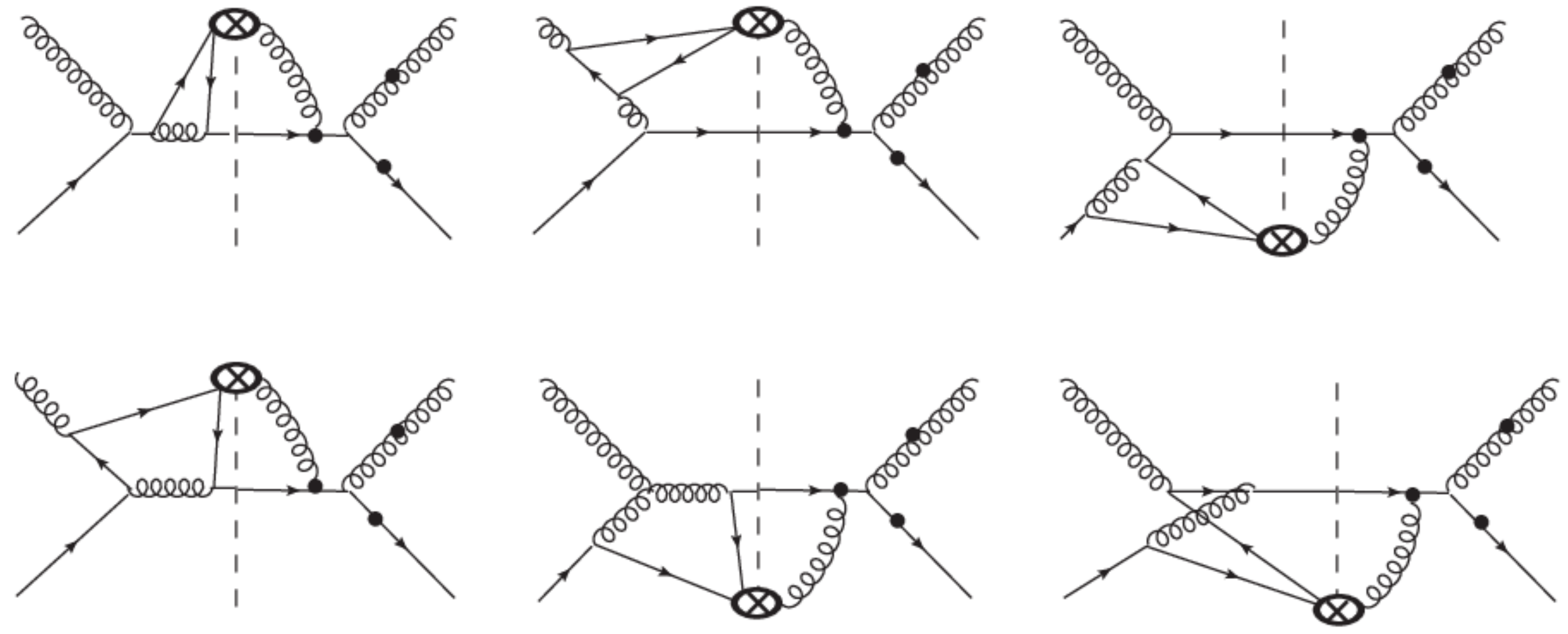}
    \caption{Diagrams for the hard part in Fig. \ref{PP_qqg}.  For 
    the upper middle diagram, a quark loop with the reversed arrow
    also needs to be included.   
    }
        \label{fig_qggq2}
      \end{minipage}
    \end{tabular}
  \end{figure}

\noindent
{(1) $qg\to gq$ channel} (Figs. \ref{fig_qggq1}, \ref{fig_qggq2})
\begin{align*}
&\hat{\sigma}_{G}=-\frac{C_F}{N}(\frac{1}{s^2}-\frac{1}{t^2})-\frac{2 (s - t)}{(s + t)^3},
&
&\hat{\sigma}_{H}=0,\nn\\
&\hat{\sigma}_{N1}=-\frac{C_F}{N}\frac{4}{s^2}-\frac{4 t^3 + 10 t^2 u + 4 t u^2 + u^3}{t u^3 (t + u)},
&
&\hat{\sigma}_{N2}=0,
&
&\hat{\sigma}_{N3}=-\hat{\sigma}_{N1},\nn\\
&\hat{\sigma}_{DN1}=-\frac{C_F}{N}(\frac{1}{s^2}+\frac{1}{t^2})+\frac{(2 t - u) (s^2 + t^2)}{2 t u^3 (t + u)},
&
&\hat{\sigma}_{DN2}=\hat{\sigma}_{DN1},
&
&\hat{\sigma}_{DN3}=-2 \hat{\sigma}_{DN1},\nn\\
&\hat{\sigma}_{O1}=0,
&
&\hat{\sigma}_{O2}=-\frac{C_F}{N}(\frac{2}{s^2}+\frac{2}{t^2})+\frac{3 (s^2 + t^2)}{s t (s + t)^2},
&
&\hat{\sigma}_{O3}=\hat{\sigma}_{O2},\nn\\
&\hat{\sigma}_{DO1}=-\frac{C_F}{N}(\frac{1}{s^2}+\frac{1}{t^2})+\frac{3 (s^2 + t^2)}{2 s t (s + t)^2},
&
&\hat{\sigma}_{DO2}=\hat{\sigma}_{DO1},
&
&\hat{\sigma}_{DO3}=2 \hat{\sigma}_{DO1},\nn
\end{align*}
\begin{align*}
&\hspace{-5cm}\hat{\sigma}_{DF1}=\frac{C_F}{N}\frac{2}{s^2}
+\frac{4 t^3 + 10 t^2 u + 4 t u^2 + u^3}{2t u^3 (t + u)},\nn\\
&\hspace{-5cm}\hat{\sigma}_{DF2}=
\frac{C_F}{N}\frac{(s^2 + t^2)}{s t (s + t)^2}
+\frac{(s + 2 t) (s^2 + t^2)}{4 s t (s + t)^2 u}
+\frac{1}{N}\frac{s}{2t (s + t)^2}-\frac{C_F}{N^2}\frac{1}{2t^2},\nn\\
&\hspace{-5cm}\hat{\sigma}_{DF3}=
-\frac{C_F}{N}\frac{(s^2 + t^2)}{s t (s + t)^2}
+\frac{(2 s + t) (s^2 + t^2)}{4 s t (s + t)^3}-\frac{1}{N}\frac{t}{2s (s + t)^2}
+\frac{C_F}{N^2}\frac{1}{2s^2},\nn
\end{align*}
\begin{eqnarray}
\hspace{-2.7cm}\hat{\sigma}_{GF1}=0,
\qquad
\hat{\sigma}_{GF2}=-\hat{\sigma}_{DF2},
\qquad
\hat{\sigma}_{GF3}=\hat{\sigma}_{DF3}.
\end{eqnarray}

\vspace{0.5cm}

\begin{figure}[H]
    \begin{tabular}{cc}
      \begin{minipage}[t]{0.43\hsize}
        \centering
       \includegraphics[width=7.0cm]{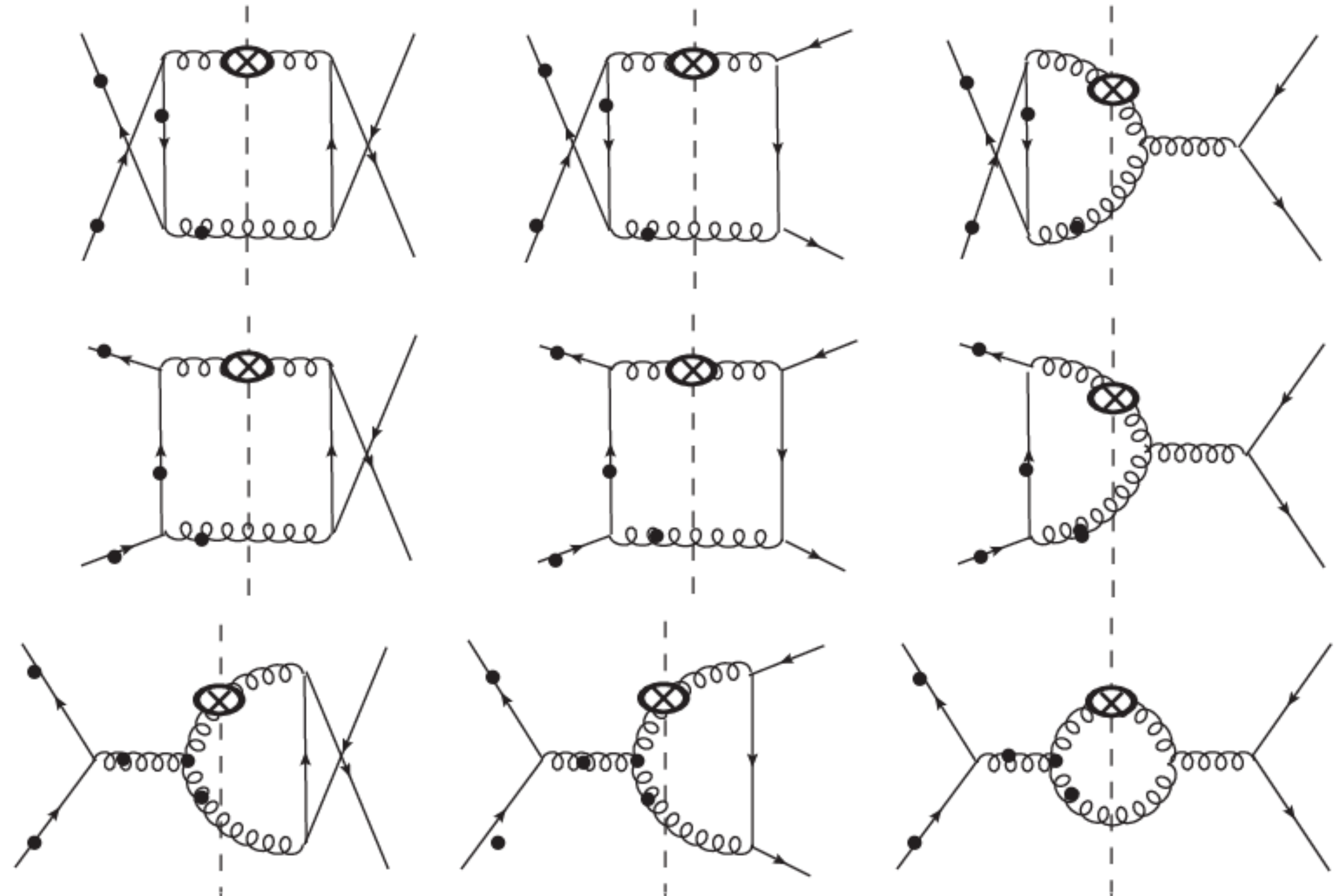}
        \caption{Diagrams in the $q\bar{q}\to gg$ channel for the hard parts $S(k)$ and $S_{L}(z',z)$
         in eq. (\ref{PPgfragForma}).  
        For $S_{L}(z',z)$, an extra gluon line connecting $\otimes$ and each of the black dots should be 
        added for each diagram. }     
        \label{fig_qqgg1}
      \end{minipage} &\hspace{0.3cm}
      \begin{minipage}[t]{0.43\hsize}
        \centering
  \includegraphics[width=8.0cm]{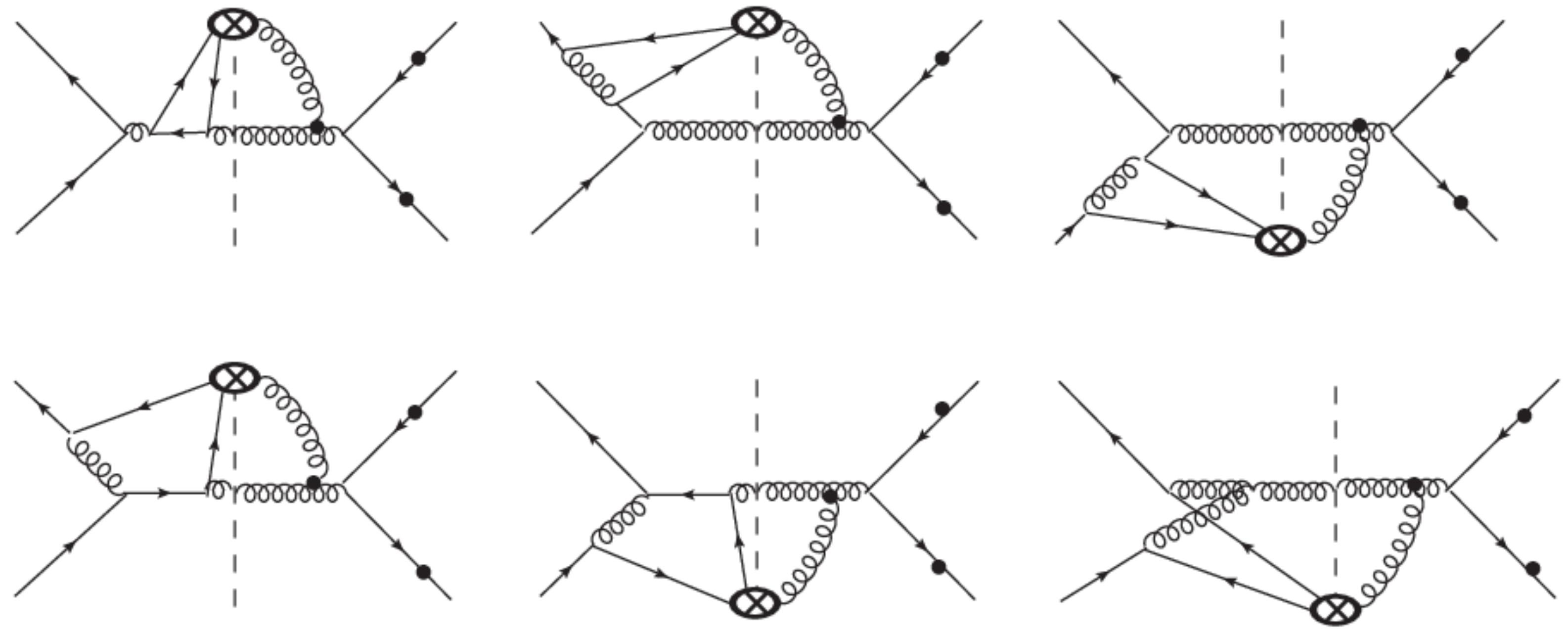}
    \caption{Diagrams for the hard part in Fig. \ref{PP_qqg}.  For 
    the upper left diagram, a quark loop with the reversed arrow also needs to be included.  }       
    \label{fig_qqgg2}
      \end{minipage}
    \end{tabular}
  \end{figure}

\noindent
{(2) $q\bar{q}\to gg$ channel} (Figs. \ref{fig_qqgg1}, \ref{fig_qqgg2})\\
\begin{align*}
&\hat{\sigma}_{G}=-\frac{C_F}{N^2}(\frac{1}{t^2}-\frac{1}{u^2})-C_F\frac{(t - u) (t^2 + u^2) (t^2 + 4 t u + u^2)}{t^2 u^2 (t + u)^3},
&
&\hat{\sigma}_{H}=0,\nn\\
&\hat{\sigma}_{N1}=\frac{C_F}{N^2}\frac{4 (t - u)}{t u (t + u)}-C_F\frac{6 (t - u) (t^2 + u^2)}{t u (t + u)^3},
&
&\hat{\sigma}_{N2}=0,
&
&\hat{\sigma}_{N3}=-\hat{\sigma}_{N1},\nn\\
&\hat{\sigma}_{DN1}=-\frac{C_F}{N^2}\frac{(t - u) (t^2 + u^2)}{t^2 u^2 (t + u)}+C_F\frac{(t^2 + u^2) (t^3 - u^3)}{t^2 u^2 (t + u)^3},
&
&\hat{\sigma}_{DN2}=\hat{\sigma}_{DN1},
&
&\hat{\sigma}_{DN3}=-2 \hat{\sigma}_{DN1},\nn
\end{align*}
\begin{align*}
&\hat{\sigma}_{O1}=0,\nn
&
&\hat{\sigma}_{O2}=\frac{C_F}{N^2}(\frac{2}{t^2}+\frac{2}{u^2})-C_F\frac{2 (t^2 + u^2) (t^2 - t u + u^2)}{t^2 u^2 (t + u)^2},
&
&\hat{\sigma}_{O3}=\hat{\sigma}_{O2},\nn
\end{align*}
\begin{align*}
&\hat{\sigma}_{DO1}=\frac{C_F}{N^2}(\frac{1}{t^2}+\frac{1}{u^2})-C_F\frac{(t^2 + u^2) (t^2 - t u + u^2)}{t^2 u^2 (t + u)^2},
&
&\hat{\sigma}_{DO2}=\hat{\sigma}_{DO1},
&
&\hat{\sigma}_{DO3}=2 \hat{\sigma}_{DO1},\nn
\end{align*}
\begin{align*}
&\hat{\sigma}_{DF1}=-\frac{C_F}{N^2}\frac{2(t - u)}{t u (t + u)}+C_F\frac{3 (t - u) (t^2 + u^2)}{t u (t + u)^3},\nn\\
&\hat{\sigma}_{DF2}=-\frac{C_F}{N^2}\frac{(t^2 + u^2)}{t u (t + u)^2}+C_F\frac{(t^2 + u^2)}{2t (t + u)^3}-\frac{C_F}{N}\frac{(t^2 + u^2)}{2t^2 (t + u)^2}+\frac{C_F}{N^3}\frac{1}{2t^2},\nn\\
&\hat{\sigma}_{DF3}=\frac{C_F}{N^2}\frac{(t^2 + u^2)}{t u (t + u)^2}-C_F\frac{(t^2 + u^2)}{2u (t + u)^3}+\frac{C_F}{N}\frac{(t^2 + u^2)}{2u^2 (t + u)^2}-\frac{C_F}{N^3}\frac{1}{2u^2},\nn\\
\end{align*}
\begin{eqnarray}
\hat{\sigma}_{GF1}=0,\qquad
\hat{\sigma}_{GF2}=-\hat{\sigma}_{DF2},\qquad
\hat{\sigma}_{GF3}=\hat{\sigma}_{DF3}.
\end{eqnarray}

\vspace{0.5cm}

\begin{figure}[h]
  \begin{center}
   \includegraphics[width=10.0cm]{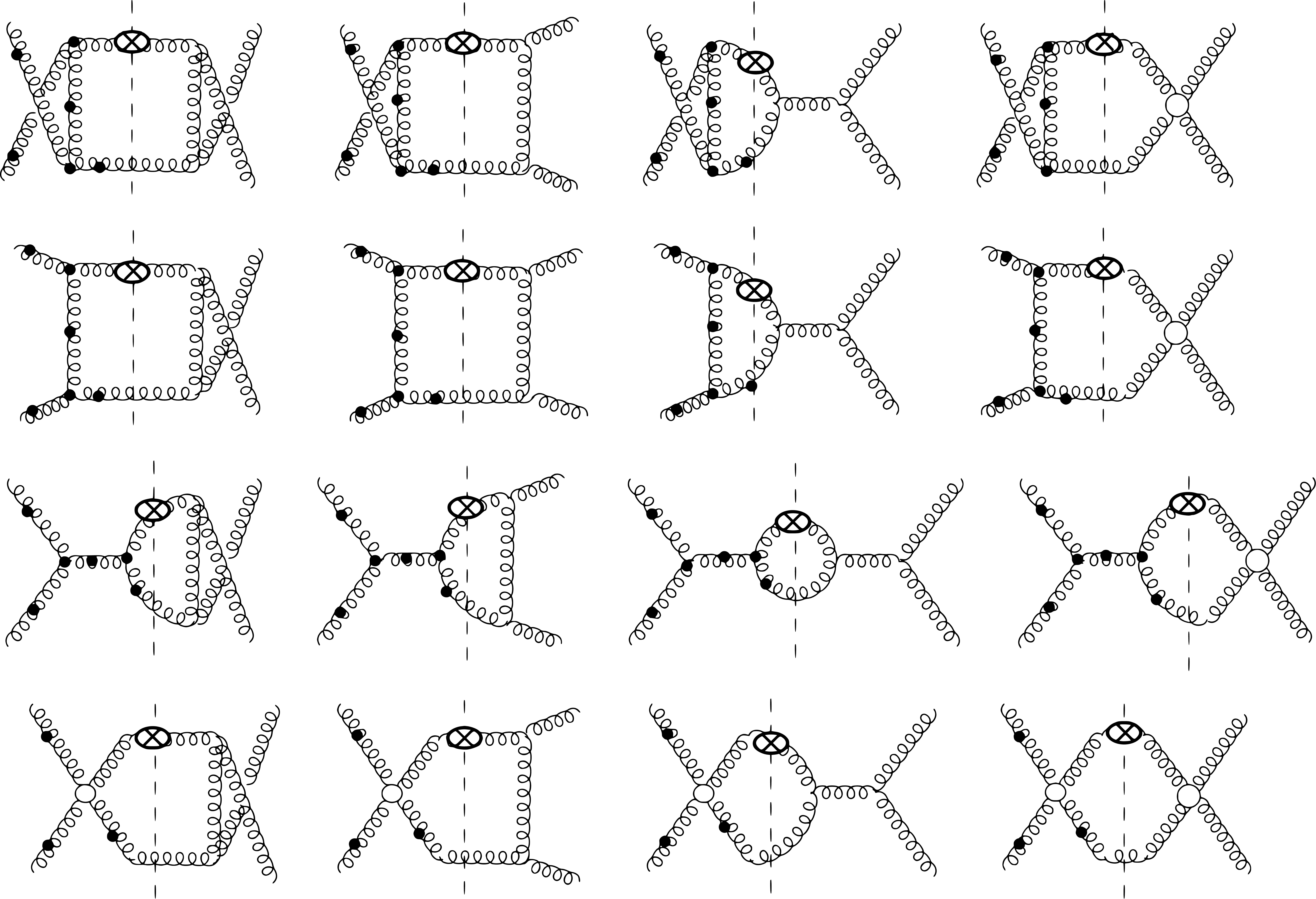}
  \end{center}
\caption{Diagrams in the $gg\to gg$ channel for the hard parts $S(k)$ and $S_{L}(z',z)$ in
eq. (\ref{PPgfragForma}).  For $S_{L}(z',z)$, 
an extra gluon line connecting $\otimes$ and each of the black dots should be added for each diagram.  
A white circle represents a four-gluon vertex, making clear the difference
from the attachment of the extra gluon line.}
 \label{fig_gggg1}
\end{figure}

\begin{figure}[h]
  \begin{center}
   \includegraphics[width=7.0cm]{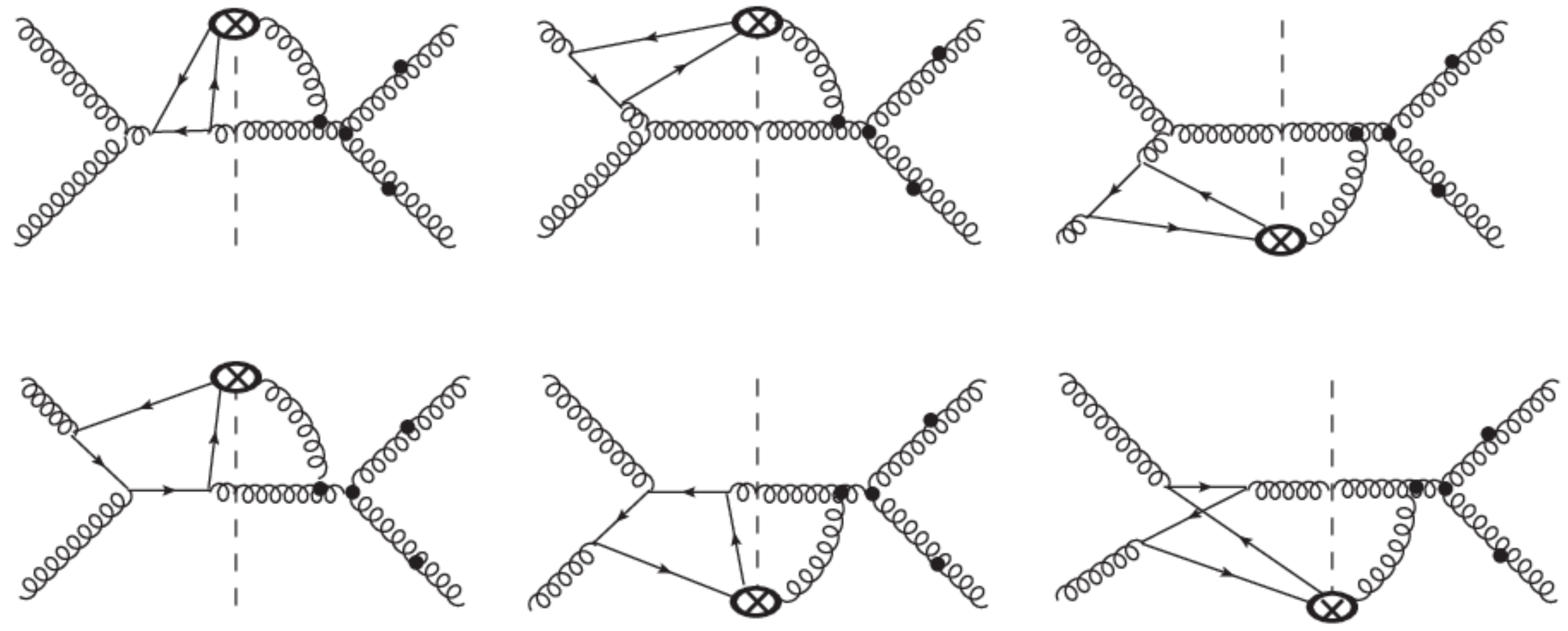}
  \end{center}
\caption{Diagrams contributing to the hard part in Fig. \ref{PP_qqg}
in the $gg\to gg$ channel.  
For all diagrams, a quark loop with the reversed arrow
also needs to be included. }
\label{fig_gggg2}
\end{figure}

\noindent
{(3) $gg\to gg$ channel} (Figs. \ref{fig_gggg1}, \ref{fig_gggg2})\\
\begin{align}
&\hat{\sigma}_{G}=-\frac{N}{C_F}\frac{2 (s^2 + s t + t^2)^2 (s - t) (2 s + t) (s + 2 t)}{s^3 t^3 (s + t)^3},
&
&\hat{\sigma}_{H}=0,\nn\\
&\hat{\sigma}_{N1}=-\frac{N}{C_F}\frac{2 (t^2 + t u + u^2)^2 (t - u) (2 t^2 + 7 t u + 2 u^2)}{t^3 u^3 (t + u)^3},
&
&\hat{\sigma}_{N2}=0,
&
&\hat{\sigma}_{N3}=-\hat{\sigma}_{N1},\nn\\
&\hat{\sigma}_{DN1}=\frac{N}{C_F}\frac{(t^2 + t u + u^2)^2 (t - u) (2 t^2 + 3 t u + 2 u^2)}{t^3 u^3 (t + u)^3},
&
&\hat{\sigma}_{DN2}=\hat{\sigma}_{DN1},
&
&\hat{\sigma}_{DN3}=-2 \hat{\sigma}_{DN1},\nn
\end{align}
\begin{align}
&\hat{\sigma}_{DF1}=
\frac{N}{C_F}\frac{(t^2 + t u + u^2)^2 (t - u) (2 t^2 + 7 t u + 2 u^2)}{t^3 u^3 (t + u)^3},\nn\\
&\hat{\sigma}_{DF2}=-\frac{N}{C_F}\frac{(t^2 + t u + u^2)^2 (t - u) (2 t + u) (t + 2 u)}{4 t^3 u^3 (t + u)^3},\nn\\
&\hat{\sigma}_{DF3}=-\frac{N}{C_F}\frac{(t^2 + t u + u^2)^2 (t - u) (2 t + u) (t + 2 u)}{4 t^3 u^3 (t + u)^3},\nn
\end{align}
\begin{align}
&\hat{\sigma}_{GF1}=0,
&
&\hat{\sigma}_{GF2}=-\hat{\sigma}_{DF2},
&
&\hat{\sigma}_{GF3}=\hat{\sigma}_{DF3}.
\end{align}


\section{Summary and conclusion}

In this paper we have studied the transverse polarization of a spin-1/2 hyperon produced
in the unpolarized proton-proton collision, $pp\to \Lambda^\uparrow X$,
within the framework of the collinear twist-3 factorization, which is relevant for the
large-$p_T$ hyperon production.  We focused on the contribution
from the twist-3 gluon FFs,
which had never been studied in previous studies.
To this end we have developed a formalism to
include all effects associated with the twist-3 gluon FFs.  
The twist-3 cross section
receives contributions from three types of the gluon FFs, i.e., intrinsic, kinematical 
and dynamical (purely gluonic and quark-antiquark-gluon type) ones.  
Applying the formalism we have calculated the LO cross section
for $pp\to \Lambda^\uparrow X$.  This completes the LO twist-3 cross section
combined with the known results for the other contributions from the
twist-3 distribution in the unpolarized proton and the twist-3 quark FFs for the hyperon.
Using the EOM relation and the LIRs for the twist-3 gluon FFs, we have shown 
that the derived cross section satisfies the color gauge invariance and the
frame independence.  
Since the formalism developed here is a general one, it can be applied to other
processes to which the twist-3 gluon FFs contribute.

\section*{Acknowledgments}

We thank Daniel Pitonyak and Andreas Metz for discussions and comments
on the manuscript.  
This work has been supported by the Grant-in-Aid for
Scientific Research from the Japanese Society of Promotion of Science
under Contract Nos.~19K03843 (Y.K.) and 18J11148 (K.Y.),
National Natural Science Foundation in China 
under grant No. 11950410495, Guangdong Natural Science Foundation under
No. 2020A1515010794
and research startup funding at South China
Normal University.



\hspace{1cm}

\appendix

\LARGE

\noindent
{\bf Appendices}

\normalsize

\section{Ward identities for the gluon fragmentation channels}\label{Ward}
\subsection{Derivation of the Ward identities}
To get a twist-3 gluon fragmentation contribution to the cross section in a gauge invariant form, 
one needs to fully utilize the Ward identities for the hard parts
(\ref{ward1})-(\ref{ward7}).  Here we present their derivation.  

\subsubsection{$q\bar{q}\to gg$ channel}
Ward identities in this channel read
\begin{eqnarray}
{(k_2-k_1)}_{\lambda}S^{\mu\nu\lambda,abc}_L(k_1,k_2)&=&
\frac{-if^{abc}}{N^2-1}S^{\mu\nu}(k_2)+G_{q\bar{q}\to gg}^{\mu\nu, abc}(k_1,k_2),
\label{wardL1}\\
{k_1}_{\mu}S^{\mu\nu\lambda,abc}_L(k_1,k_2)&=&\frac{if^{abc}}{N^2-1}S^{\lambda\nu}(k_2)+G_{q\bar{q}\to gg}^{\lambda\nu, cab}(k_2-k_1,k_2),
\label{wardL2}\\
{k_2}_{\nu}S^{\mu\nu\lambda,abc}_L(k_1,k_2)&=&0,
\label{wardL3}
\end{eqnarray}
\begin{eqnarray}
{(k_2-k_1)}_{\lambda}S^{\mu\nu\lambda,abc}_R(k_1,k_2)&=&
\frac{if^{abc}}{N^2-1}S^{\mu\nu}(k_1)
+\left(G_{q\bar{q}\to gg}^{\nu\mu, bac}(k_2,k_1)\right)^\star, 
\label{wardR1}
\\
{k_2}_{\nu}S^{\mu\nu\lambda,abc}_R(k_1,k_2)&=&
\frac{if^{abc}}{N^2-1}S^{\mu\lambda}(k_1)
+\left(G_{q\bar{q}\to gg}^{\lambda\mu, cab}(k_1-k_2,k_1)\right)^\star,
\label{wardR2}\\
{k_1}_{\mu}S^{\mu\nu\lambda,abc}_R(k_1,k_2)&=&0,
\label{wardR3}
\end{eqnarray}
where
$S^{\mu\nu\lambda,abc}_{L(R)}(k_1,k_2)$ represents a hard part 
for a three-gluon correlation functions
which has two gluon legs in 
the left (right) of the final-state cut, and
$S^{\mu\nu}(k)\equiv S^{\mu\nu}_{ab}(k)\delta^{ab}$ represents a hard part 
for a two-gluon correlation functions.
$G$-terms in the right hand side of these equations are ghost terms.  
\begin{figure}[h]
\centering
\includegraphics[width=15cm]{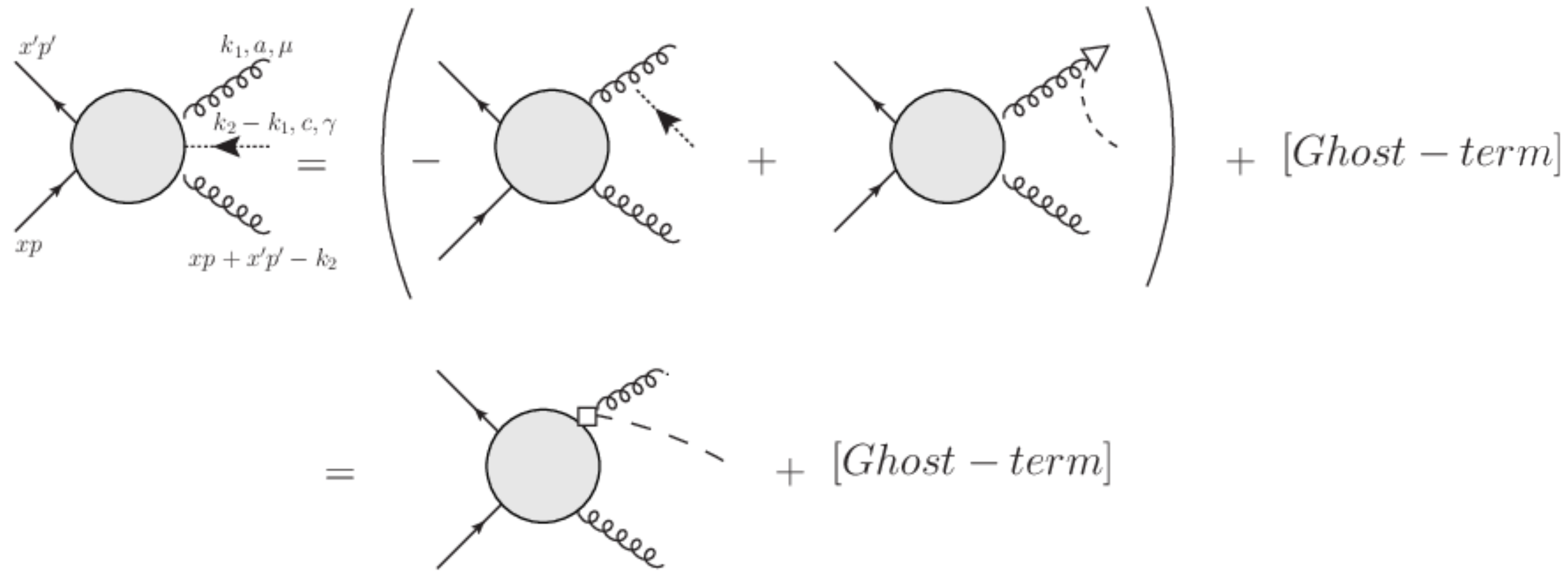}
\caption{
Ward identity (\ref{wardL1}) for the
$q\bar{q}\to gg$ channel.  The figure shows the left side of the cut, representing the 
attachment of the scalar polarized
gluon (dotted line) with the momentum $k_2-k_1$.  
Note that the hard part represented in the LHS does not contain the diagram in which 
the scalar polarized gluon is directly attached to the gluon line fragmenting into the final hadron, 
hence it appears in the first term of the RHS.
For the meaning of the notations in the figure, see Appendix A.2.}
\label{wardL1_fig}
\end{figure}
Figure \ref{wardL1_fig} shows the Ward identity ({\ref{wardL1}}), which states that
the attachment of the scalar polarized gluon with the momentum $k_2-k_1$ to the hard part
$S_{L\,\mu\nu\lambda}^{abc}(k_1,k_2)$ can be decomposed into the
two-body hard part and the ghost terms.  
To identify the ghost terms, we consider the diagrams in Fig. \ref{All3g}.  
The first term in the LHS of the figure represents the hard part 
$S_{L\,\mu\nu\lambda}^{abc}(k_1,k_2)$
and the
second term 
represents the attachment of the scalar polarized gluon to the gluon line fragmenting into $\Lambda^\uparrow$.  
The LO diagrams can be classified into three types as shown in the RHS of this figure.  
\begin{figure}[h]
\centering
  \includegraphics[width=15cm]{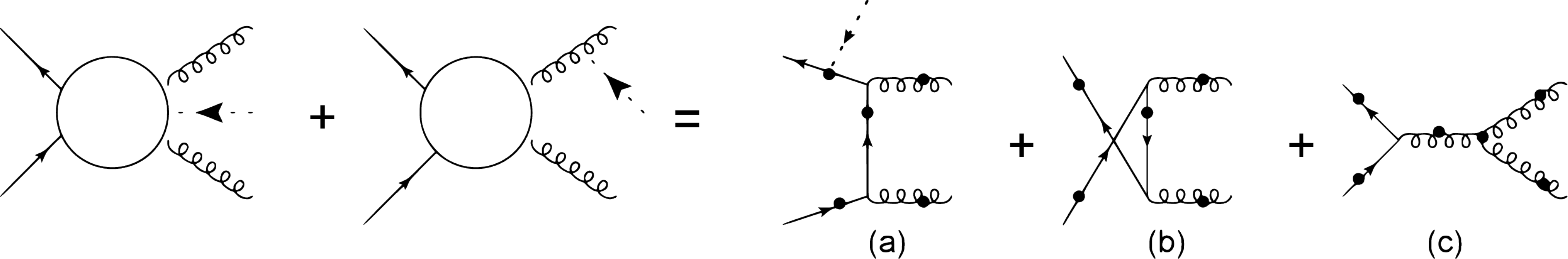}
\caption{Attachment of the scalar polarized gluon of momentum $k_2-k_1$ to $qq\to gg$ 
diagrams.  In the RHS, it is implied that the scalar polarized gluon is attached one of
the black dots in all possible ways. }
\label{All3g}
\end{figure}
Using the tree level Ward identities (See \ref{separate}), each diagram in Fig. \ref{All3g} can be 
decomposed into several pieces,
and some of them cancel each others owing to the on-shell condition of the external lines
(See \ref{cancel}).
Taking these facts into account we rearrange each term in the RHS of  Fig. \ref{All3g}.

\begin{figure}[h]
\centering
\includegraphics[width=13cm]{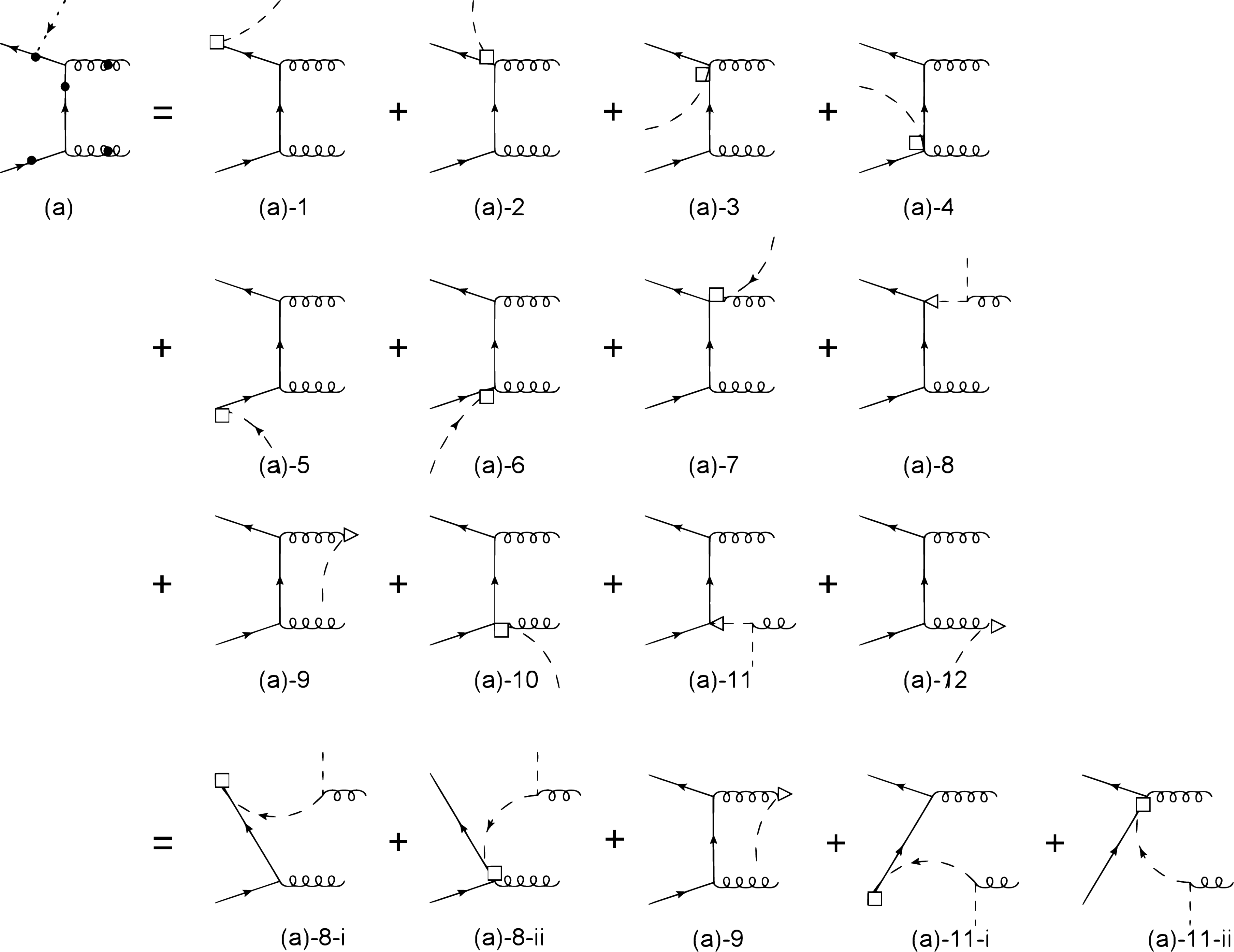}
\caption{Decomposition of Fig. \ref{All3g} (a) and the resulting terms due to the cancellation among diagrams.  
Diagrams (a)-1, (a)-5 and (a)-12 vanish owing to the 
on-shell condition.  Combination of (a)-2, (a)-3 and (a)-7 cancels that of (a)-4, (a)-6 and (a)-10
owing to (\ref{can_q}).}
\label{a}
\end{figure}

First, diagrams in Fig. \ref{All3g} (a) can be rewritten as in Fig. \ref{a}.  

\clearpage

\begin{figure}[h]
\centering
\includegraphics[width=14cm]{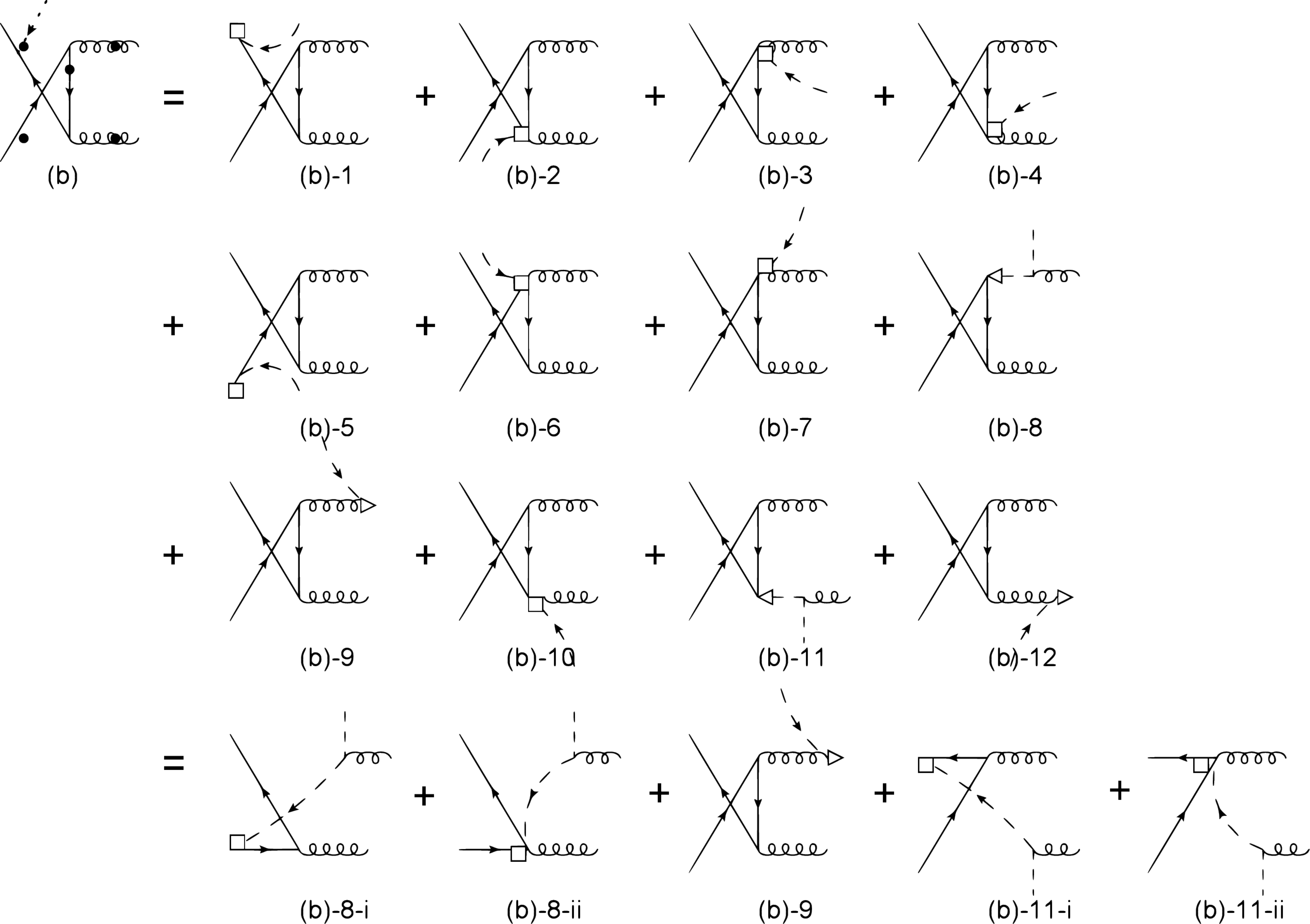}
\caption{Decomposition of Fig. \ref{All3g} (b) and the
resulting terms due to the cancellation among diagrams. 
Diagrams (b)-1, (b)-5 and (b)-12 vanish owing to the on-shell condition. 
Combination of
(b)-2, (b)-4 and (b)-10 cancels that of (b)-3, (b)-6 and (b)-7 owing to (\ref{can_q}).  
}
\label{b}
\end{figure}

Similarly Fig. \ref{All3g} (b) can be rewritten as in Fig. \ref{b}.  

\clearpage

\begin{figure}[h]
\centering
\includegraphics[width=15cm]{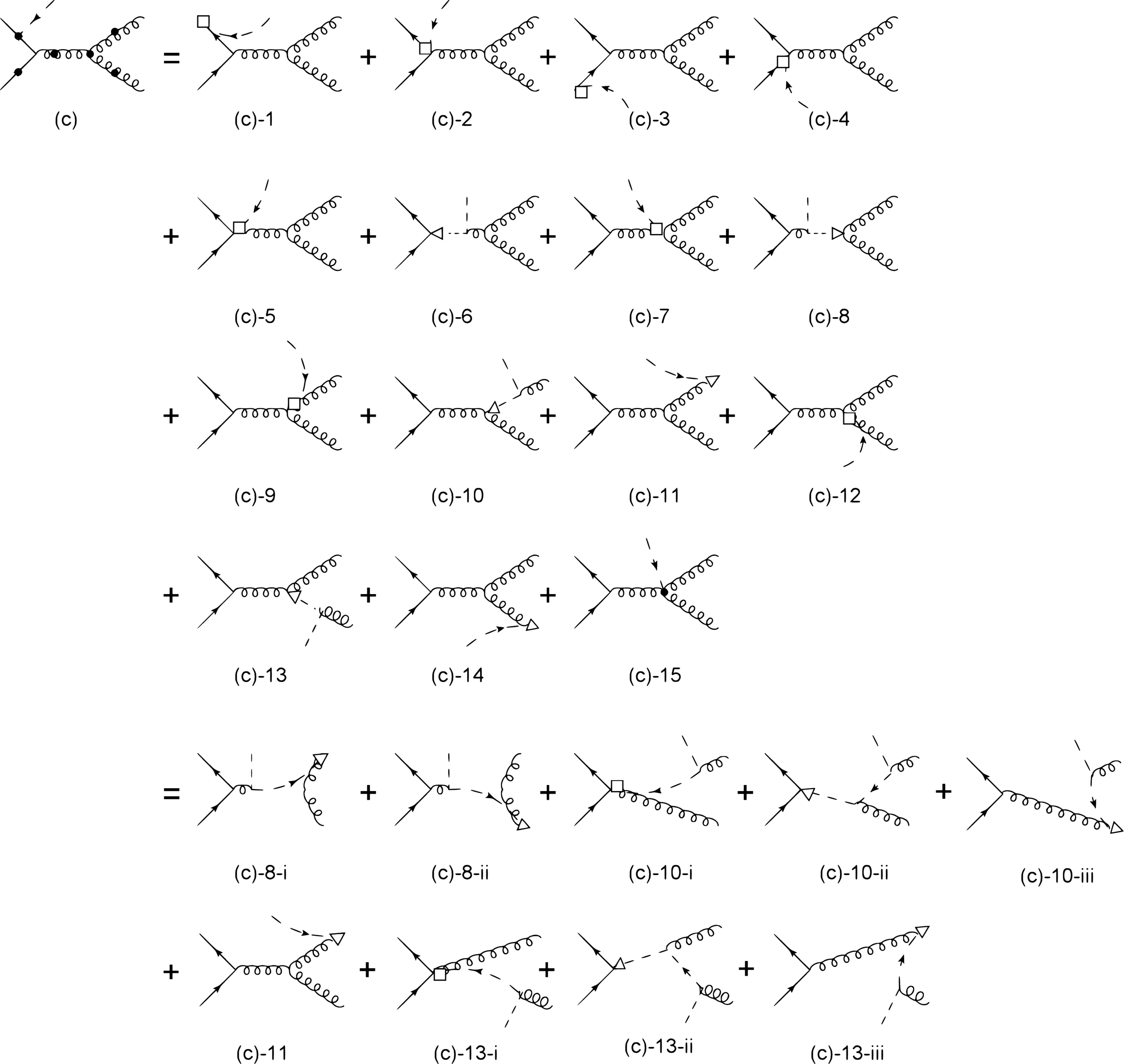}
\caption{
Decomposition of Fig. \ref{All3g} (c) and the resulting
terms due to the cancellation among the diagrams. Contributions from (c)-1, (c)-3, (c)-6 and (c)-14
vanish due to the on-shell conditions. Diagrams (c)-2, (c)-4 and (c)-5 cancel 
each others by (\ref{can_q}). 
Diagrams (c)-7, (c)-9, (c)-12 and (c)-15 cancel each others by 
(\ref{can_g}).
}
\label{c}
\end{figure}

Likewise Fig. \ref{All3g} (c) can be rewritten as in Fig. \ref{c}.  
\clearpage

\begin{figure}[h]
\centering
\includegraphics[width=10cm]{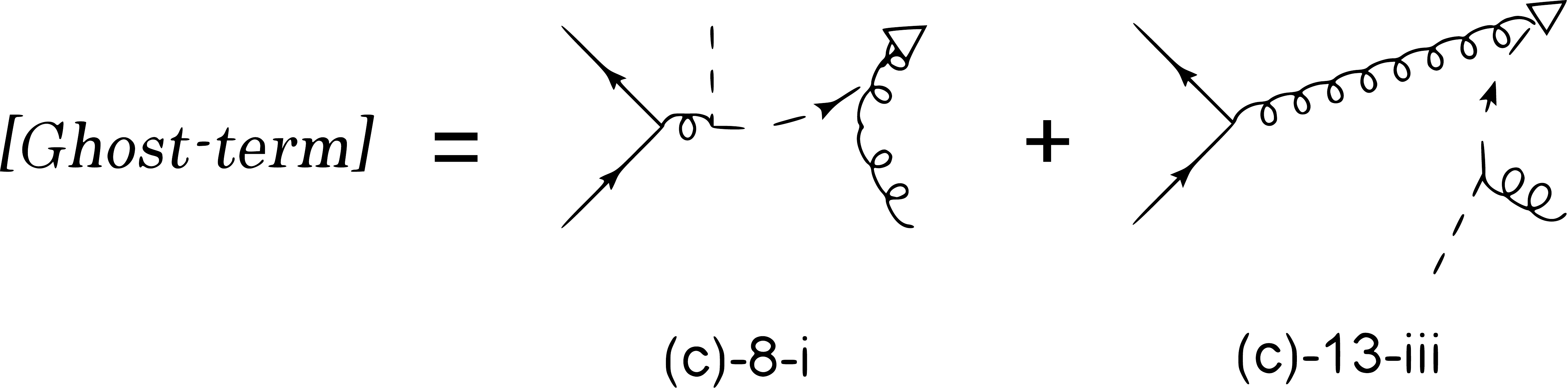}
\caption{Diagrams for the ghost-term (\ref{qqG}) in the
$q\bar{q}\to gg$ channel.}
\label{GhostTerm1}
\end{figure}

Using the on-shell condition, we find that diagrams
(a)-8-i, (a)-11-i, (b)-8-i, (b)-11-i and (c)-8-ii vanish, and diagrams
(c)-10-ii, (c)-10-iii and (c)-13-ii also vanish. 
In addition, contribution from (a)-8-ii, (b)-8-ii and (c)-10-i cancels among them, 
and likewise for the combination of (a)-11-ii, (b)-11-ii and (c)-13-i by (\ref{can_q}). 
Therefore, besides 
(a)-9, (b)-9 and (c)-11 (see Fig. \ref{wardL1_fig}), 
remaining contributions 
(c)-8-i and (c)-13-iii become the ghost term in (\ref{wardL1}).  
The ghost term takes the following structure in the
$q\bar{q}\to gg$ channel\footnote{In these appendices, we follow the convention of \cite{MutaQCD}
for the Feynman rule.}. 
\begin{eqnarray}
G_{q\bar{q}\to gg}^{\mu\nu, abc}(k_1,k_2)&=&-\Big{\{}(-if^{cef})(-if^{fda})d^{\mu\sigma}(k_1)
\frac{{(k_2-k_1)}_{\rho}}{(xp+x'p'-k_1)^2}
\nn\\
&+&(-if^{aef})(-if^{fdc})d^{\mu}_{\ \rho}(k_1)\frac{{(k_2-k_1)}^{\sigma}}{(xp+x'p'-(k_2-k_1))^2}
\Big{\}}
H^d_\sigma(xp,x'p'){1\over (xp+x'p')^2}
\nn\\
&\times&{P}^{\rho\gamma}(xp+x'p'-k_2)\times\left(M^{\nu,be}_\gamma(k_2)\right)^\star, 
\label{qqG}
\end{eqnarray}
where $d^{\mu\sigma}(k_1)\equiv k_1^2g^{\mu\sigma}-k_1^{\mu}k_1^{\sigma}$,
$H^d_\sigma(xp,x'p')$ represents the quark-line part in the left of the cut,
which is connected to the right of the cut $\left(M^{\nu,be}_\gamma(k_2)\right)^\star$, 
and the gluon's polarization tensor is given by
\begin{equation}
{P}^{\rho\gamma}(xp+x'p'-k_2)\equiv -g^{\rho\gamma}
+{(xp+x'p'-k_2)^{\rho} k_1^{\gamma}
+(xp+x'p'-k_2)^{\gamma} k_1^{\rho} \over (xp+x'p'-k_2)\cdot k_1}.
\end{equation}  
Diagrammatically the ghost term (\ref{qqG}) can be written as in Fig. \ref{GhostTerm1}.  
\\
This completes the derivation of the Ward identity in the $q\bar{q}\to gg$ channel.  


\subsubsection{
$qg\to gq$ channel}
Ward identities in this channel are given by 
(\ref{wardL1})-(\ref{wardR3}) except that the ghost terms
$G_{q\bar{q}\to gg}^{\mu\nu, abc}(k_1,k_2)$ are replaced by 
$G_{qg\to gq}^{\mu\nu, abc}(k_1,k_2)$ which takes the following structure:
\begin{eqnarray}
G_{qg\to gq}^{\mu\nu, abc}(k_1,k_2)&=&-\Big{\{}(-if^{cef})(-if^{fda})d^{\mu\alpha}(k_1)
\frac{{(k_2-k_1)}^{\lambda}}{(x'p'-k_1)^2}
\nn\\
&+&(-if^{aef})(-if^{fdc})d^{\mu\lambda}(k_1)\frac{{(k_2-k_1)}^{\alpha}}{(x'p'-(k_2-k_1))^2}
\Big{\}}H'^e_\lambda(xp,xp+x'p'-k_2)
\nn\\
&\times&{1\over (x'p'-k_2)^2}\left(M^{\nu,bd}_\alpha(k_2)\right)^\star,
\end{eqnarray}
where $H'^e_\lambda(xp,xp+x'p'-k_2)$ is the quark-line part connected to the left of the cut
$\left(M^{\nu,bd}_\alpha(k_2)\right)^\star$.  


\subsubsection{
$gg\to gg$ channel}
\begin{figure}[H]
\centering
\includegraphics[width=14cm]{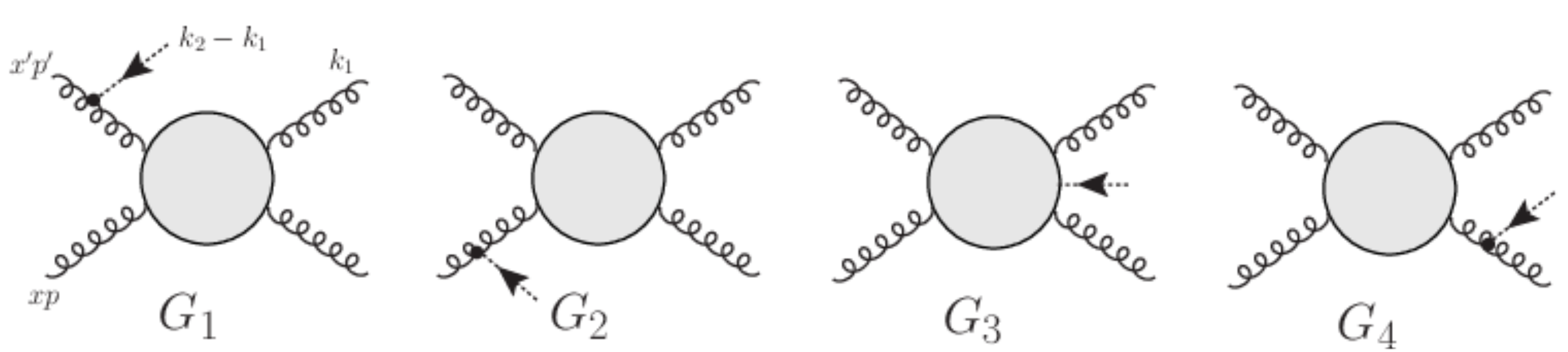}
\caption{Diagrammatic representation of the ghost terms in the 
$gg\to gg$ channel.  Depending on the position of attachment of the scalar polarized gluon, 
diagrams can be classified into 4 types.  Figure $G_3$ represents diagrams 
in which the scalar polarized gluon is attached to an internal propagator.  }
\label{ggG}
\end{figure}

Ward identities in this channel are given by
(\ref{wardL1})-(\ref{wardR3}) in which
$G_{q\bar{q}\to gg}^{\mu\nu, abc}(k_1,k_2)$ is replaced by the following
$G_{gg\to gg}^{\mu\nu, abc}(k_1,k_2)$:  
\begin{eqnarray}
G_{gg\to gg}^{\mu\nu, abc}(k_1,k_2)&\equiv&G_{1{gg\to gg}}^{\mu\nu, abc}(k_1,k_2)
+G_{2{gg\to gg}}^{\mu\nu, abc}(k_1,k_2)+G_{3{gg\to gg}}^{\mu\nu, abc}(k_1,k_2)
+G_{4{gg\to gg}}^{\mu\nu, abc}(k_1,k_2),
\label{ggGeq}
\end{eqnarray}
where $G_{1{gg\to gg}}$ through $G_{4{gg\to gg}}$ are the ghost terms which occur from four types 
of diagrams shown in Fig. \ref{ggG} for the hard scattering amplitudes in this channel.
Each term in (\ref{ggGeq}) takes the following structures: 
\begin{eqnarray}
G_{1{gg\to gg}}^{\mu\nu, abc}(k_1,k_2)&=&(-if^{cdg})\frac{{(k_2-k_1)}^\alpha}{(x'p'-(k_2-k_1))^2}
\Big{\{}(-if^{gfh})(-if^{hea})d^{\mu\rho}(k_1)\frac{{(x'p'-(k_2-k_1))}^{\beta}}{(xp+x'p'-(k_2-k_1))^2}\nn\\
&+&(-if^{geh})(-if^{hfa})d^{\mu\beta}(k_1)\frac{{(x'p'-(k_2-k_1))}_{\rho}}{(xp-k_1)^2}\nn\\
&-&(-if^{gha})V^{\beta, hfe}_{\sigma\rho}d^{\mu\sigma}(k_1)\frac{1}{(x'p'-k_2)^2}\Big{\}}\nn\\
&\times&{P}^{\rho\gamma}(xp+x'p'-k_2)\times
\left(M^{be, df}_{\nu\gamma, \alpha\beta}(k_2)\right)^\star,
\label{gg_G1}
\end{eqnarray}
\begin{eqnarray}
G_{2{gg\to gg}}^{\mu\nu, abc}(k_1,k_2)&=&(-if^{cfg})\frac{{(k_2-k_1)}^\beta}{(xp-(k_2-k_1))^2}
\Big{\{}(-if^{gdh})(-if^{hea})d^{\mu\rho}(k_1)\frac{{(xp-(k_2-k_1))}^{\alpha}}{(xp+x'p'-(k_2-k_1))^2}\nn\\
&-&(-if^{gha})V^{\alpha, hde}_{\sigma\rho}d^{\mu\sigma}(k_1)\frac{1}{(xp-k_2)^2}\nn\\
&+&(-if^{geh})(-if^{hda})d^{\mu\alpha}(k_1)\frac{{(xp-(k_2-k_1))}_{\rho}}{(x'p'-k_1)^2}\Big{\}}\nn\\
&\times&{P}^{\rho\gamma}(xp+x'p'-k_2)\times
\left(M^{be, df}_{\nu\gamma, \alpha\beta}(k_2)\right)^\star,
\label{gg_G2}
\end{eqnarray}
\begin{eqnarray}
G_{3{gg\to gg}}^{\mu\nu, abc}(k_1,k_2)&=&(-if^{cgh}) {(k_2-k_1)}^\sigma
\Big{\{}-(-if^{hea})V^{\alpha\beta, dfg}_{\sigma}d^{\mu\rho}(k_1)
\frac{1}{(xp+x'p'-(k_2-k_1))^2}\frac{1}{(xp+x'p')^2}\nn\\
&+&(-if^{gfa})V^{\alpha\rho, hde}_{\sigma}d^{\mu\beta}(k_1)
\frac{1}{(xp-k_1)^2}\frac{1}{{(xp-k_2)}^2}\nn\\
&+&(-if^{gda})V^{\beta\rho, hfe}_{\sigma}d^{\mu\alpha}(k_1)\frac{1}{(x'p'-k_2)^2}
\frac{1}{(x'p'-k_1)^2}\Big{\}}\nn\\
&\times&{P}_\rho^{\gamma}(xp+x'p'-k_2)\times
\left(M^{be, df}_{\nu\gamma, \alpha\beta}(k_2)\right)^\star,
\label{gg_G3}
\end{eqnarray}
and 
\begin{eqnarray}
G_{4{gg\to gg}}^{\mu\nu, abc}(k_1,k_2)&=&(-if^{ceg})\frac{{(k_2-k_1)}_\rho}{(xp+x'p'-k_1)^2}\Big{\{}-(-if^{gha})V^{\alpha\beta, dfh}_{\sigma}d^{\mu\sigma}(k_1)\frac{1}{(xp+x'p')^2}\nn\\
&+&(-if^{gdh})(-if^{hfa})d^{\mu\beta}(k_1)\frac{{(k_1-xp)}^{\alpha}}{(xp-k_1)^2}\nn\\
&+&(-if^{gfh})(-if^{hda})d^{\mu\alpha}(k_1)\frac{{(k_1-x'p')}^{\beta}}{(x'p'-k_1)^2}\Big{\}}\nn\\
&\times&{P}^{\rho\gamma}(xp+x'p'-k_2)\times
\left(M^{be, df}_{\nu\gamma, \alpha\beta}(k_2)\right)^\star, 
\label{gg_G4}
\end{eqnarray}
where
$V^{\beta, hfe}_{\sigma\rho}$ represents an appropriate three-gluon vertex in each channel.

\subsection{Decomposition of vertices}\label{separate}
\begin{figure}[h]
\centering
\includegraphics[width=10cm]{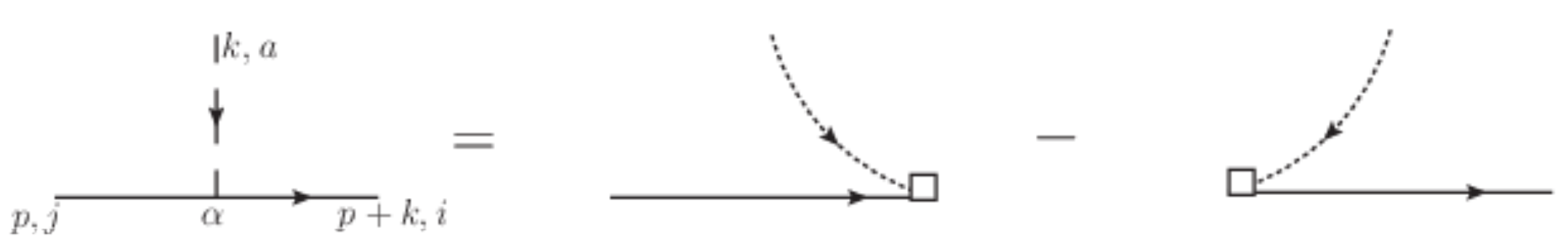}
\caption{Attachment of a scalar polarized gluon to a
quark-gluon vertex corresponding to (\ref{qg_vtx}). }
\label{qg_vtx_fig}
\end{figure}

Attachment of the scalar polarized gluon to a quark or gluon line can be decomposed 
as follows:  

\noindent
(i) quark-gluon vertex (Fig. \ref{qg_vtx_fig}): 
\begin{eqnarray}
k^{\alpha}\cdot (\gamma_\alpha)_{ij}T^a=(\pslash+\kslash)_{ij}T^a-(\pslash)_{ij} T^a,
\label{qg_vtx}
\end{eqnarray}
where $T^a$ is the generator of color SU(N) group.

\begin{figure}[h]
\centering
\includegraphics[width=10cm]{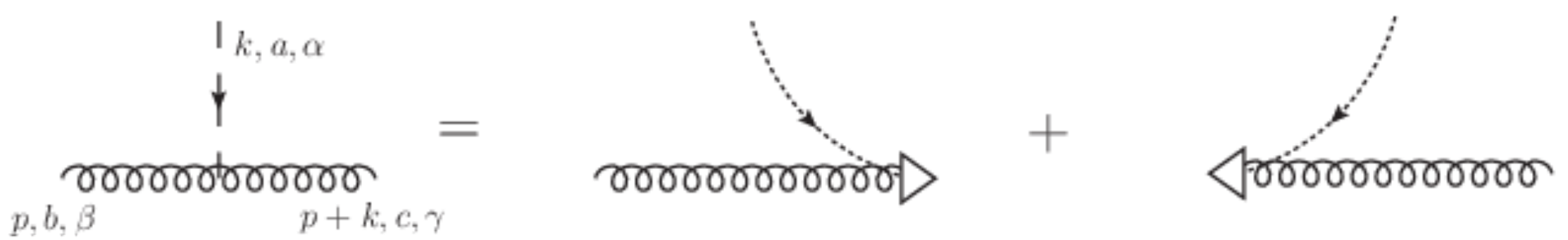}
\caption{Attachment of a scalar polarized gluon through a 3-gluon vertex
corresponding to (\ref{3g_vtx}).}
\label{3g_vtx_fig}
\end{figure}

\begin{figure}[h]
\centering
 \includegraphics[width=10cm]{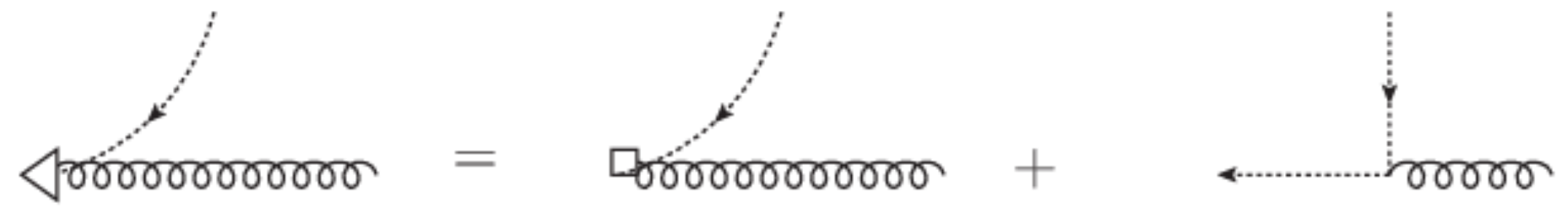}
\caption{Decomposition of $d^{\beta\gamma}$ in (\ref{d_dec}).}
\label{d_dec_fig}
\end{figure}

\noindent
(ii) 3-gluon vertex (Fig. \ref{3g_vtx_fig}): 
\begin{eqnarray}
k^\alpha\cdot V^{abc}_{\alpha\beta\gamma}(k, p, -p-k)={\it F}^{abc}d_{\beta\gamma}(p+k)
+{\it F}^{acb}d_{\beta\gamma}(p),
\label{3g_vtx}
\end{eqnarray}
where $V^{abc}_{\alpha\beta\gamma}(p_1, p_2, p_3)\equiv F^{abc} 
\left[(p_1-p_2)_{\gamma}g_{\alpha\beta}+(p_2-p_3)_{\alpha}g_{\beta\gamma}
+(p_3-p_1)_{\beta}g_{\alpha\gamma}\right]$
and $F^{abc}\equiv -i f^{abc}$.  
Furthermore, $d^{\beta\gamma}$ in the above equations is decomposed as (Fig. \ref{d_dec_fig})
\begin{eqnarray}
{\it F}^{acb}d^{\beta\gamma}(p)=p^2 g^{\beta\gamma}{\it F}^{acb}+(-p^{\beta}p^{\gamma}){\it F}^{acb}
\label{d_dec}.  
\end{eqnarray}

\newpage
\subsection{Cancellation among vertices}\label{cancel}
After decomposition of vertices, the following cancellation holds among vertices.  

\begin{figure}[h]
\centering
  \includegraphics[width=12cm]{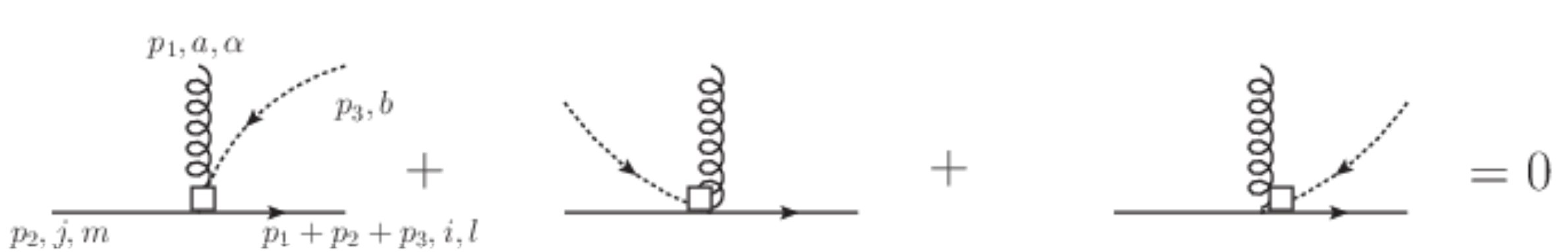}
\caption{Cancelation for quark-gluon vertices (\ref{can_q}).}
\label{can_q_fig}
\end{figure}

\noindent
(i) Quark-gluon vertex (Fig. \ref{can_q_fig})
\begin{eqnarray}
\left[-(T^aT^b)_{lm}+(T^bT^a)_{lm}+{\it F^{alb}}(T^l)_{lm}\right](\gamma^\alpha)_{ij}=0.  
\label{can_q}
\end{eqnarray}

\begin{figure}[h]
\centering
  \includegraphics[width=16cm]{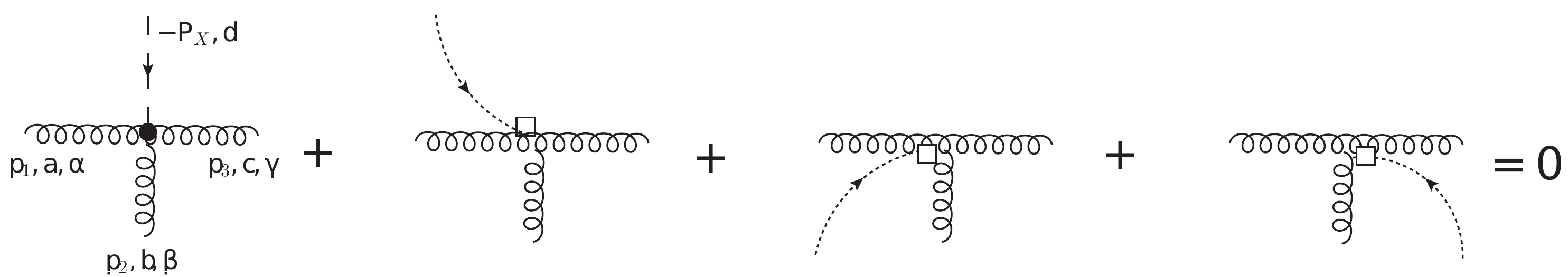}
  \caption{Cancellation among 3- and 4-gluon vertices (\ref{can_g}).}
\label{can_g_fig}
\end{figure}

\noindent
(ii) Purely gluon vertex (Fig. \ref{can_g_fig}) 
\begin{eqnarray}
(-P_X)^{\sigma}\cdot {\rm W}^{abcd}_{\alpha\beta\gamma\sigma}
&+&{\it F^{ald}}V^{lbc}_{\alpha\beta\gamma}(-p_2-p_3, p_2, p_3)\nn\\
&+&{\it F^{bld}}V^{alc}_{\alpha\beta\gamma}(p_1, -p_1-p_3, p_3)
+{\it F^{cld}}V^{abl}_{\alpha\beta\gamma}(p_1, p_2, -p_1-p_2)=0,
\label{can_g}
\end{eqnarray}
where $P_X\equiv p_1+p_2+p_3$.  

\section{Color gauge invariance of the twist-3 cross section}

Here we prove that the twist-3 cross section in (\ref{PPgfragForma}) supplemented by the
quark-antiquark-gluon contribution shown in Fig. 2
is color gauge invariant owing to the EOM relation (\ref{FFDFodd1}).  
We illustrate this property for the $qg\to gq$ channel.   The proof for 
other channels is essentially the same.
The cross section in this channel can be written as
\begin{eqnarray}
E_{h}\frac{d\sigma(p,p',P_h;S_\perp)}{d^3P_h}=
\frac{1}{16\pi^2S_E}\int\frac{dx}{x}f_1(x)
\int\frac{dx'}{x'}G(x')\int{dz}\left(-{1\over 2}g_\perp^{\rho\sigma}(p')\right)W_{\rho\sigma}(xp,x'p,P_h/z),\nn\\
\end{eqnarray}
where $G(x')$ is the unpolarized gluon distribution in the proton with momentum $p'$,
and the Lorentz indices $\rho$ and $\sigma$ are contracted 
with 
$g_\perp^{\rho\sigma}(p')\equiv g^{\rho\sigma} - p'^\rho n'^\sigma -p'^\sigma n'^\rho$
where $n'$ is the usual lightlike vector satisfying $p'\cdot n'=1$ 
to extract the contribution from $G(x')$.  From (\ref{PPgfragForma}), we can read off 
$W_{\rho\sigma}(xp,x'p',P_h/z)$ as
\begin{eqnarray}
&&W_{\rho\sigma}(xp,x'p',P_h/z)=
{\Omega^\mu_{\ \alpha}}{\Omega^\nu_{\ \beta}}
{\rm Tr}
\left[
\hat{\Gamma}^{\alpha\beta}(z)S_{\mu\nu,\rho\sigma}(P_h/z)
\right]
\nn\\
&& 
-i\,{\Omega^\mu_{\ \alpha}}{\Omega^\nu_{\ \beta}}{\Omega^\lambda_{\ \gamma}}
{\rm Tr}
\left[
\hat{\Gamma}_{\del}^{\alpha\beta\gamma}(z)
\left.\frac{\del S_{\mu\nu,\rho\sigma}(k)}{\del k^\lambda}\right|_{c.l.}
\right]
+
{\Re}\left\{
i\,
{\Omega^\mu_{\ \alpha}}{\Omega^\nu_{\ \beta}}{\Omega^\lambda_{\ \gamma}}
{1\over z} \int \frac{dz'}{z'}\,
\left(\frac{1}{1/z-1/z'}\right)\right.\nn\\
&&\qquad\left.
\times
{\rm Tr}
\left[
\left(
-\frac{if^{abc}}{N}
\hat{\Gamma}_{FA}^{\alpha\beta\gamma}\left(\ozd,\oz\right)
+d^{abc}\frac{N}{N^2-4}
\hat{\Gamma}_{FS}^{\alpha\beta\gamma}\left(\ozd,\oz\right)
\right)
S^{L,abc}_{\mu\nu\lambda,\rho\sigma}(z',z)\right]
\right\} \nn\\
&&+w^{q\bar{q}g}_{\rho\sigma}(xp,x'p',P_h/z),
\label{Wtensor}
\end{eqnarray}
where the hard part $S^{ab}_{\mu\nu,\rho\sigma}(k)$ 
is related to the hard part
$S^{ab}_{\mu\nu}(k)$ in (\ref{PPgfragForma}) by 
$\left(-{1\over 2}g_\perp^{\rho\sigma}(p')\right)S^{ab}_{\mu\nu,\rho\sigma}(k)
=S^{ab}_{\mu\nu}(k)$ and likewise for $S^{abc}_{\mu\nu\lambda,\rho\sigma}(k)$, and
$w^{q\bar{q}g}_{\rho\sigma}(xp,x'p',P_h/z)$ represents the contribution from Fig. 2.  
Then the color gauge invariance of the cross section implies 
\beq
x'p'^\rho W_{\rho\sigma}(xp,x'p',P_h/z)=x'p'^\sigma W_{\rho\sigma}(xp,x'p',P_h/z)=0. 
\label{cgaugeinv}
\eeq
To show (\ref{cgaugeinv}), we use the EOM relation (\ref{FFDFodd1}) and
eliminate the intrinsic twist-3 FF $\Delta \widehat{G}_{3\bar{T}}(z)$ in 
the first term of the RHS of (\ref{Wtensor}).
Then $W_{\rho\sigma}$ can be decomposed into three pieces:
\beq
W_{\rho\sigma}(xp,x'p',P_h/z)=W^{({\rm i})}_{\rho\sigma}(xp,x'p',P_h/z)+
W^{({\rm ii})}_{\rho\sigma}(xp,x'p',P_h/z)+W^{({\rm iii})}_{\rho\sigma}(xp,x'p',P_h/z),
\label{Wdecomp}
\eeq
where $W^{({\rm i})}_{\rho\sigma}$ represents the contribution from the dynamical
3-body correlation function in (\ref{gFraDA}) and (\ref{gFraDS}), 
$W^{({\rm ii})}_{\rho\sigma}$ represents the one from
the dynamical $q\bar{q}g$-correlation function in (\ref{FFtilde}) 
and $W^{({\rm iii})}_{\rho\sigma}$ represents
the one from the kinematical
FFs in (\ref{gFraK}).  To show (\ref{cgaugeinv}), it suffices to prove that each term
of (\ref{Wdecomp}) separately satisfies (\ref{cgaugeinv}).

\subsection{Contribution from dynamical FFs: $W^{({\rm i})}_{\rho\sigma}$}

\begin{figure}[h]
\centering 
\includegraphics[width=14cm]{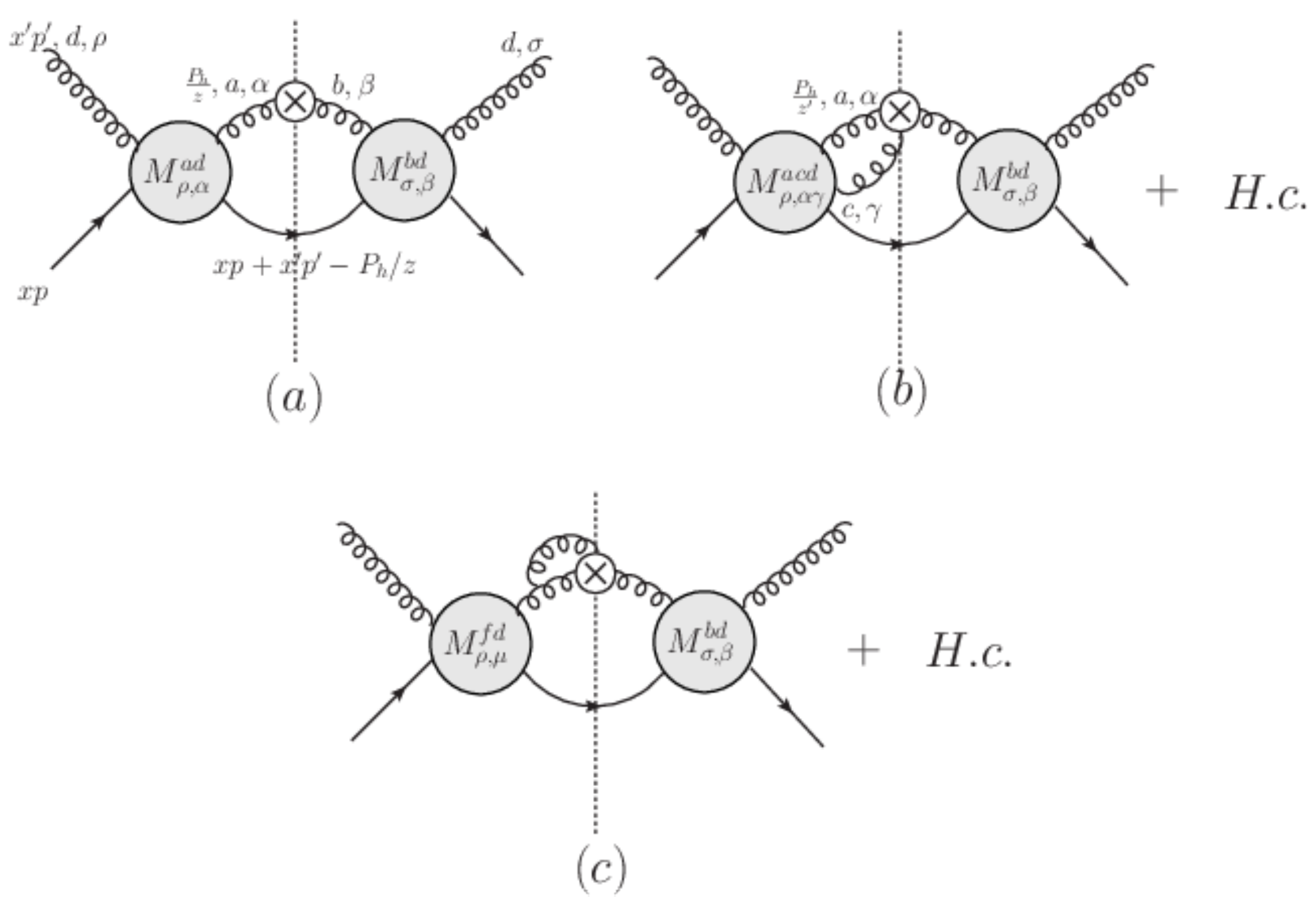}
\caption{Diagrams contributing to $W^{({\rm i})}_{\rho\sigma}$.    
$M^{ad}_{\rho, \alpha}$ and
$M^{acd}_{\rho, \alpha\gamma}$, respectively, represent the $qg\to gq$ and $qg\to ggq$ scattering amplitudes, 
which constitute the hard cross section as shown in the figure.}
\label{part_fig1}
\end{figure}

\begin{figure}[h]
\centering
\includegraphics[width=7cm]{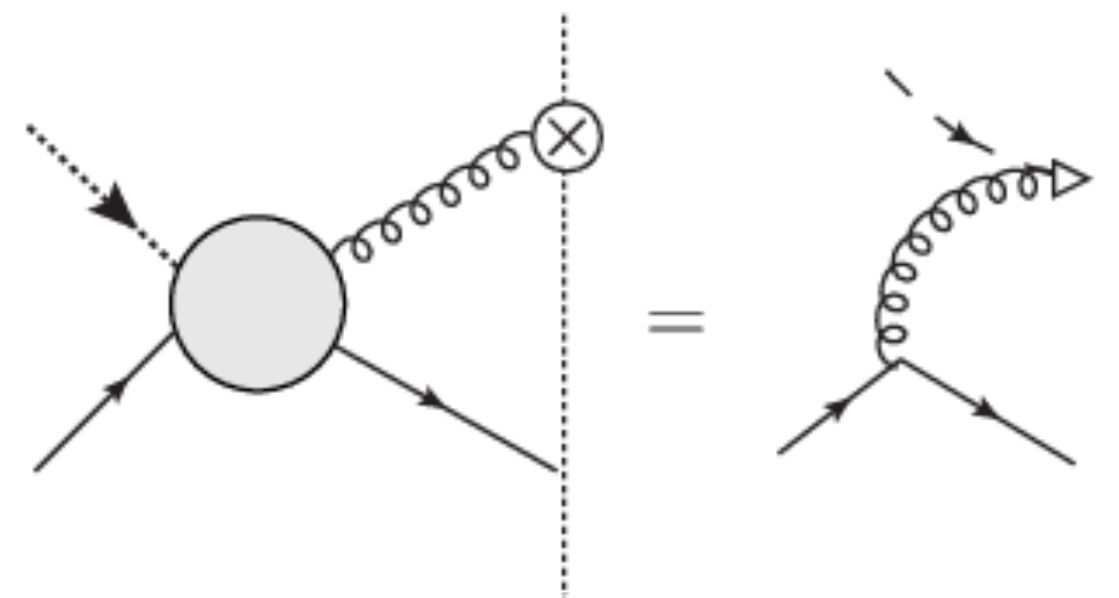}
\caption{Diagrammatic representation for $x'p'^{\rho}W^{(a)}_{\rho\sigma}$.}
\label{a_part}
\end{figure}

We first show $W^{({\rm i})}_{\rho\sigma}$ satisfies (\ref{cgaugeinv}).  
Relevant diagrams are shown in Figs. \ref{part_fig1} (a), (b) and (c).  
Here Fig. \ref{part_fig1} (b) is meant to contain diagrams including those of Fig. \ref{part_fig1} (c),
which is not a part of partonic cross sections for the dynamical FFs. 
Therefore,
after using the EOM relations, we get the following combination of the hard cross section, 
\begin{eqnarray}
W^{({\rm i})}_{\rho\sigma}=
W^{(a)}_{\rho\sigma}+
\left(W^{(b)L}_{\rho\sigma}-W^{(c)L}_{\rho\sigma}\right)
+\left(W^{(b)R}_{\rho\sigma}-W^{(c)R}_{\rho\sigma}\right),
\label{part1}
\end{eqnarray}
where the hard part with index $L$ indicates diagrams in Fig. \ref{part_fig1} (b) and (c), and that
with $R$ indicates their hermitian conjugate diagrams.  
We define the amplitude $M^{ad}_{\rho, \alpha}$ for $qg\to gq$ scattering and 
$M^{acd}_{\rho, \alpha\gamma}$ for $qg\to ggq$ scattering as shown in Fig. \ref{part_fig1}.  
With the physical polarizations for the Lorentz indices $\alpha$ and $\gamma$,
these amplitudes satisfy the following relations, 
\begin{eqnarray}
x'p'^{\rho}M^{ad}_{\rho, \alpha}=0\label{per_van1},\qquad
x'p'^{\rho}M^{acd}_{\rho, \alpha\gamma}=0.  
\label{per_van2}
\end{eqnarray}
This implies
\begin{eqnarray}
x'p'^{\rho}W^{(b)L}_{\rho\sigma}=0,\qquad
x'p'^{\rho}W^{(b)R}_{\rho\sigma}=0,\qquad
x'p'^{\rho}W^{(c)R}_{\rho\sigma}=0.  
\end{eqnarray}\\
We thus have
\begin{eqnarray}
x'p'^{\rho}W^{({\rm i})}_{\rho\sigma}=x'p'^{\rho}W^{(a)}_{\rho\sigma}-x'p'^{\rho}W^{(c)L}_{\rho\sigma}.  
\label{remnant}
\end{eqnarray}
The first term in the RHS of (\ref{remnant}) reads
\begin{eqnarray}
x'p'^{\rho}W^{(a)}_{\rho\sigma}&=&
\frac{F^{ade}}{(x'p'-P_h/z)^2}d^{\alpha\rho}(P_h/z)
\times{\rm Tr}\left[\Pxslash\gamma_\rho\,\,\pslash \,T^e M^{\star, ad}_{\sigma, \beta}\right]\nn\\
&\times&\left[-w^\alpha\epsilon^{\beta P_h w S_\perp}-
w^\beta\epsilon^{\alpha P_h w S_\perp}\right]\left.{\Delta \hat{G}_{3\bar{T}}(z)}\right|_{\rm 3-gluon}\nn\\
&=&\frac{1}{z}\epsilon^{\beta P_h w S_\perp}\frac{F^{ade}}{(x'p'-P_h/z)^2}\times{\rm Tr}\left[\Pxslash \Phslash\,\,\pslash \,T^e M^{\star, ad}_{\sigma, \beta}\right]\left.\frac{\Delta \hat{G}_{3\bar{T}}(z)}{z}\right|_{\rm 3-gluon},
\label{a_calc}
\end{eqnarray}
where $P_X=xp+x'p'-P_h/z$, $F^{abc}\equiv -i f^{abc}$
and $d^{\mu\sigma}(k)\equiv k^2g^{\mu\sigma}-k^{\mu}k^{\sigma}$ and
we have denoted the $\widehat{N}_i$-contribution to $\Delta \hat{G}_{3\bar{T}}(z)$
in the RHS of (\ref{FFDFodd1}) by
$\Delta \hat{G}_{3\bar{T}}(z)|_{\rm 3-gluon}$.  
This result is shown in Fig. \ref{a_part}.  

\begin{figure}[h]
\centering
\includegraphics[width=7cm]{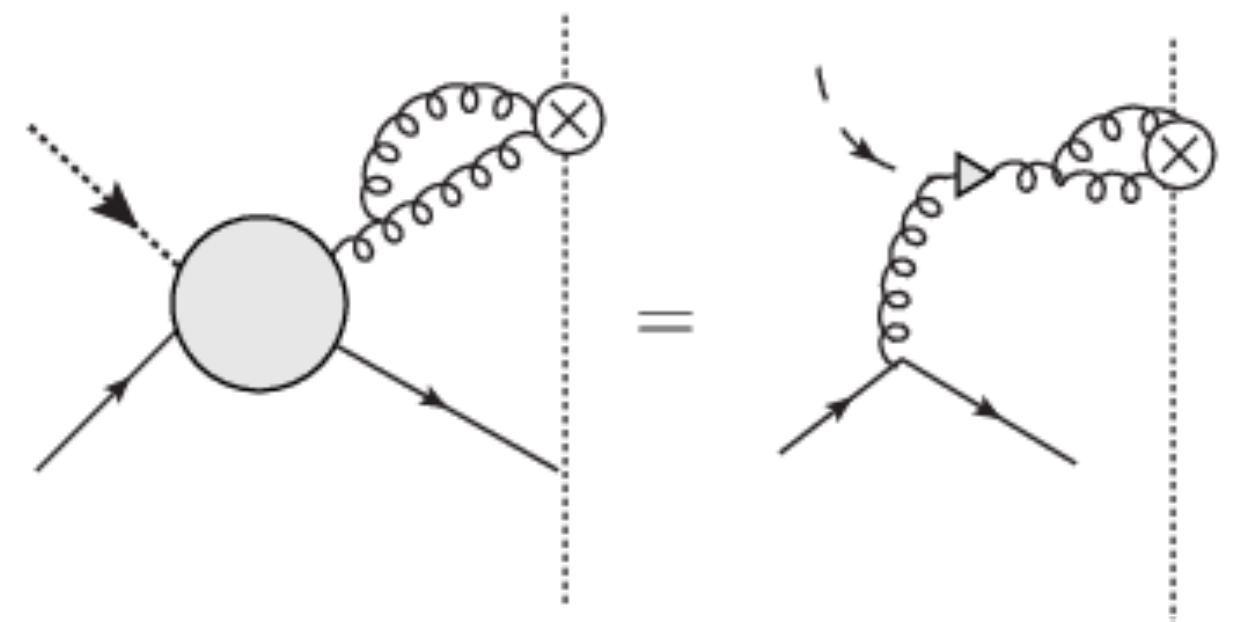}
\caption{Diagrammatic representation for
$x'p'^{\rho}W^{(c)L}_{\rho\sigma}$ in the $qg\to gq$ channel.}
\label{c_part}
\end{figure}

Next we consider the second term in the RHS of (\ref{remnant}).   
Since its hard part is proportional to $f^{abc}$, the contribution from 
$\hat{\Gamma}_{FS}^{\alpha\beta\gamma}$ drops.  The contribution from 
$\hat{\Gamma}_{AS}^{\alpha\beta\gamma}$ can be written as
\begin{eqnarray}
&&x'p'^{\rho}W^{(c)L}_{\rho\sigma}\nn\\
&&\sim
-\frac{1}{2}z\epsilon^{\beta P_h w S_\perp}\frac{F^{ade}}{(x'p'-P_h/z)^2}
\int d\left(\frac{1}{z'}\right) \frac{z'}{z^2} \frac{1}{1-z/z'}\nn\\
&&\times\left(1-2\frac{z}{z'}\right){\rm Tr}\left[\Pxslash \Phslash\,\,\pslash \,T^e
M^{\star, ad}_{\sigma, \beta}\right]\times\left[-\Nhat_{2}\left(\oz-\ozd, \oz\right)+
\Nhat_{2}\left(\ozd, \oz\right)+2\Nhat_{1}\left(\ozd, \oz\right)\right]\nn\\
&&=-\frac{1}{2}z\epsilon^{\beta P_h w S_\perp}\frac{F^{ade}}{(x'p'-P_h/z)^2}
{\rm Tr}\left[\Pxslash \Phslash\,\,\pslash \,T^e M^{\star, ad}_{\sigma, \beta}\right]\nn\\
&&\times\int d\left(\frac{1}{z'}\right) \frac{1}{z} \left\{-\frac{1}{1-z/z'}
+\frac{z'}{z}\right\}\left[-\Nhat_{2}\left(\oz-\ozd, \oz\right)+\Nhat_{2}\left(\ozd, \oz\right)+
2\Nhat_{1}\left(\ozd, \oz\right)\right]\nn\\
&&=-\frac{1}{2}z\epsilon^{\beta P_h w S_\perp}\frac{F^{ade}}{(x'p'-P_h/z)^2}
{\rm Tr}\left[\Pxslash \Phslash\,\,\pslash \,T^e M^{\star, ad}_{\sigma, \beta}\right]\nn\\
&&\times\int d\left(\frac{1}{z'}\right) \frac{1}{z} \left(-2\times\frac{1}{1-z/z'}\right)
\left[-\Nhat_{2}\left(\oz-\ozd, \oz\right)+\Nhat_{2}\left(\ozd, \oz\right)+2\Nhat_{1}
\left(\ozd, \oz\right)\right]\nn\\
&&=\frac{1}{z}\epsilon^{\beta P_h w S_\perp}\frac{F^{ade}}{(x'p'-P_h/z)^2}
{\rm Tr}\left[\Pxslash \Phslash\,\,\pslash \,T^e M^{\star, ad}_{\sigma, \beta}\right]\nn\\
&&\times\int d\left(\frac{1}{z'}\right) \frac{1}{1/z-1/z'}
\left[-\Nhat_{2}\left(\oz-\ozd, \oz\right)+\Nhat_{2}\left(\ozd, \oz\right)+2\Nhat_{1}
\left(\ozd, \oz\right)\right],
\label{c_calc}
\end{eqnarray}
which can be diagrammatically written as Fig. \ref{c_part}.  
Using the EOM relation (\ref{FFDFodd1}), one finds (\ref{a_calc})
is equal to (\ref{c_calc}), which implies (\ref{remnant})=0.   
This completes the proof for the relation 
$x'p'^\rho W^{({\rm i})}_{\rho\sigma}(xp,x'p',P_h/z)=0$.  

\subsection{Contribution from quark-antiquark-gluon FF: $W^{({\rm ii})}_{\rho\sigma}$}

\begin{figure}[h]
\centering
  \includegraphics[width=12cm]{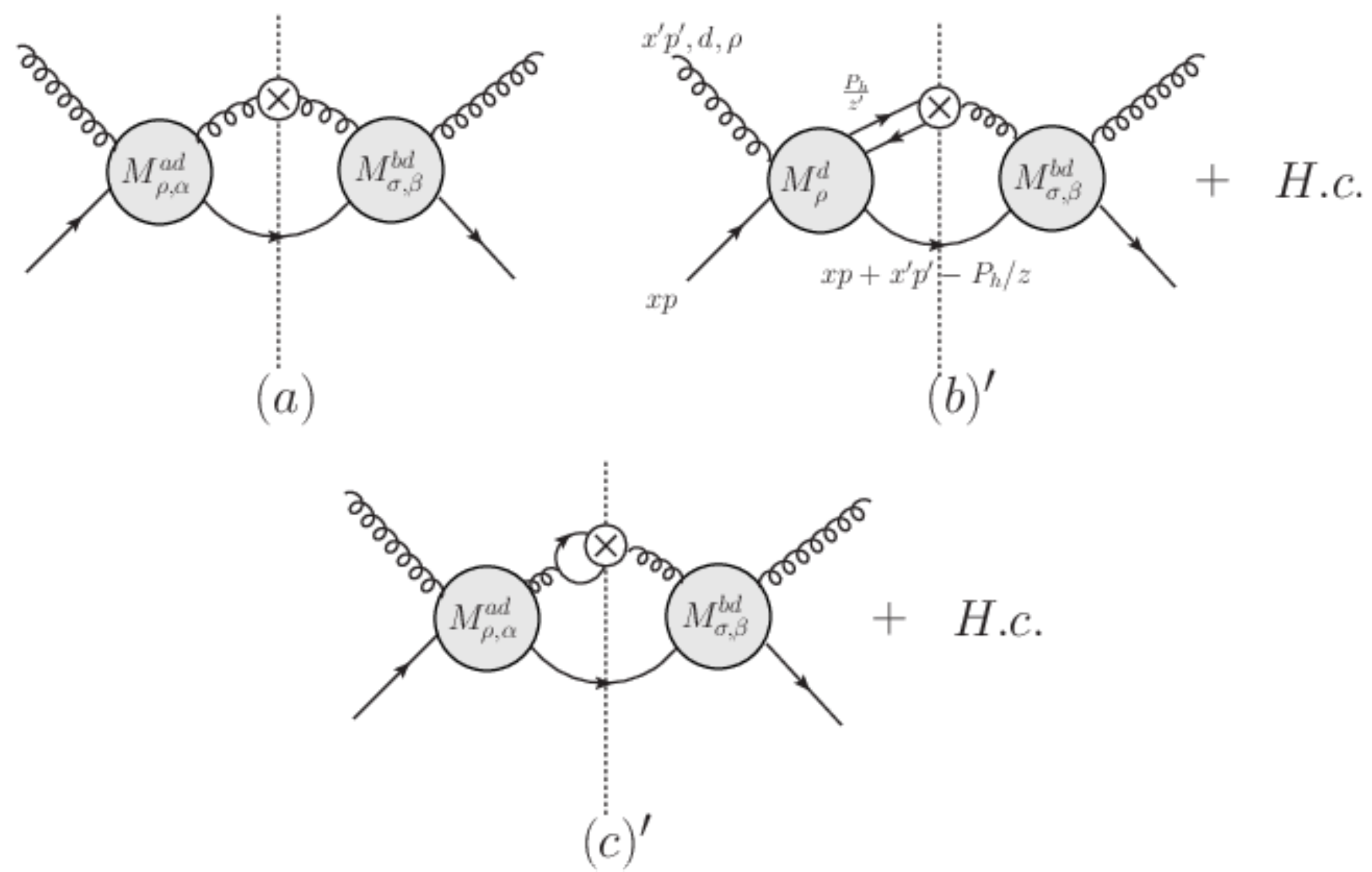}
\caption{Diagrams contributing to $W^{({\rm iii})}_{\rho\sigma}$.  
$M^{d}_{\rho}$ represents the 
$gq\to q\bar{q}q$ scattering amplitude. }
\label{part_fig2}
\end{figure}

Next we show $W^{({\rm ii})}_{\rho\sigma}$ shown 
in Fig. \ref{part_fig2} satisfies (\ref{cgaugeinv}).  
Fig. \ref{part_fig2} (b)' is defined to include the contribution of the type Fig. \ref{part_fig2} (c)'
which is not a part of the hard cross section for a quark-antiquark-gluon contribution.  
Accordingly the corresponding hard cross section is
written as
\begin{eqnarray}
W^{({\rm iii})}_{\rho\sigma}=W^{(a)}_{\rho\sigma}+
\left(W^{(b)'L}_{\rho\sigma}-W^{(c)'L}_{\rho\sigma}\right)+
\left(W^{(b)'R}_{\rho\sigma}-W^{(c)'R}_{\rho\sigma}\right).  
\label{part2}
\end{eqnarray}

\begin{figure}[h]
\centering
\includegraphics[width=8cm]{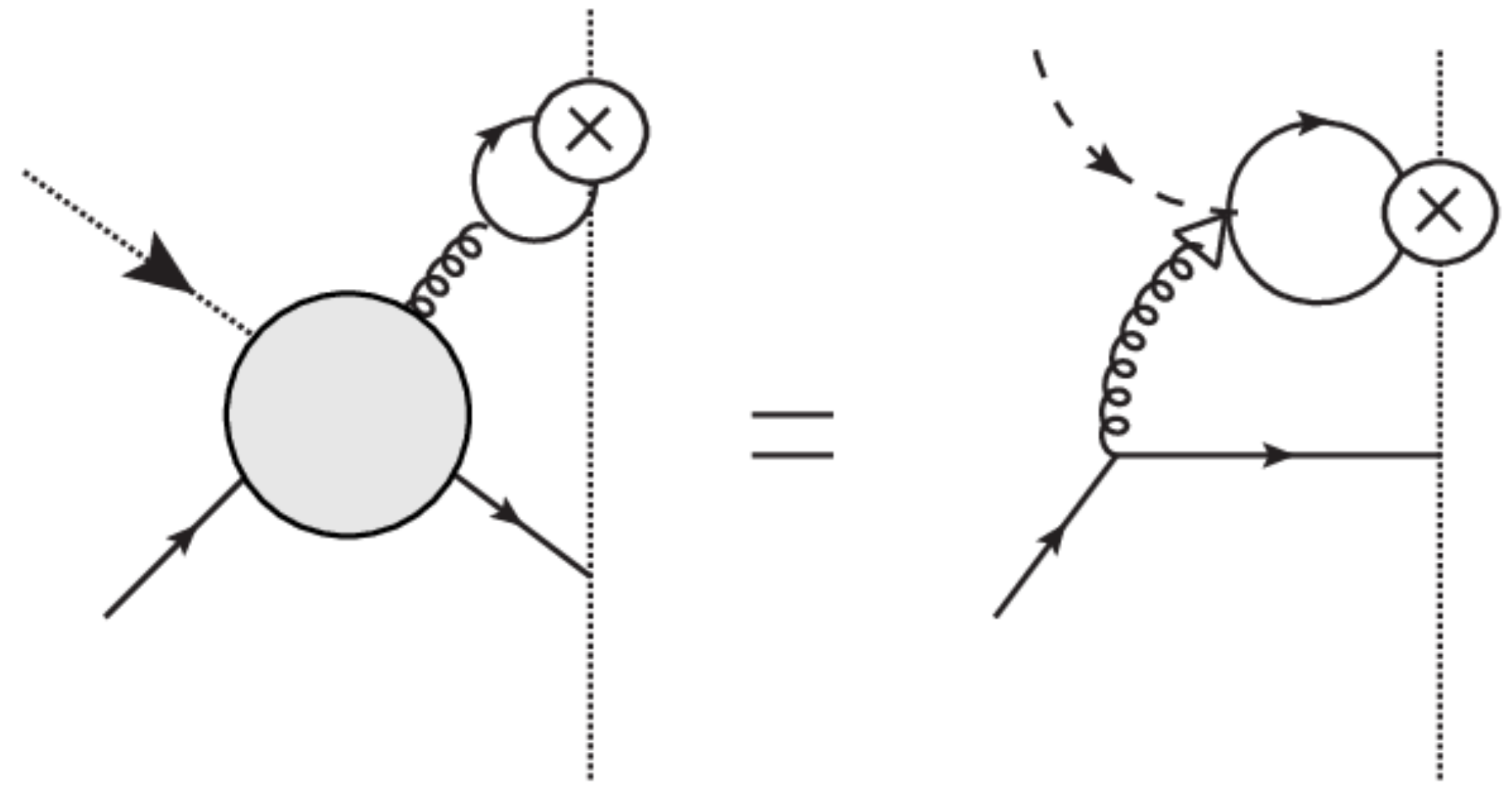}
\caption{Diagrammatic representation of 
$x'p'^{\rho} W^{(c)'L}_{\rho\sigma}$. }
\label{c'_part}
\end{figure}

\noindent
Taking into account of the relation (\ref{per_van1}), we have
\begin{eqnarray}
x'p'^{\rho} W^{(b)'L}_{\rho\sigma}=0,\qquad
x'p'^{\rho} W^{(b)'R}_{\rho\sigma}=0,\qquad
x'p'^{\rho} W^{(c)'R}_{\rho\sigma}=0.
\end{eqnarray}
Therefore we obtain
\begin{eqnarray}
x'p'^\rho W^{({\rm iii})}_{\rho\sigma}=x'p'^{\rho} W^{(a)}_{\rho\sigma} -   x'p'^{\rho} W^{(c)'L}_{\rho\sigma}.
\label{remnant2}
\end{eqnarray}
The second term in the RHS of this equation can be written as
\begin{eqnarray}
x'p'^{\rho} W^{(c)'L}_{\rho\sigma}&\sim&-\frac{1}{z}\frac{2}{C_F}
\epsilon^{\beta P_h w S_
\perp}\frac{F^{ade}}{(x'p'-P_h/z)^2}\times{\rm Tr}\left[\Pxslash\Phslash\,\,\pslash \,T^e
M^{\star, ad}_{\sigma, \beta}\right]\widetilde{D}_{FT},
\label{c'_calc}
\end{eqnarray}
which is diagrammatically written as Fig. \ref{c'_part}.  
We again find the coefficient of $(2/C_F)\widetilde{D}_{FT}$ in (\ref{c'_calc}) 
is equal to the coefficient 
of $\Delta \hat{G}_{3\bar{T}}(z)/z|_{\rm 3-gluon}$
in (\ref{a_calc}),
which shows (\ref{remnant2}) vanishes owing to the EOM relation (\ref{FFDFodd1}).  

\subsection{Contribution from kinematical FFs: $W^{({\rm iii})}_{\rho\sigma}$}
Using the explicit forms of the intrinsic and kinematical FFs in (\ref{gFraI}) and (\ref{gFraK}), 
$W^{({\rm iii})}_{\rho\sigma}(xp,x'p',P_h/z)$ can be written as
\beq
&&W^{({\rm iii})}_{\rho\sigma}(xp,x'p',P_h/z)={M_h\over 2}{\rm Tr}\left[
\left(\epsilon^{P_hwS_\perp\mu}w^\nu + \epsilon^{P_hwS_\perp\nu}w^\mu\right)
S_{\mu\nu,\rho\sigma}({P_h/ z})
\right]
z\,\left( \widehat{G}_T^{(1)}(z)+ \Delta\widehat{H}_T^{(1)}(z)\right)\nn\\
&&\qquad\qquad
-{M_h\over 2}{\rm Tr}\left[
g_\perp^{\mu\nu}\epsilon^{P_hwS_\perp\lambda}
\left.{\partial S_{\mu\nu,\rho\sigma}(k)\over \partial k^\lambda}\right|_{c.l.}\right]
\widehat{G}_T^{(1)}(z)
\nn\\
&&\qquad\qquad-{M_h\over 8} {\rm Tr}\left[
\left( \epsilon^{P_hwS_\perp\{\mu}g_\perp^{\nu\}\lambda}+
\epsilon^{P_hw\lambda\{\mu}S_\perp^{\nu\}}
\right) \left.{\partial S_{\mu\nu,\rho\sigma}(k)\over \partial k^\lambda}\right|_{c.l.}\right]
\Delta \widehat{H}_T^{(1)}(z).
\label{B31}
\eeq
with $g_\perp^{\mu\nu}=g^{\mu\nu}-P_h^\mu w^\nu -P_h^\nu w^\mu$.
To prove $x'p'^\rho W^{(iii)}_{\rho\sigma}(xp,x'p',P_h/z)=0$, one needs to show
the coefficients of $\widehat{G}_T^{(1)}(z)$ and $\Delta \widehat{H}_T^{(1)}(z)$ in (\ref{B31})
satisfy this property.  
To this end we first note that one can set $k$ on-shell, 
$k=(k^+,\, k^-,\,\vec{k}_\perp)=(k^+, {\vec{k}_\perp^2\over 2k^+},\vec{k}_\perp)$,  
in $S_{\mu\nu,\rho\sigma}(k)$ by regarding $k^-$ as a dependent variable, 
since we take
the collinear limit $k\to P_h/z$ after taking the derivative.
We also introduce the following tensors for an on-shell $k$:
\beq
&&A_1^{\mu\nu\lambda}(k)=\epsilon^{P_h w S_\perp \lambda} g_\perp^{\mu\nu}(k),\\
&&A_2^{\mu\nu\lambda}(k)={1\over k\cdot w}\left(
\epsilon^{kwS_\perp\mu}g_\perp^{\nu\lambda}(k)+
\epsilon^{kwS_\perp\nu}g_\perp^{\mu\lambda}(k)+ 
\epsilon^{kw\lambda\mu}S_\perp^{\nu}(k)+
\epsilon^{kw\lambda\nu}S_\perp^{\mu}(k)\right),
\eeq
where
\beq
g_\perp^{\mu\nu}(k) = g^{\mu\nu} - {k^\mu w^\nu + k^\nu w^\mu \over k\cdot w},\qquad
S_\perp^\mu (k) = S_\perp^\mu -{k\cdot S_\perp \over k\cdot w} w^\mu.
\eeq
Since $g_\perp^{\mu\nu}(k)k_\mu=0$ and $S_\perp^\mu (k) k_\mu=0$, 
$A_{1,2}^{\mu\nu\lambda}(k) k_\mu=A_{1,2}^{\mu\nu\lambda}(k) k_\nu=0$.  
$A_{1,2}^{\mu\nu\lambda}(k)$ also satisfy the following relations: 
\beq
&&
A_1^{\mu\nu\lambda}(P_h/z)= \epsilon^{P_h w S_\perp \lambda} g_\perp^{\mu\nu},\\
&& A_2^{\mu\nu\lambda}(P_h/z)= \epsilon^{P_hwS_\perp\{\mu}g_\perp^{\nu\}\lambda}+
\epsilon^{P_hw\lambda\{\mu}S_\perp^{\nu\}},\\
&&\left.{\partial A_1^{\mu\nu\lambda}(k) \over \partial k^\lambda}\right|_{k=P_h/z} =
-z\left( \epsilon^{P_h w S_\perp \mu} w^\nu+ \epsilon^{P_h w S_\perp \nu} w^\mu \right),\\
&&\left.{\partial A_2^{\mu\nu\lambda}(k) \over \partial k^\lambda}\right|_{k=P_h/z} =
-4z\left( \epsilon^{P_h w S_\perp \mu} w^\nu+ \epsilon^{P_h w S_\perp \nu} w^\mu \right).
\eeq
From these relations, we have
\beq
A_{1,2}^{\mu\nu\lambda}(k) S_{\mu\nu,\rho\sigma}(k) x'p'^\rho=0.
\label{B32}
\eeq
This is because the Lorentz indices $\mu$ and $\nu$ provide
physical polarizations for the on-shell gluon with momentum $k$.
By taking the derivative of (\ref{B32}) with respect to $k^\lambda$ and 
setting $k\to P_h/z$, it is easy to see
$x'p'^\rho W^{({\rm iii})}_{\rho\sigma}(xp,x'p',P_h/z)=
x'p'^\sigma W^{({\rm iii})}_{\rho\sigma}(xp,x'p',P_h/z)=0$, which completes
the proof.

\section{Separation of the 3-body partonic cross sections 
based on $z'$-dependence}\label{z'_dep}
\begin{figure}[h]
\centering
\includegraphics[width=6cm]{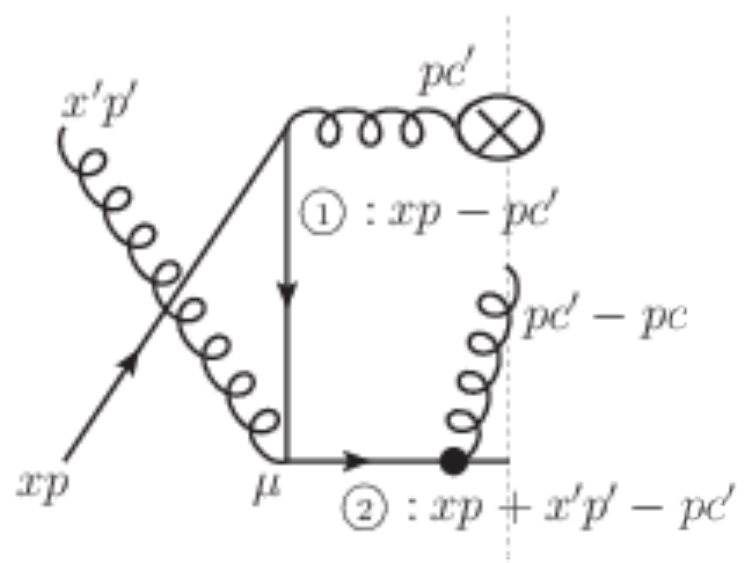}
\caption{An example of diagrams which contribute
to $qg\to gq$ channel.  Notation 
$pc\equiv P_h/z$ and $pc'\equiv P_h/z'$ are used. }
\label{z_depe}
\end{figure}

Here we discuss separation of the 3-body cross section 
(\ref{PPgfragForma3}) based on $z'$-dependence.  
Inserting (\ref{gFraDA}) into  (\ref{PPgfragForma}), we write the cross section 
for $\Im\Nhat_{i}$ as
\beq
&&E_{h}\frac{d\sigma(p,p',P_h;S_\perp)}{d^3P_h}\sim
\int\left(\frac{1}{z}\right) z^2\int d\left(\frac{1}{z'}\right) \delta((xp+x'p'-P_h/z)^2)\nn\\ 
&&\qquad\qquad\qquad\times\frac{z'}{z}P\left(\frac{1}{1/z-1/z'}\right) 
\sum_{i=1}^3 \Im\Nhat_{i} \hat{\sigma}(\ozd,\oz),
\eeq
where $\hat{\sigma}(\ozd,\oz)$ is a partonic hard cross section defined by
\beq
&&\Re\left[-i{\Omega^\mu_\alpha}{\Omega^\nu_\beta}{\Omega^\lambda_\gamma}
\frac{if^{abc}}{N}
\hat{\Gamma}_{FA}^{\alpha\beta\gamma}\left(\ozd,\oz\right)
S^L_{\mu\nu\lambda,abc}(z',z))\right]\nn\\
&&\qquad\qquad\qquad=\delta((xp+x'p'-P_h/z)^2)\sum_{i=1}^3 \Im\Nhat_{i} 
\hat{\sigma}(\ozd,\oz), 
\eeq
to which 
diagrams shown in Fig. \ref{z_depe} contribute to  in the 
$qg\to gq$ channel. 
\\
The amplitude in Fig. \ref{z_depe} has two propagators with $z'$-dependence
(\textcircled{\small 1} and \textcircled{\small 2}).  
For these two propagators, $z'$-dependence is written as follows: 
\beq
\mbox{\textcircled{\footnotesize 1}: }\frac{x\pslash-\Phslash/z'}{(xp-P_h/z')^2}
=\underbrace{-\frac{\Phslash/z}{\hat{t}}}_{\mbox{\textcircled{\small 1}}-{\rm I}}+\frac{z'}{z}\underbrace{\frac{x\pslash}{\hat{t}}}_{\mbox{\textcircled{\small 1}}-{\rm II}}, 
\eeq
and
\beq
\mbox{\textcircled{\footnotesize 2}: }\frac{x\pslash+x'\pslash'-\Phslash/z'}{(xp+x'p'-P_h/z')^2}=\underbrace{\frac{\Phslash/z}{\hat{s}}}_{\mbox{\textcircled{\small 2}}-{\rm I}}+\frac{1}{1-z/z'}\underbrace{\frac{x\pslash+x'\pslash'-\Phslash/z}{\hat{s}}}_{\mbox{\textcircled{\small 2}}-{\rm II}}.
\eeq
In these two equations, we call each term in the RHS 
\textcircled{\small 1}-{\rm I}, \textcircled{\small 1}-{\rm II}, \textcircled{\small 2}-{\rm I} and\textcircled{\small 2}-{\rm II} as shown in the figure.  
Then, depending on the combination of the propagators, 
one can separate the 
$z'$-dependence of the cross section as
\beq
E_{h}\frac{d\sigma(p,p',P_h;S_\perp)}{d^3P_h}&\sim&
\int\left(\frac{1}{z}\right)z^2 \int d\left(\frac{1}{z'}\right) \delta((xp+x'p'-P_h/z)^2) \nn\\
&\times& \frac{z'}{z}P\left(\frac{1}{1/z-1/z'}\right) \sum_{i=1}^3 \Im\Nhat_{i} \left[\hat{\sigma}_{p1}+\frac{1}{1-z/z'}\hat{\sigma}_{p2}+\frac{z'}{z}\hat{\sigma}_{p3}\right], 
\eeq
where each partonic cross section is defined as
\beq
\hat{\sigma}_{p1}&&\sim \cdots  {\rm Tr}[\cdots (\mbox{\textcircled{\small 2}}-{\rm I})\gamma^{\mu}(\mbox{\textcircled{\small 1}}-{\rm I}) \cdots]  \cdots,\\
\hat{\sigma}_{p2}&&\sim \cdots  {\rm Tr}[\cdots (\mbox{\textcircled{\small 2}}-{\rm II})\gamma^{\mu}(\mbox{\textcircled{\small 1}}-{\rm I}) \cdots] + \cdots {\rm Tr}[\cdots (\mbox{\textcircled{\small 2}}-{\rm II})\gamma^{\mu}(\mbox{\textcircled{\small 1}}-{\rm II}) \cdots]  \cdots,\\
\hat{\sigma}_{p3}&&\sim \cdots  {\rm Tr}[\cdots (\mbox{\textcircled{\small 2}}-{\rm I})\gamma^{\mu}(\mbox{\textcircled{\small 1}}-{\rm II}) \cdots]+ \cdots {\rm Tr}[\cdots (\mbox{\textcircled{\small 2}}-{\rm II})\gamma^{\mu}(\mbox{\textcircled{\small 1}}-{\rm II}) \cdots],  \cdots
\eeq
and we used the relation $z'/z\times1/(1-z/z')=z'/z+1/(1-z/z')$.  

Taking into account of the overall factor
$z'/z\times1/(1/z-1/z')$ and rearranging the $z'$-dependence, one obtains
the cross section in the following form:  
\beq
E_{h}\frac{d\sigma(p,p',P_h;S_\perp)}{d^3P_h}&\sim&
\int\left(\frac{1}{z}\right) z^2\int d\left(\frac{1}{z'}\right) \delta((xp+x'p'-P_h/z)^2)\nn\\
&\times&  \sum_{i=1}^3 \Im\Nhat_{i} 
\left[
\frac{1}{1/z-1/z'}\hat{\sigma}_{1}+\frac{1}{z}\frac{1}{(1/z-1/z')^2}\hat{\sigma}_{2}+{z'}\hat{\sigma}_{3}+\frac{z'^2}{z}\hat{\sigma}_{4}\right], 
\eeq
where
\beq
\hat{\sigma}_{1}&=& \hat{\sigma}_{p1}+\hat{\sigma}_{p2}+\hat{\sigma}_{p3},\\
\hat{\sigma}_{2}&=& \hat{\sigma}_{p2},\\
\hat{\sigma}_{3}&=& \hat{\sigma}_{p1}+\hat{\sigma}_{p2}+\hat{\sigma}_{p3},\\
\hat{\sigma}_{4}&=& \hat{\sigma}_{p3}. 
\eeq
This form is the expression used in (\ref{PPgfragForma3}).  One can thus separate different
$z'$-dependence of the cross section.  

\newpage


\begin{thebibliography}{99}
\bibitem{Bunce:1976yb} 
 G.~Bunce, R.~Handler, R.~March, P.~Martin, L.~Pondrom, M.~Sheaff, K.~J.~Heller and O.~Overseth {\it et al.},
Phys.\ Rev.\ Lett.\ {\bf 36}, 1113 (1976).

\bibitem{Heller:1978ty} 
K.~J.~Heller {\it et al.},
Phys.\ Rev.\ Lett.\ {\bf 41}, 607 (1978)
Erratum: Phys.\ Rev.\ Lett.\ {\bf 45}, 1043 (1980).

\bibitem{Erhan:1979xm} 
S.~Erhan {\it et al.},
Phys.\ Lett.\ {\bf 82B}, 301 (1979).

\bibitem{Heller:1983ia} 
K.~J.~Heller {\it et al.},
Phys.\ Rev.\ Lett.\ {\bf 51}, 2025 (1983).

\bibitem{Lundberg:1989hw} 
B.~Lundberg {\it et al.},
Phys.\ Rev.\ D {\bf 40}, 3557 (1989).

\bibitem{Yuldashev:1990az} 
B.~S.~Yuldashev {\it et al.},
Phys.\ Rev.\ D {\bf 43}, 2792 (1991).

\bibitem{Ramberg:1994tk} 
E.~J.~Ramberg {\it et al.},
Phys.\ Lett.\ B {\bf 338}, 403 (1994).

\bibitem{Fanti:1998px} 
V.~Fanti {\it et al.},
Eur.\ Phys.\ J.\ C {\bf 6}, 265 (1999).

\bibitem{Abt:2006da} 
 I.~Abt {\it et al.} [HERA-B Collaboration],
Phys.\ Lett.\ B {\bf 638}, 415 (2006)
[hep-ex/0603047].

\bibitem{Aaij:2013oxa} 
R.~Aaij {\it et al.} [LHCb Collaboration],
Phys.\ Lett.\ B {\bf 724}, 27 (2013)

\bibitem{ATLAS:2014ona} 
G.~Aad {\it et al.} [ATLAS Collaboration],
Phys.\ Rev.\ D {\bf 91}, 032004 (2015)

\bibitem{Aston:1981em} 
  D.~Aston {\it et al.} 
  Nucl.\ Phys.\ B {\bf 195}, 189 (1982).

\bibitem{Abe:1983jy} 
  K.~Abe {\it et al.} 
  Phys.\ Rev.\ D {\bf 29}, 1877 (1984).

\bibitem{Airapetian:2007mx} 
  A.~Airapetian {\it et al.} [HERMES Collaboration],
  Phys.\ Rev.\ D {\bf 76}, 092008 (2007)

\bibitem{Airapetian:2014tyc} 
  A.~Airapetian {\it et al.} [HERMES Collaboration],
  Phys.\ Rev.\ D {\bf 90}, no. 7, 072007 (2014)

\bibitem{Ackerstaff:1997nh} 
  K.~Ackerstaff {\it et al.} [OPAL Collaboration],
  Eur.\ Phys.\ J.\ C {\bf 2}, 49 (1998)
  [hep-ex/9708027].

\bibitem{Abdesselam:2016nym} 
  A.~Abdesselam {\it et al.} [Belle Collaboration],
  arXiv:1611.06648 [hep-ex].

\bibitem{Adams:1991rw} 
D.~L.~Adams {\it et al.} [E581 and E704 Collaborations],
Phys.\ Lett.\ B {\bf 261}, 201 (1991);

\bibitem{Adams:1991cs} 
D.~L.~Adams {\it et al.} [E704 Collaboration],
Phys.\ Lett.\ B {\bf 264}, 462 (1991).

\bibitem{Adams:2003fx}
  J.~Adams {\it et al.} [STAR Collaboration],
  Phys.\ Rev.\ Lett.\  {\bf 92}, 171801 (2004).

  
\bibitem{Adamczyk:2012xd} 
  L.~Adamczyk {\it et al.}  [STAR Collaboration],
  Phys.\ Rev.\ D {\bf 86}, 051101 (2012).

\bibitem{:2008mi} 
  I.~Arsene {\it et al.}  [BRAHMS Collaboration],
  Phys.\ Rev.\ Lett.\  {\bf 101}, 042001 (2008).
    
\bibitem{Adare:2013ekj} 
  A.~Adare {\it et al.}  [PHENIX Collaboration],
  arXiv:1312.1995 [hep-ex].


\bibitem{Airapetian:2013bim}
A.~Airapetian \textit{et al.} [HERMES],
Phys. Lett. B \textbf{728} (2014), 183-190
doi:10.1016/j.physletb.2013.11.021
[arXiv:1310.5070 [hep-ex]].

\bibitem{Allada:2013nsw}
K.~Allada \textit{et al.} [Jefferson Lab Hall A],
Phys. Rev. C \textbf{89}, no.4, 042201 (2014)
doi:10.1103/PhysRevC.89.042201
[arXiv:1311.1866 [nucl-ex]].


\bibitem{Kane:1978nd} 
  G.~L.~Kane, J.~Pumplin and W.~Repko,
  Phys.\ Rev.\ Lett.\  {\bf 41}, 1689 (1978).


\bibitem{Qiu:1991wg} 
  J.~w.~Qiu and G.~F.~Sterman,
  Nucl.\ Phys.\ B {\bf 378}, 52 (1992).
  doi:10.1016/0550-3213(92)90003-T

\bibitem{Qiu:1998ia} 
  J.-w.~Qiu and G.~F.~Sterman,
  Phys.\ Rev.\ D {\bf 59}, 014004 (1999)
 [hep-ph/9806356].
  
\bibitem{Kanazawa:2000hz} 
  Y.~Kanazawa and Y.~Koike,
  Phys.\ Lett.\ B {\bf 478}, 121 (2000)
  [hep-ph/0001021]; 
%
  Phys.\ Lett.\ B {\bf 490}, 99 (2000)
  [hep-ph/0007272].
  
\bibitem{Ji:2006vf} 
  X.~Ji, J.~w.~Qiu, W.~Vogelsang and F.~Yuan,
  Phys.\ Rev.\ D {\bf 73}, 094017 (2006)
  [hep-ph/0604023].
  
\bibitem{Kouvaris:2006zy}
  C.~Kouvaris, J.~W.~Qiu, W.~Vogelsang and F.~Yuan,
  Phys.\ Rev.\  D {\bf 74}, 114013 (2006)
 [arXiv:hep-ph/0609238].
   
\bibitem{Koike:2007rq} 
  Y.~Koike and K.~Tanaka,
  Phys.\ Rev.\ D {\bf 76}, 011502 (2007)
  [hep-ph/0703169].

\bibitem{Koike:2009ge} 
  Y.~Koike and T.~Tomita,
  Phys.\ Lett.\ B {\bf 675}, 181 (2009)
 [arXiv:0903.1923 [hep-ph]].


\bibitem{Koike:2011b} 
  Y.~Koike and S.~Yoshida,
  Phys.\ Rev.\ D {\bf 84}, 014026 (2011).
  [arXiv:1104.3943[hep-ph]]
  
\bibitem{Koike:2011nx} 
  Y.~Koike and S.~Yoshida,
  Phys.\ Rev.\ D {\bf 85}, 034030 (2012)
  [arXiv:1112.1161 [hep-ph]].
  
\bibitem{Kang:2010zzb} 
  Z.~B.~Kang, F.~Yuan and J.~Zhou,
  Phys.\ Lett.\ B {\bf 691}, 243 (2010)
  [arXiv:1002.0399 [hep-ph]].

\bibitem{Kanazawa:2011er} 
  K.~Kanazawa and Y.~Koike,
  Phys.\ Lett.\ B {\bf 701}, 576 (2011)
  [arXiv:1105.1036 [hep-ph]].
  
\bibitem{Metz:2012ct} 
  A.~Metz and D.~Pitonyak,
  Phys.\ Lett.\ B {\bf 723}, 365 (2013)
  Erratum: Phys.\ Lett.\ B {\bf 762}, 549 (2016)
  [arXiv:1212.5037 [hep-ph]].

   
\bibitem{Beppu:2013uda} 
  H.~Beppu, K.~Kanazawa, Y.~Koike and S.~Yoshida,
  Phys.\ Rev.\ D {\bf 89}, 034029 (2014)
  [arXiv:1312.6862 [hep-ph]].

\bibitem{Kanazawa:2014nea} 
  K.~Kanazawa, Y.~Koike, A.~Metz and D.~Pitonyak,
  Phys.\ Rev.\ D {\bf 91}, no. 1, 014013 (2015)
  [arXiv:1410.3448 [hep-ph]].

\bibitem{Kanazawa:2014dca} 
K.~Kanazawa, Y.~Koike, A.~Metz and D.~Pitonyak,
Phys.\ Rev.\ D {\bf 89}, 111501(R) (2014)

\bibitem{Gamberg:2017gle} 
L.~Gamberg, Z.~B.~Kang, D.~Pitonyak and A.~Prokudin,
Phys.\ Lett.\ B {\bf 770} 242 (2017)

\bibitem{Kanazawa:2001a} 
Y.~Kanazawa and Y.~Koike,
Phys.\ Rev.\ D {\bf 64}, 034019 (2001)

\bibitem{Zhou:2008} 
J.~Zhou, F.~Yuan and Z.~T.~Liang,
 5Phys.\ Rev.\ D {\bf 78}, 114008 (2008)

\bibitem{Koike:2015zya} 
Y.~Koike, K.~Yabe and S.~Yoshida,
Phys.\ Rev.\ D {\bf 92}, 094011 (2015)

\bibitem{Koike:2017fxr} 
  Y.~Koike, A.~Metz, D.~Pitonyak, K.~Yabe and S.~Yoshida,
  Phys.\ Rev.\ D {\bf 95}, no. 11, 114013 (2017)
  doi:10.1103/PhysRevD.95.114013
  [arXiv:1703.09399 [hep-ph]].


\bibitem{Yabe:2019awq} 
  K.~Yabe, Y.~Koike, A.~Metz, D.~Pitonyak and S.~Yoshida,
  JPS Conf.\ Proc.\  {\bf 26}, 021016 (2019).
  doi:10.7566/JPSCP.26.021016

\bibitem{Kenta:2019bxd} 
  Kenta Yabe, Y.~Koike, A.~Metz, D.~Pitonyak and S.~Yoshida,
  PoS SPIN {\bf 2018}, 192 (2019).
  doi:10.22323/1.346.0192
  
\bibitem{Koike:2019zxc}
Y.~Koike, K.~Yabe and S.~Yoshida,
Phys. Rev. D \textbf{101}, no.5, 054017 (2020)

\bibitem{Kanazawa:2014tda} 
K.~Kanazawa, A.~Metz, D.~Pitonyak and M.~Schlegel,
Phys.\ Lett.\ B {\bf 742}, 340 (2015)

\bibitem{Kanazawa:2015jxa} 
K.~Kanazawa, A.~Metz, D.~Pitonyak and M.~Schlegel,
Phys.\ Lett.\ B {\bf 744}, 385 (2015)

\bibitem{Kanazawa:2015ajw} 
K.~Kanazawa, Y.~Koike, A.~Metz, D.~Pitonyak and M.~Schlegel,
Phys.\ Rev.\ D {\bf 93}, 054024 (2016)
  
\bibitem{Mulders:2000sh} 
  P.~J.~Mulders and J.~Rodrigues,
  Phys.\ Rev.\ D {\bf 63}, 094021 (2001)
  doi:10.1103/PhysRevD.63.094021
  [hep-ph/0009343].
  
  
\bibitem{Gamberg:2018fwy} 
  L.~Gamberg, Z.~B.~Kang, D.~Pitonyak, M.~Schlegel and S.~Yoshida,
  JHEP {\bf 1901}, 111 (2019)
  doi:10.1007/JHEP01(2019)111
  [arXiv:1810.08645 [hep-ph]].

\bibitem{Kanazawa:2013uia} 
K.~Kanazawa and Y.~Koike,
Phys.\ Rev.\ D {\bf 88}, 074022 (2013)

\bibitem{Hatta:2013wsa} 
Y.~Hatta, K.~Kanazawa and S.~Yoshida,
Phys.\ Rev.\ D {\bf 88}, 014037 (2013)

\bibitem{MutaQCD}
T. Muta, {\it Foundations of Quantum Chromodynamics} (Third ed., World Scientific, 2010)

\end{thebibliography}
\end{document}